\documentclass[preprint]{elsarticle}
\usepackage{lineno,hyperref}
\modulolinenumbers[5]

\usepackage{amssymb}
\usepackage{rotating} 
\usepackage{subfigure}
\usepackage{makecell}

\usepackage{appendix}
\usepackage{comment}
\usepackage{graphics}
\usepackage{adjustbox}
\usepackage{amsmath}

\usepackage{diagbox, hhline, booktabs}
    \usepackage[table, x11names, svgnames]{xcolor}
  
\usepackage{textcomp}

\usepackage{algorithm}
\usepackage{algpseudocode}
\usepackage{adjustbox}

\usepackage{chngcntr}
\usepackage{mathtools, nccmath}

\usepackage{epstopdf} 
\usepackage{mathtools}
\usepackage{float}

\usepackage{algorithmicx}
\usepackage[tone]{tipa}

\usepackage{graphics}
\usepackage[
n, 
advantage,
operators,
sets,
adversary,
landau,
probability,
notions,
logic ,
ff ,
mm,
primitives,
events ,
complexity ,
oracles ,
asymptotics ,
keys
] {cryptocode}

\usepackage{textcomp}

\newtheorem{theorem}{Theorem}

\newtheorem{definition}{Definition}

\usepackage{booktabs}
\usepackage{longtable}
\usepackage{makecell}

\usepackage{pifont}
\newcommand{\cmark}{\ding{51}}%
\newcommand{\xmark}{\ding{55}}%

\usepackage{caption}
\usepackage{epstopdf}

\usepackage{subfigure}

\usepackage{tablefootnote}
\usepackage{mathrsfs}

\usepackage{multirow}

\usepackage{mathtools}

\journal{Preprint submitted to Journal of LATEX Templates}

\begin{document}

\begin{frontmatter}



\title{A Context-Aware Information-Based Clone Node Attack Detection Scheme in Internet of Things }


\author[1]{Khizar Hameed \corref{mycorrespondingauthor}}
\ead{hameed.khizar@utas.edu.au}
\author[1]{Saurabh Garg}
\ead{saurabh.garg@utas.edu.au}
\author[1]{Muhammad Bilal Amin}
\ead{bilal.amin@utas.edu.au}
\author[1]{Byeong Kang}
\ead{byeong.kang@utas.edu.au}
\author[2]{Abid Khan}
\ead{abk15@aber.ac.uk}

\address[1]{Discipline of ICT, School of Technology, Environments, and Design,  University of Tasmania, Australia}
\address[2]{Department of Computer Science, Aberystwyth University, Wales, United Kingdom}

\cortext[mycorrespondingauthor]{Corresponding author}

\begin{abstract}
The rapidly expanding nature of the Internet of Things (IoT) networks is beginning to attract interest across a range of applications, including smart homes, smart transportation, smart health, and industrial contexts such as smart robotics. This cutting-edge technology enables individuals to track and control their integrated environment in real-time and remotely via a thousand IoT devices comprised of sensors and actuators that actively participate in sensing, processing, storing, and sharing information. Nonetheless, IoT devices are frequently deployed in hostile environments, wherein adversaries attempt to capture and breach them in order to seize control of the entire network. One such example of potentially malicious behaviour is the cloning of IoT devices, in which an attacker can physically capture the devices, obtain some sensitive information, duplicate the devices, and intelligently deploy them in desired locations to conduct various insider attacks. A device cloning attack on IoT networks is a significant security concern since it allows for selective forwarding, sink-hole, black-hole, and warm-hole attacks. To address this issue, this paper provides an efficient scheme for detecting clone node attack on IoT networks that makes use of semantic information about IoT devices known as context information sensed from the deployed environment to locate them securely. We design the location proof mechanism by combining location proofs and batch verification of the extended elliptic curve digital signature technique (ECDSA*) to accelerate the verification process at selected trusted nodes. We demonstrate the security of our proposed scheme and its resilience to secure clone node attack detection by conducting a comprehensive security analysis of our proposed scheme. The performance analysis and experimental results compared with existing studies suggest that our proposed scheme provides a high degree of detection accuracy with minimal detection time and significantly reduces the computation, communication and storage overhead.

\end{abstract}



\begin{keyword}



\texttt{Internet of things} \sep \texttt{clone node attack} \sep \texttt{clone detection} \sep \texttt{replica node detection} \sep \texttt{context-aware information}, \texttt{location proof} \sep \texttt{security analysis}

\end{keyword}

\end{frontmatter}



\section{Introduction}

The Internet of Things (IoT) is an emerging and promising network paradigm, consisting of a large number of devices that provide people and objects with the means to interact, communicate and share data for multiple purposes \cite{al2015internet}. These devices are heterogeneous and are deployed in a versatile environment to gather data and information sent to some managing authorities, i.e. clouds for analysis or improve decision-making \cite{diaz2016state}. For example, IoT-based smart home features a variety of automated devices, such as smart refrigerator, thermostat, doorbells, security alarms, and so on., which allow homeowners to control and manage their homes remotely, and to inform them when suspicious activity occurs in their absence \cite{alaa2017review}. Few other prominent IoT-based applications are smart cities \cite{zanella2014internet}, smart vehicles \cite{guerrero2015integration}, smart health \cite{qadri2020future}, and IoT-based industries \cite{wollschlaeger2017future}.

Apart from providing quality services in several day-to-day routine activities, IoT devices have experienced a large number of attacks ranging from security threats to privacy concerns due to the limitations of their functional capabilities (i.e. computing, storage, power), heterogeneous design, restricted features and availability, and lack of advanced security protocols \cite{yang2017survey}. For example, IoT devices are primarily non-tempered resistant and versatile, so malicious attackers can easily compromise the authentication mechanisms and control highly available devices of the IoT network \cite{frustaci2017evaluating}. One such type of attack on IoT devices is a clone-node attack, also referred to as a device replication attack \cite{parno2005distributed}.

\begin{figure}[!htp]
\centering
\includegraphics[width=12cm, height=8cm]{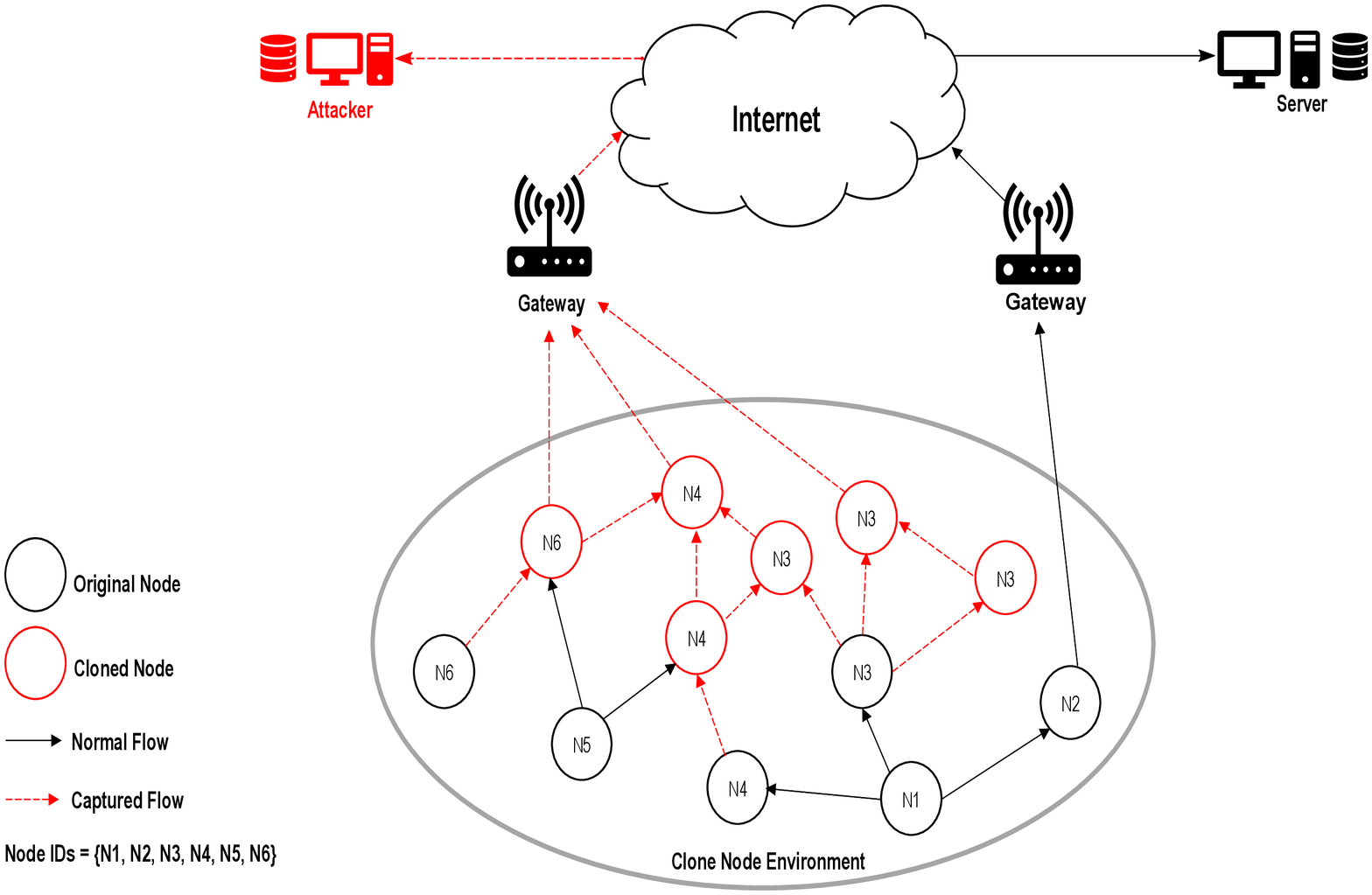}
\caption{A Scenario of Clone Node Attack and its Influence in IoT}
\label{attackscenario}
\end{figure}

In a clone-node attack, the attacker can capture the physical device(s) from the IoT network by extracting their secret credential, including ID, public and private keys. There are several steps to exploiting this vulnerability which involve capturing the physical device, obtaining the secret credentials, modifying its function, and placing it back in the network at some other location \cite{numan2020systematic}. In most cases, the IoT devices designed and assembled by untrusted security partners, which lack updated security firmware and outdated certificates, may lead to a cause of the clone-node attack \cite{becher2006tampering}. Thus, a clone-node attack is regarded as the most severe attack in which clone nodes claim to be legitimate nodes with the same credentials as original nodes. The clone node attack can also be used as a part of other malicious attacks within the IoT network, such as a selective forwarding attack, wormhole attack, and blackhole attack \cite{raza2013svelte}. A scenario of clone node attack and its influence on IoT-based networks is illustrated in Fig. \ref{attackscenario}.

A most straightforward way to mitigate the risk of a clone-node attack on an IoT network is that every device sends an authentication message consisting of its information and all the information of the neighbour devices to a base station using some data forwarding protocol. The base station verifies the message's authenticity by using a secret key shared with the particular node \cite{xing2008real}. However, this forwarding information approach incurs extra communication overhead because of redundant information of the same neighbouring nodes already sent by other nodes to the base station. As a result, this procedure creates additional computing overhead during the verification process. Although witness-based techniques are also used to detect clone node attack, these rely on public-key cryptography, which may not be feasible for low constrained devices. In addition, some clone node attack detection mechanisms rely on secret key sharing, which allows adversaries to intercept, drop some messages, and create their authentication messages \cite{lou2012single}.

An effective clone node attack detection mechanism must accomplish the following objectives to secure IoT devices from adversaries. For example, a clone node detection mechanism must achieve a high \textit{detection rate} of attacks in the system, which can be achieved by matching the traffic patterns or attack patterns with the incoming traffic in order to track the adversary's activities. Implementing a secure database with consistent and up-to-date device-related record and attack patterns can also result in a high detection rate of clone node attack. The \textit{detection time} is also an important objective for developing the cloned node detection system since it facilitates detecting malicious activities performed by adversaries in the shortest amount of time. A robust framework for detecting clone nodes attack on IoT systems must detect attacks in real-time. Along with achieving a high detection rate and a low detection time, one of the main objectives of developing an effective clone node attack detection mechanism is to minimise the \textit{false positive rate}, which occurs when the system incorrectly interprets normal behaviour as an attack, triggering false alarms in the system and causing significant issues. Furthermore, since IoT networks are often composed of heterogeneous devices that are resource-constrained, they are unable to perform computationally intensive operations or store large amounts of data. As a result, when designing the cloned node detection mechanism, one must carefully consider the \textit{computation, storage, and energy limitations} of IoT devices.

To address the shortcomings of existing clone node attack detection mechanisms and achieve the objectives above, we propose a novel and efficient clone node attack detection mechanism in IoT networks that leverage context-aware information of IoT devices. Context information is the semantic information that enables users to comprehend the networking environment and subsequently locate network entities via the relationships of entities with the environment. We develop a location-proof system that works in conjunction with context-aware modalities information to locate IoT devices securely. To achieve computational efficiency in the location proof system (LPS), we used batch verification of ECDSA* (an upgraded version of the elliptic curve digital signature algorithm) to accelerate the verification process of location proofs at selected trusted nodes than verifying them at a single base station. In addition, we conducted an extensive security analysis, outlining the prerequisites and several security needs for ECC and the possibility of several types of attacks on the signatures and hashes used in our proposed scheme. Our simulation results indicated that the proposed scheme is robust towards attack detection in a timely manner and significantly reduces computation, communication, and storage overhead.

To design the context-aware information-based clone node attack detection mechanism in IoT, the following are the main contributions to this paper. 

\begin{itemize}
    \item We emphasise the importance of addressing security issues associated with existing IoT device authentication mechanisms that frequently result in node cloning attacks on IoT networks.
    \item We explore the objectives for designing an efficient clone node attack detection mechanism, with the aim of implementing an efficient context-aware information-based mechanism for detecting clone node attack on IoT networks.
    \item We design a location proof mechanism in conjunction with batch verification of ECDSA* to accelerate the verification process at selected trusted nodes.
    \item We conduct an extensive security analysis, outlining the prerequisites and several security needs for ECC and the possibility of several types of attacks.
    \item We compare our proposed scheme to existing clone node attack detection mechanisms, proving its efficiency in detecting attacks with minimal detection time and highlighting its importance in terms of computation, communication, and storage overhead in IoT networks.

\end{itemize}

The rest of the paper is organised as follows: Section \ref{sec:background} presents the preliminaries as a background to our work and summarise the existing clone node attack detection mechanisms developed for IoT. In section \ref{sec:methodology}, we present a detailed description of our proposed methodology, including network and attacker models, an enhanced elliptic curve digital signature technique, and the selection of trusted nodes. Section \ref{sec:locationproof} discusses the mechanism for localisation employed in the LPS, including the proposed algorithm and execution flow. In section \ref{sec:securityanalysis}, we conduct a security analysis of our proposed method, including multiple ECC security requirements and attacks on hashes and digital signatures. The simulation results and analysis of our proposed scheme are then presented in section \ref{simulations}. Finally, in section \ref{sec:conclusion}, we summarise our work and describe future work.

\section{Background and Related Work} \label{sec:background}

This section describes the following concepts as a preliminary or background to our work, such as batch verification, context-aware systems and LPSs, and then provides an overview of existing state-of-the-art clone-node detection mechanisms and their limitations in the related work sub-section.

\subsection{Preliminaries}

\subsubsection{Batch Verification}
With the proliferation of IoT applications ranging from home to large-scale industrial setups and adopting advanced communication protocols, security has evolved into an integral component of IoT networks. To guarantee the security of IoT devices and the data they transmit, digital signatures are used to verify the authenticity of the devices or messages during communication \cite{alamer2020efficient}. However, since IoT network scalability is critical in most applications, verifying individual IoT device signatures is a time-consuming process that is not recommended in IoT systems. However, by verifying digital signatures in batches, it is possible to reduce verification time significantly. 

Batch Verification is a concept that involves verifying multiple signatures simultaneously in order to minimise the time taken to validate each signature submitted by thousands of sensor nodes in large-scale IoT networks. The advantage of employing the batch verification principle is that it can significantly reduce the computation load and time required for IoT devices by performing multiple digital signatures. As a result, this concept is advantageous in an IoT setting where nodes have limited computing power and run in real-time \cite{kittur2017batch, hu2019autonomous}. Numerous batch verification methods have been proposed for several cryptographic protocols, including the digital signature algorithm (DSA) \cite{naccache1994can} and the rivest-shamir-adleman (RSA) \cite{harn1998batch}. However, these protocols are not appropriate to IoT devices due to their resource scarcity. Recent years have seen a surge of interest in the use of ECDSA in IoT applications, owing to the smaller key and signature sizes needed by ECDSA compared to DSA and RSA. However, the implementation of batch verification of these protocols is not immediately applicable to ECDSA signatures \cite{karati2012batch}.

\subsubsection{Context-Aware Systems}
IoT networks comprise a generous amount of connected devices stacked with a certain degree of intelligence that collects data from their surroundings. These IoT devices are expected to produce many data, which often requires careful identification, interpretation, and analysis. A context-aware system is an exemplary IoT environment where the system is deployed to keep track of surrounding objects and also provide timely feedback to the user, and their related applications \cite{abowd1999towards,perera2013context}. 

Context-aware systems have been implemented in IoT environments to sense the landscaping operational eco-system and react appropriately to the user and application. These systems analyse and translate data produced by IoT devices into contextual information, providing a high level of understanding of semantic data used for machine-integrated setup. 
In such systems, the server transforms the information stored in devices into a higher-level type and analyses it semantically before acting on it to perform further operations. Therefore, context information is the foundation that allows users to understand the networking environment and then locate the network entities by leveraging the relationships between the entities and the environment \cite{de2020context}.  

Context information is used to describe the state of an environment, which is typically composed of the following primary entities: users, an application, a location, or a device. There are essentially two forms of contextual information used in the IoT environments, such as primary and secondary. In the primary context, the information is specified as name, time, location, activity, etc.  Furthermore, the secondary context is any information that can be measured by using the primary context. For example, the sensor deployed in the smart city environment is capable of monitoring vehicle position information along with other information, such as vehicle identification, activity, time, and so on \cite{sezer2017context}.

\subsubsection{Location proof system}
Location-based services are primarily used to support a variety of services in diverse IoT-based applications, such as tracking and monitoring patients, tracking the location of vehicles on the road, and determining the actual position of an individual \cite{zafar2020location}. However, it is critical and a difficult problem in most IoT-based applications to establish trust for the physical presence of any users and its operated devices at a specific location and time using an efficient, accurate, and robust method \cite{krishna2020location}.

To resolve this issue, LPSs are being developed in IoT-based applications, which provide a means for creating and sharing digitally signed context data to provide evidence of a particular user's location at any point instance at a given time. Such systems provide secure proof of a user's location, in which LPS first verify a location of a device at a given time and then grant the additional access needed for a particular device \cite{sun2017efficient}. Additionally, it may also facilitate the process of establishing evidence for a variety of location-based scenarios, including a single location, travel route history, and event summaries such as running and walking \cite{zafar2020mobchain}.

\subsection{Related Work}
Due to the popularity of IoT-based specific applications in recent years, there has been an increase in interest from researchers and academia in finding security solutions for protecting IoT devices and the data shared between them. A clone node attack detection mechanism has proven to be an effective way to protect IoT devices from adversaries. The detection methods attempt to locate the abnormal activities performed by the attacker, which resulted in the creation of duplicate nodes in the network. For example, a unique ID is associated with two different locations. A few other methods used to detect the cloned node attacks are received signal strength, witness finding, random key predistribution, location verification, and calculating distance measurement using the Euclidean distance algorithm \cite{numan2020systematic}. 


\begin{table}[!htp]
\centering
\scriptsize
\caption{A Comparison of Existing Clone Node Attack Detection Mechanisms in Internet of Things }
\begin{adjustbox}{width=1.1\textwidth,center}
\begin{tabular}{|c|c|c|c|c|c|c|c|c|}
\hline
\multirow{0}{*}{\textbf{\thead{Ref}}} & \multirow{0}{*}{\textbf{\thead{clone node attack \\ Detection \\ Technique}}} & \multirow{0}{*}{\textbf{\thead{Network \\ Type}}} & \multirow{0}{*}{\textbf{Strengths}} & \multirow{0}{*}{\textbf{Weaknesses}} & \multicolumn{4}{c|}{\textbf{\thead{
Evaluation \\ Parameters}}} \\ \cline{6-9} 
                  &                   &                   &                   &                   &    \textbf{DP}  & \textbf{CO}       &  \textbf{SO} &\textbf{CMO}     \\ \hline

 \cite{xing2008real}          &   \thead{Create fingerprints \\  of each node \\ by analysing \\ the social features \\ of neighbourhood \\ nodes}               &      Static             &  \thead{Significantly \\ decrease computation,\\ communication, and \\ storage overhead \\ while maintaining a \\ high probability \\ of detection}                 &  \thead{Performs additional \\ processing to \\ generate fingerprints\\ on both \\ sensors and \\ base station}                 &    \cmark   & \cmark       & \cmark &     \cmark \\ \hline

 \cite{parno2005distributed}          &   \thead{Witness \\ finding \\ technique}               &      Static             &         \thead{Detection of \\ globally aware \\ distributed \\node replicas \\ with optimum efficiency \\ characteristics}          &     \thead{High storage \\ overhead }              &      \cmark   &     \cmark  &\cmark &    \cmark \\ \hline
                  
\cite{brooks2007}          &   \thead{Random key \\ predistribution}               &      Static             & \thead{The hypothesis \\ testing approach and \\ bloom filters \\allow the safe and \\ efficient collection \\ of key usage \\ data}                  &        \thead{ Proposed mechanism\\  is limited \\ to detection probability \\ and \\ communication\\ overhead, relying \\ entirely on symmetric \\ cryptography}          &    \cmark   &     \xmark  &\xmark &    \cmark  \\ \hline
        
\cite{choi2007set}            &   \thead{Perform set \\ operations \\ (union and \\ interaction) \\on exclusive \\ subsets in \\ the network}  &       Static            &    \thead{An adversary's \\ exclusive subset is \\ unpredictable due \\ to the tree\\ structure and \\ randomisation}               &       \thead{High computation \\ overhead}            &      \cmark & \cmark&   \cmark    &    \cmark   \\ \hline

\cite{raza2013svelte}          & \thead{Utilize the \\ mapper function \\ to assign \\ each node a \\ unique ID\\  and rank \\ information}                  &  Static                  & \thead{The mapper \\ function efficiently \\ identifies anomalies\\ between node \\ IDs and their ranks}                  &     \thead{An attacker \\ can seize a witness \\ node to isolate\\ itself from the \\ network's other \\ replicas}              &   \cmark    & \cmark       & \cmark  &\xmark    \\ \hline
        
         \cite{alsaedi2017detecting}        &  \thead{Extensive node \\ energy \\ consumption \\ smoothing}                  &     Static              &  \thead{The trust algorithm \\ includes multi-\\stage identity \\ and location verification}                 &    \thead{Restricted the \\ scope of its \\ application to \\ computation and \\ security analysis}               &      \cmark &  \xmark     & \cmark& \cmark      \\ \hline
          \cite{rikli2016lightweight}      & \thead{Trust-profiling \\ of sensor nodes}                  &   Static                &    \thead{Attribution analysis\\  detects all \\ forms of \\ potential malicious \\ nodes with \\ contact behavior-\\based data}               &    \thead{Introduces   an \\ additional \\ computational  \\ overhead for large \\ networks  with \\ hundreds  of  nodes}               & \cmark      &  \cmark     & \xmark     & \xmark \\ \hline
         \cite{shanmugam2020two}       &  \thead{Fingerprint-based \\ zero-knowledge \\ mechanism}                 &    Static               &  \thead{The fingerprint algorithm \\ increases the \\ detection rate \\ with minimal time}                 &    \thead{Increase the \\ computation and \\ communication \\ overhead}               &   \cmark    &\xmark    &\xmark   & \xmark      \\ \hline

     \cite{yu2013localized}   &  \thead{Localised  \\ algorithms \\ (XED, EDD)}                 &     Mobile              & \thead{Localized detection, \\ Efficiency and \\ effectiveness, Network-wide \\ synchronization and \\ revocation \\ avoidance}                   &   \thead{High computation \\ overhead}                &     \cmark  &    \cmark   &    \cmark & \cmark \\ \hline

 \cite{zhou2016improved}              &  \thead{ Sequential tests of \\ statistical hypotheses}                 &   Mobile                &      \thead{Composed of \\ random and \\ sequential \\ measures decreases the\\ false-positive \\ and false-negative \\ instances}             &    \thead{Increases computation \\ and storage \\ on the witness node \\ for detecting \\ cloned devices}               &   \cmark    & \cmark      & \cmark    & \cmark \\ \hline
           \cite{lee2018mdsclone}       &   Multidimensional scaling                &    \thead{Both}    
           
           & \thead{Requires zero \\ geographical location \\ information, \\ improving the entire  \\ detection process }
                           &   \thead{Restricted in \\ scope of \\ achieving security}                & \cmark      & \cmark      & \cmark & \cmark    \\ \hline
\end{tabular}
    
\end{adjustbox}
 {\raggedright \textbf{DP} =  Detection Probability, \textbf{CO} = Computation Overhead, \textbf{SO} = Storage Overhead,  \textbf{CMO} = Communication Overhead \par}
\label{Tab:comparison}
\end{table}

Each detection mechanism is intended to detect clone node attack in the shortest amount of time with the least amount of damage; however, this is not a simple task due to the legal identity of nodes and other environmental factors such as location. In general, detection mechanisms according to their detection ability are categorised into two different position patterns in the deployed network: static and mobile. Since most nodes deployed in an IoT setting adhere to a static position pattern due to their fixed locations, it becomes easier to detect node replicas or clones by matching their identities. On the other hand, in comparison to static position patterns, detecting clone nodes in a mobility pattern is more difficult because nodes are not fixed in one location and often move in the network; thus, even if identity information is matched, it is difficult to conclude that a clone of the ID is found in another location \cite{shaukat2014node}.

In static networks, a commonly used technique for detecting cloned node attack is to compare the information of all neighbouring nodes, in which each node compares its information to the information of all neighbouring nodes and then determines whether or not there is any inconsistency in the stored information \cite{xing2008real}. Witness finding is another widely used technique for detecting clone nodes attack in static networks. In particular, the idea of using the witness finding technique is to find the existence of clone's nodes and decide what proportion of nodes are cloned based on their positions being conflicted with others in the network. Parno et al. \cite{parno2005distributed} suggested a witness-based scheme for detecting replica nodes in which several witness nodes were randomly chosen or selected along the network's forwarding path. Thus, each sensor node kept track of the identity of neighbouring nodes and the location of several witness nodes. However, the scheme is inefficient for the majority of sensor nodes with limited memory since it is based on public-key cryptography, which allows each sensor to store the public keys of the other nodes, increasing the memory overhead. 

Brooks et al. \cite{brooks2007} employ the principle of random key predistribution to detect clone attacks in sensor networks, in which the keys accessible on clone nodes and their maximum usage for authentication determine the attacker's presence in the network. The process recovered from a cloning attack by eliminating ties using cloned keys upon identification of clone nodes. Choi et al. \cite{choi2007set} proposed a clone node attack detection mechanism for sensor networks to reduce computation and storage overhead by using various set operations such as union and interactions. The main idea of the proposed scheme is to detect clones by calculating set functions (intersection and union) on exclusive subsets in the network in order to securely define additional exclusive unit subsets across neighbours. Nonetheless, this method allows an attacker to obtain the secret information of the sensor nodes, which can then be used to conduct network-wide insider attacks.

Raza et al. \cite{raza2013svelte} suggested a clone node attack detection protocol for the 6LoWPAN networks called SVELTE to detect unusual behaviours, such as cloning and selective forwarding. In this protocol, each node, including its parents and neighbours, is allocated with a unique ID and the rank information using a mapper function called the 6LoWPAN Mapper. Later, the mapper feature employed by the witness node can identify anomalies between the node IDs and their ranks. However, if an attacker can capture a witness node, this information is shielded from the other replicas in the network. Alsaedi et al. \cite{alsaedi2017detecting} developed a multi-level-based cloning mechanism to verify the identity and positions of nodes using the exponential smoothing trust algorithm. In this process, the energy consumption of the nodes suggested the difference between the legitimate nodes and the cloned nodes, which means that nodes with higher energy consumption in the network are considered to be cloned nodes. However, this approach significantly increased the energy consumption of routine workload due to the additional computation required to differentiate the cloned and original nodes.

Rikli et al. \cite{rikli2016lightweight} proposed a trust-profiling based mechanism that detects cloned nodes attack in wireless sensor networks. In this method, trust values called threshold values are determined for the subsequent adjacent nodes, and then these values are used to decide whether the observed values are less than the calculated threshold values. This method, however, introduces an additional computational overhead for large networks with hundreds of nodes, as each sensor node determines the threshold value for their neighbouring nodes and then compares it to the threshold values of others for detection purposes. Shanmugam et al. \cite{shanmugam2020two} proposed a method for detecting clone nodes in an IoT-enabled smart cities environment using a fingerprint-based zero-knowledge mechanism for two-level authentication of sensor devices. The base station calculates the unique fingerprint for each node using information about its immediate surroundings, which is defined by a superimposed s-distinct code matrix. Although this work claims a high detection rate by comparing the fingerprints of each device to the stored information at the base station, this method increases the computation overhead on the base station if a large number of sensor nodes need to verify their fingerprints simultaneously. Additionally, this work restricts the sensor device pattern to a static network, increasing the communication overhead between cluster nodes and base station.

While defending against node replication attacks is critical, very few solutions have been proposed in mobile networks compared to the comprehensive research on defending against node replication attacks in static networks. A limited number of techniques are used to protect mobile sensor networks against clone node attack, including localised algorithms, witness finding, and multidimensional scaling.

For instance, Yu et al. \cite{yu2013localized} suggest a similar method based on localisation algorithms (e.g. XED and EDD), in which each node in the network communicates with only its one-hop neighbours, as opposed to a distributed algorithm, which only implies that nodes accomplish the job independently of the base station. With this method, each node can enforce network-wide revocation of the clone nodes without overwhelming the entire network with revocation messages. Similarly to the witness finding strategy in static networks, Zhou et al. \cite{zhou2016improved} proposed a distributed and management technique for detecting mobile replicas that tolerates node failures by forwarding sensor node location claims to collect samples only when the relevant witnesses meet. Sequential tests based on statistical hypotheses are used to detect the cloned node in accordance with the witness nodes, significantly reducing the routeing overhead and false positive/negative rate for detection; however, this technique imposes additional computation and storage overhead on the witness node acting as the main node, collecting samples and forwarding information to the base station for detection of clone nodes. Lee et al. \cite{lee2018mdsclone} suggested a multidimensional scaling-based approach for detecting clone nodes in IoT networks that works for both static and mobile networks. This approach creates a network map by using the nodes' relative neighbourhood-distance information. This method generates the network map using the relative neighbourhood-distance information for the nodes and distributing the total computational load of such information across multiple base stations with increased computational capacity. Since this approach aims to achieve a high detection rate for clone nodes in the network, it is restricted in the scope of computational, communication, and storage overhead for IoT devices and base stations.

To overcome the limitations of existing solutions for detecting clone node attack, such as the most use of static networks and a high detection rate, and the computational and communication overhead on the network as compared in Table \ref{Tab:comparison}, we propose an efficient scheme that makes effective use of the context information of IoT devices to detect clone node attack. We used the LPS concept to determine the exact location of IoT devices in a mobile network environment. To achieve computational efficiency, we use the batch verification principle of ECDSA* to verify the signatures of IoT devices in less time, intending to achieve a high detection rate with low computation overhead.

\section{Proposed Methodology}\label{sec:methodology}
This section describes our proposed methodology for detecting clone node attack on IoT networks using sensed context-aware information and location-based services provided by an LPS. First, we create a network model of our proposed scheme that provides an overview of the cloned node detection mechanism for an IoT network composed of different entities such as the original node, clone node, and gateway node, as well as an explanation of their working mechanism and the assumptions underlying them. In the attacker model, we outline the attacker's assumptions and capabilities for carrying out malicious actions in an IoT network. Furthermore, we summarise the batch verification protocol for ECDSA*, which aims to aid our proposed detection scheme with its underlying algorithms. 

\subsection{Network Model}

\begin{figure}[!htp]
\centering
\includegraphics[width=12cm, height=8cm]{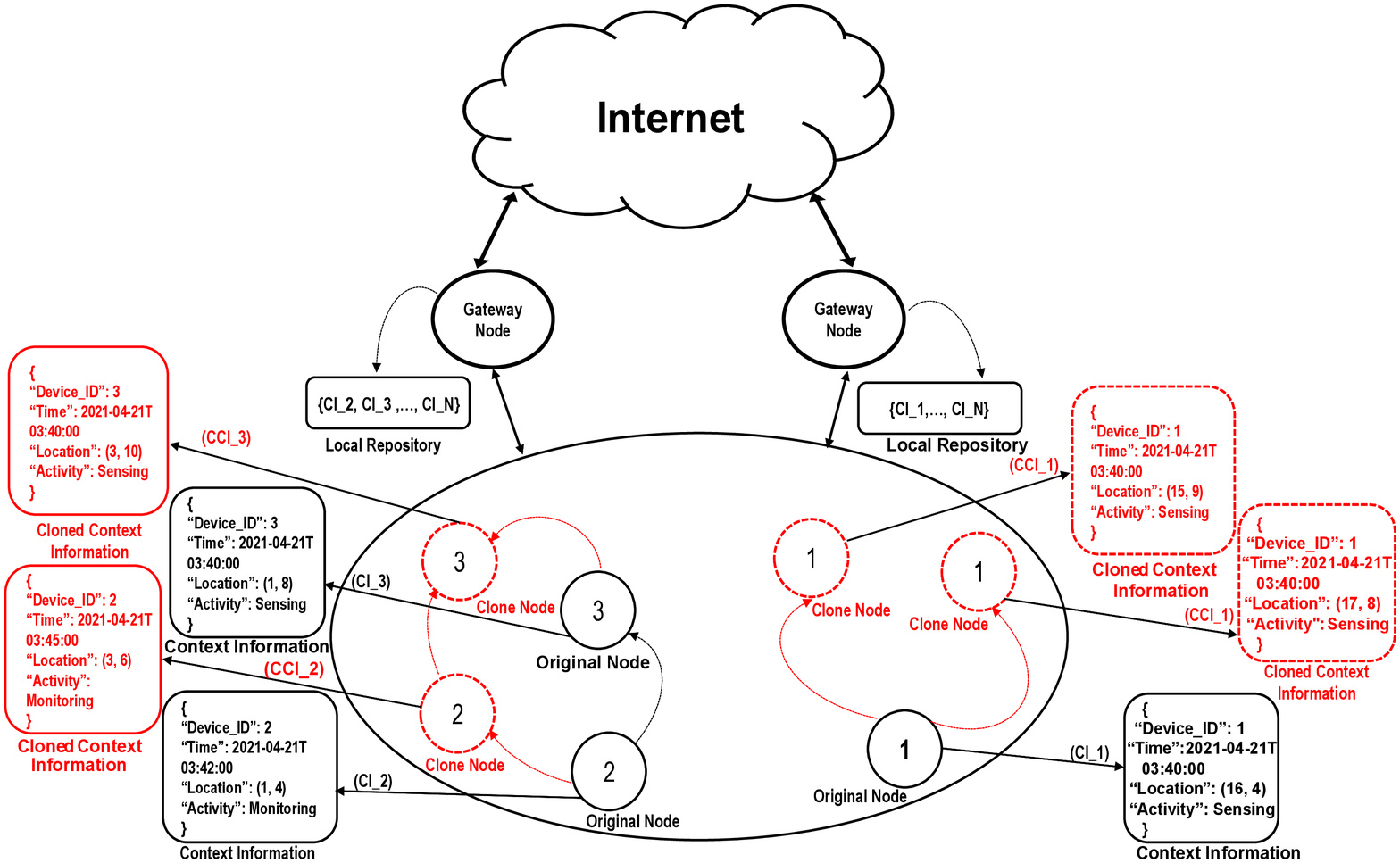}
\caption{Network Model}
\label{networkmodel}
\end{figure}

We consider an IoT network composed of a large number of heterogeneous resource-constrained IoT devices connected to gateway nodes in order to facilitate their interaction and provision of a variety of services. Furthermore, IoT devices adopt a spatial mobility pattern, allowing them to move freely within a specific geographic region. The primary objective of the proposed scheme is to detect clone node attack in IoT networks efficiently by leveraging context-aware information sensed by IoT devices and to achieve a high detection rate with minimum storage and communication overhead. Fig. \ref{networkmodel} depicts the network model of the proposed context-aware clone node attack detection mechanism in the IoT environment.

The network model is represented as an undirected graph \textit{G} = (\textit{V}, \textit{E}), where \textit{V} and \textit{E} denote a set of nodes and edges, respectively. We employ a connected graph model in which an edge exists between nodes \textit{u} and \textit{v}, (\textit{u}, \textit{v}) $\in$ \textit{E}, if the Gaussian distribution between $\textit{u}$ and $\textit{v}$ such as $|\textbf{u} - \textit{v}|$ $\leq$ 1. In our proposed network model, we have the following communicating components: original nodes, clone nodes, and gateway nodes. The following are the specifics for each component:

\subsubsection{Original Nodes}
In our network model, an original node is referred to as a ``IoT device''. An IoT network is described as a collection of heterogeneous, resource-constrained IoT devices equipped with sensors and actuators that allow data transmission to other nodes and communication through the internet. Each node in the network exhibits mobile behaviour distributed geographically using a random way point (RWP) mobility model. The criteria for choosing this model allow devices to move around and provide appropriate patterns to geographic regions used for most real-time applications. To minimise the computational complexity of the mobility pattern, each node in the network can have a maximum of ``\textit{p}'' neighbours, where ``\textit{p}'' belongs to the interval \{1, N-1\}, and N is the total number of nodes in the network.

Further, each node is associated with a specific piece of information known as context-information gathered from the deployed environment. The context information is often expressed as semantic information used to uniquely identify objects (e.g., IoT devices) about the deployed environment and is easily comprehended and interpreted by humans. The following parameters pertain to the context information (\textit{CI}) sensed by the original nodes in our proposed network model:

\begin{itemize}
    \item \textbf{ID:}  \textit{ID}  is a unique identifier that is guaranteed to be distinct from all other identifiers used for objects and a particular purpose. It is one of the important context information parameters associated with a given collection of nodes. Each node has a unique identity ranging from 1 to \textit{n}, which is defined as a set of nodes \{$N_{1}$, $N_{2}$, $N_{3}$, \dots, $N_{n}$\}.

    \item \textbf{Time:} In an IoT context, each device generates data observations that are followed by a timestamp. This timestamp is critical in the analysis section because it defines when data is collected and allows for statistical and time-series analysis. There is no single format that imposes data and time serialisation requirements for data collection. However, we used the \textit{ISO8601} format for DateTime representation in our proposed model, which is defined as \textit{YYYY-MM-DDTHH:mm:ss.sssZ}.
    
    \item \textbf{Location:} The data obtained by IoT devices must be attributed to the device's relative location at the time of collection. Indeed, when location data is paired with timestamps, companies can determine when and where anything is or was. As a result, we defined graph \textit{G} = (\textit{V}, \textit{E}) as two-dimensional, which means that \textit{G} is a Euclidean graph in which each node has a coordinate ($x_{i}$, $y_{i}$) in a two-dimensional space. The coordinates ($x_{i}$, $y_{i}$) indicate the position of node \textit{i} within the defined object plane.
    
    \item \textbf{Activity:} Since IoT devices are equipped with sensors and actuators, they can perform a wide range of activities over the network, including data sensing, computation, temperature control, and data transmission. Additionally, some powerful IoT devices can perform a limited range of data filtering and analysis functions.
\end{itemize}

Each original node is equipped with a pair of keys: a private key (e.g., $K_{pr}$) that is kept secret and a public key (e.g., $K_{pb}$) that is accessible to everyone. The private key $K_{pr}$ is used to sign the context information as digital content in our proposed scheme. To identify itself as the issuer of the context information, the IoT device employs its own secretly private key.

\subsubsection{Clone Node}
A clone node is a replica of the original node in the network model that the attacker physically captures to control a more significant portion of the setup and perform malicious actions over it. Several steps are involved in exploiting this vulnerability, including capturing the physical device, extracting the hidden credentials, modifying its function, and reinstating it into the network at a different location. In our network model, the attacker can create clone nodes by capturing and copying their context information, which is referred to as captured context information (\textit{CCI}).
\subsubsection{Gateway Node}
A gateway is a device that acts as an interface between IoT devices and other systems, such as the cloud. IoT gateways can be physical or virtual devices that collect data from IoT devices and send it to the cloud for processing and storage. Since IoT devices cannot communicate directly over the internet, they are typically connected via a gateway.

In our proposed network model, a gateway node has many responsibilities such as (i) provide the interaction between IoT devices and the external server, (ii) ensure the verification process of location proofs signatures (iii) it possesses the keys material, such as the public key used for signature verification, for all deployed IoT devices, as well as his own key pair, e.g., public and private keys (iv) it keeps track of all deployed IoT devices and their associated context information, referred to as \{$CI_{1}$, $CI_{2}$, $CI_{3}$, \dots, $CI_{N}$\}.

\subsubsection{Assumptions about the Network Model}
\begin{itemize}
    \item The Gateway node is a trusted third party which provides the secure interaction between IoT devices and system.
    \item The Gateway node maintains its role as the central trusted authority.
    \item The communication medium utilises a symmetric routing pattern, which refers to the same path taken by data movement between original nodes and gateway nodes and vice-versa.
\end{itemize}

\subsection{Attacker Model}
In our attacker model, we consider an environment in which an attacker can obtain physical parameters from IoT devices, copy them and replicate or clone legitimate nodes in order to attack the IoT network. Since cloned nodes have hidden credentials for authentication and encryption purposes, stolen credentials can be used to interrupt network operations and launch numerous attacks inside the network.

\begin{itemize}
    \item The \textit{attacker's capability} is limited to the extent that it can compromise a small number of nodes that is less than the total number of original nodes. 
    \item An attacker has \textit{complete control} over the compromised nodes, which are limited in number.
    \item An attacker with malicious intent is capable of \textit{dropping or misdirecting information} before forwarding it to the base station.
    \item An attacker can place IoT devices in \textit{strategic locations}, such as redirecting traffic to a specific server.
    \item An attacker may \textit{manipulate the detection mechanism} in order to remain undetected in the detection scenario.
    \item Cloned nodes can also collaborate with other cloned nodes by exchanging their cloned identifiers extracted from the original nodes.
    
\end{itemize}

\subsection{Enhanced Elliptic Curve Digital Signature Algorithm (ECDSA*)}
Elliptic Curve Digital Signature Algorithm (ECDSA) is a digital signature algorithm that utilises elliptic curve cryptography (ECC) to derive the keys. ECC is a subset of public-key cryptography focused on the elliptic curves derived from the algebraic structure over finite fields. Although ECDSA performs similarly to other signature algorithms such as DSA and RSA, it is more efficient and robust due to its ECC foundation, which needs smaller keys to provide equivalent security. Using the ECDSA has proven effective in improving speed and reliability while improving the performance and strength of the overall ECC algorithm. Additionally, a version of ECDSA is proposed called ECDSA* that takes less time to compute and validate signatures than ECDSA. However, more significantly, ECDSA* increases the size of the signature in comparison to ECDSA without compromising stability \cite{kittur2017batch}. 

\subsubsection{ECDSA* Batch Verification}

Batch verification is a technique for validating several digital signatures in less time than it takes to validate them individually. In this technique, the signer produces \textit{t} signatures by interacting with the verifier, and the verifier validates all of these \textit{t} signatures at the same time.

ECDSA is a common digital signature algorithm in IoT since it offers the same level of security as public-key cryptography but uses smaller key sizes and ensures the authenticity of devices and data communication between them. Therefore, in our work, we aimed at ECDSA* signatures as a means of verifying location proof signatures produced by IoT devices. ECDSA* signatures are a variant of ECDSA signatures that offer 40\% more efficiency in verification without compromising security. 
Similarly to the ECDSA, the ECDSA* requires the implementation of the following algorithms: (i) key generation, (ii) signature generation and (iii) signature verification. The followings outline the implementation and descriptions of these algorithms.

\begin{itemize}
    \item \textbf{ECDSA* Key Generation Algorithm:} The algorithm \ref{alg:keygeneration} demonstrate the working of the ECDSA* key generation mechanism, including public and private keys. The algorithm for key generation generates a public and private key pair for use in the signing and verification processes. It is important to note that the key generation algorithm for ECDSA and ECDSA* works similarly. This algorithm takes standard domain parameters as a set such as \{p, \textit{E}, \textit{P}, \textit{n}, \textit{h}\}. These parameters are listed in greater detail below.
\begin{itemize}
    
\item \textit{p} = The order in which the prime field  \(\mathbb{F}_p\) exists
        \item \textit{E (a,b)} =  An elliptic curve $y^{2} = x^{3} + ax+b$ defined over thhe prime field \(\mathbb{F}_p\)
        \item \textit{P} = A non-zero random base point in \textit{E} (\(\mathbb{F}_p\)) 
        \item \textit{n} = The ordinal value of \textit{P}, which is normally a prime number.
        \item \textit{h} = The co-factor $  = \dfrac{|E(\mathbb{F}_p)  |}{n} $
         \end{itemize}

The private key can be computed by selecting a random integer \textit{d} from the range  \textit{d} = \{1 , \textit{n}-1\}. While, the public key \textit{Q} is calculated by multiplying the private key \textit{d} by a non-zero random base point \textit{P}.

\begin{algorithm}[hbt!]
\caption{ECDSA* Key Generation}\label{alg:keygeneration}
\textbf{Input:} Domain Parameters: \{p, \textit{E}, \textit{P}, \textit{n}, \textit{h} \} \\
\textbf{Output:} Key Pairs: Public key \textit{Q} and private key \textit{d}

\begin{algorithmic}[1]
\Procedure{Key Generation} {p, \textit{E}, \textit{P}, \textit{n}, \textit{h}}
\State \parbox[t]{\dimexpr\linewidth-\algorithmicindent}{%
Choose \textit{P} of order n for an elliptic curve E(\(\mathbb{F}_{p}\)), where \textit{P}  $\in$  E(\(\mathbb{F}_{p}\))}
\State Generate private key \textit{d}, where \textit{d} = \{1 , \textit{n}-1\}
\State Compute \textit{Q} = \textit{d}P
\EndProcedure
\end{algorithmic}
\end{algorithm}

\item \textbf{ECDSA* Signature Generation Algorithm:} The signing process is performed to generate the actual digital signature. Even though the algorithm for generating signatures is similar to that used by ECDSA, the signature forward to the verifier is different for ECDSA and ECDSA* signature schemes. Algorithm \ref{alg:signaturegeneration} illustrates the ECDSA* Signature Generation procedure, which begins with the following parameters as inputs: message \textit{m}, hash function \textit{H}, and domain parameters such as \textit{P}, and outputs the signature (\textit{r}, \textit{s}) for each participant. In IoT, for example, each different signature is generated for each device for verification. The signature generation process in this algorithm begins with the selection of the \textit{k} parameters as a random integer between 1 and n-1. Following that, the coordinates \textit{X} are determined by multiplying the random integer \textit{k} by the random point \textit{P}. The hash function \textit{H} (in this case, SHA-1) takes the message m and produces a hash value in the form of a digest string value, which is then converted to an integer \textit{e}. Finally, a signature value \textit{s} is calculated by taking the inverse of \textit{k} random integers and multiplying the sum of integer \textit{e} and private key \textit{d} by \textit{r}. The final output of ECDSA* signature generation is a pair, such as  (\textit{r}, \textit{s}).

\begin{algorithm}[hbt!]

\caption{ECDSA* Signature Generation}\label{alg:signaturegeneration}
\textbf{Input:} Message \textit{m}, Private Key \textit{d}, Hash Function \textit{H}, Domain Parameters \{\textit{P}\}  \\
\textbf{Output:} Signature (\textit{r}, \textit{s})
 
\begin{algorithmic}[1]
\Procedure{Signature Generation}{$m,d,H, {P}$}
\State Select a random integer \textit{k}, where 1 $\leq$	\textit{k} $\leq$ \textit{n} - 1
\State Compute \textit{X} = \textit{kP} \Comment{\textit{X}= coordinates (\textit{x},\textit{y})}
\State \parbox[t]{\dimexpr\linewidth-\algorithmicindent}{%
Compute \textit{r} = \textit{x} mod \textit{n} \Comment{If \textit{r} = 0, repeat the process with the next k}}

\State \parbox[t]{\dimexpr\linewidth-\algorithmicindent}{%
Compute \textit{H(m)} and convert it to an integer \textit{e} \Comment{Hash function e.g., SHA-1}}
\State \parbox[t]{\dimexpr\linewidth-\algorithmicindent}{%
Compute \textit{s} = \textit{$k^{-1}$} (\textit{e} + \textit{d}\textit{r}) mod \textit{n} \Comment{If s = 0, repeat the process with the next k.}}

\EndProcedure
\end{algorithmic}
 \end{algorithm}

\item \textbf{ECDSA* Signature Verification Algorithm:} The signature verification process is used to verify the signatures sent by the signer with his/her public key. The verification process depends on the signature size; the lengthy the signature is, the more time-consuming. Hence the signature verification algorithms are a little different since the signature size is different.

Algorithm \ref{alg:verification} illustrates the process of ECDSA* signature verification. This algorithm requires the following inputs: a signature value (\textit{r}, \textit{s}) that must be validated and a public key \textit{Q}. However, the signature verification output is in the form of a binary decision, such as accept or reject. The signature verification process begins by determining if the signature values \textit{r} and \textit{s} belong to the interval [1, \textit{n}-1] or not. Following that, the hash \textit{H} function calculates the hash value of the message \textit{m} for comparison purposes. Similarly to the signature generation algorithm, the hash value is converted into an integer \textit{e}. By taking the modulus of the inverse value of the signature, an integer value \textit{w} is generated. Following that, two coordinates, $u_{1}$ and $u_{2}$, are determined by multiplying the integers \textit{e} and \textit{r} by the value \textit{w}, respectively. An \textit{X} value is generated by combining the multiplications of \textit{P} and \textit{Q} by the calculated coordinates ($u_{1}$, $u_{2}$) from the previous step. If X = $\mathcal{O}$, the signature will be rejected; otherwise, it will be accepted if and only if $\upsilon$ = \textit{r}.

\begin{algorithm}[hbt!]
\caption{ECDSA* Signature Verification}\label{alg:verification}
\textbf{Input:} ECDSA* Signature (\textit{r}, \textit{s}), Public Key \textit{Q} \\
\textbf{Output:} Accept or Reject Signature (\textit{r}, \textit{s})

\begin{algorithmic}[1]
\Procedure{Signature Verification}{Signature (\textit{r}, \textit{s}), Public Key \textit{Q}}
\State \parbox[t]{\dimexpr\linewidth-\algorithmicindent}{%
Determine that \textit{r} and \textit{s} are both integers in the range [1, \textit{n}-1]}
\State  
\parbox[t]{\dimexpr\linewidth-\algorithmicindent}{%
Calculate \textit{H(m)} and convert it to an integer \textit{e} \Comment{Hash function e.g., SHA-1}
}
\State Compute \textit{w} = \textit{$s^{-1}$} mod \textit{n}

\State Compute $u_1$ = \textit{e}\textit{w }mod \textit{n} and $u_2$ = \textit{r}\textit{w }mod \textit{n}
\State Compute \textit{X} = $u_1$P + $u_2$Q \Comment{\textit{X}= coordinates ($u_1$,$u_2$)}

\If  {\textit{X} = $\mathcal{O}$  } 
    \State Reject signature 
        \Else
        \State Accept signature if \textit{$\upsilon$} = \textit{r} \Comment{$\upsilon$ = $u_{1}$ mod \textit{n}}
\EndIf

\EndProcedure
\end{algorithmic}
\end{algorithm}
    
\end{itemize}

\subsection{Selection of Trusted Nodes}
Batch verification significantly reduces the time required to validate the signature of each IoT device on gateway nodes. In IoT networks, gateway nodes are considered to have greater computing power than resource-constrained IoT devices. However, along with providing internet connectivity through cloud services, gateway nodes in an IoT network often have additional responsibilities such as data preprocessing, data aggregation, and running protocols to ensure the security of connected IoT devices. Considering these responsibilities, it is a critical requirement for batch verification to minimise the amount of work at gateway nodes. One approach is to offload the signature verification task from the gateway node to a few other IoT devices without compromising security.

However, one of the critical tasks is to select the trusted IoT devices for signature verification from the pool of available IoT devices in the network. To facilitate this, we proposed the model to select the trusted IoT devices from available IoT devices for signature verification task. Numerous others trusted models for selecting nodes to perform computation tasks have also been proposed, some of which are cloud-based and platform-dependent. In contrast, others concentrate on selecting individuals based on their trust values \cite{khalil2021identification, altaf2020robust, qolomany2020trust}. For instance, one model \cite{kittur2020trust} is proposed for selecting trusted IoT devices for signature verification. Since this proposed model selects IoT devices based on physical and security criteria, it imposes additional computation overhead on the overall system by selecting IoT devices that satisfy these two critical conditions. 

Given that IoT devices usually have limited processing power, memory, and battery capacity, it is important to select trusted IoT devices and distribute the load to them to perform operations similar to those previously performed by the gateway nodes. Thus, we aim to distinguish trustworthy IoT devices from other IoT devices that exhibit the ECDSA* signature verification process in a more efficient and timely manner. Our trusted device selection model focuses on selecting trusted IoT devices based on device profiling capabilities such as processing capacity, sensed context information, and confidence credibility.

For example, in a particular IoT network area, the devices are denoted by $D$ = \{$D_{1}$, $D_{2}$, $D_{3}$, \dots, $D_{n}$\} where \textit{D} = 1,2,3, \dots, \textit{n}. The entire network device area is classified into two types of devices. One type is simple devices such as $D_{s}$ = \{$D_{1}$, $D_{2}$, $D_{3}$, \dots, $D_{n}$\}, which are responsible for signature generation and must be verified; the other type is trustworthy devices such as $D_{t}$ = \{$D_{1}$, $D_{2}$, $D_{3}$, \dots, $D_{p}$\}, which are trusted devices responsible for batch signature verification. The number of trusted devices is less than the number of original devices, such as \textit{p} < \textit{n}.

In confidence credibility, numerous known trust factors are calculated whenever a transaction happens between simple IoT devices \textit{$D_{s}$} and trusted IoT devices \textit{$D_{t}$} to evaluate an IoT device's trustworthiness. The confidence measure for the \textit{p} dimension are respectively ${C_{1}}_{D_{s}, D_{t}}$, ${C_{2}}_{D_{s}, D_{t}}$, ${C_{3}}_{D_{s}, D_{t}}$, \dots, ${C_{p}}_{D_{s}, D_{t}}$. To calculate the confidence \textit{C}, we divide the confidence credibility for an individual IoT device into two measures: implicit confidence and explicit confidence. In implicit confidence (\textit{IC}), trust is determined by examining an independent reputation of IoT device.  In explicit confidence (\textit{EC}), the trust placed in the suggestions of other nodes based on their prior experiences. The overall confidence of both implicit and explicit measures for the given devices in the IoT network is determined in Eq. \ref{eq1}.

\begin{equation} \label{eq1}
 C_{D_{s}, D_{t}} = \alpha_{i} * IC_{D_{s}, D_{t}} +  \beta_{i} * EC_{D_{s}, D_{t}}
  \end{equation}

Eq. \ref{eq1} provides the formula for calculating total confidence. It is composed of two parts: $IC_{D_{s}, D_{t}}$ represents the implicit trust between a simple IoT device and a trusted IoT device, and $EC_{D_{s}, D_{t}}$ represents the explicit trust between a simple IoT device and a trusted IoT device. $\alpha$ and $\beta$ are the total weighted factors for \textit{IC} and \textit{EC}, respectively, and the total weighted factor $\tau$ is determined as $\tau$ = $\alpha$ + $\beta$.

Both weighted factors $\alpha_{i}$ and $\beta_{i}$ satisfy the following equations independently, as shown in Eq. \ref{eq2} and Eq. \ref{eq3}

 \begin{equation} \label{eq2}
     0 \leq  \alpha_{i} \leq 1 ,  \sum_{i=1}^{p}= 1 
 \end{equation}

 \begin{equation} \label{eq3}
     0 \leq  \beta_{i} \leq 1 , \sum_{i=1}^{p}= 1 
 \end{equation}

The implicit confidence for a trusted IoT system is measured by adding the implicit confidence for a location as sensed context information to the device's feedback. If no transaction history exists for IoT devices $D_{s}, D_{t}$, an initial value is allocated to implicit confidence; however, if there is no association between IoT devices $D_{s}, D_{t}$, the default value of 0.5 is used. The implicit confidence interval for a set of devices at specific locations is computed using Eq. \ref{eq4}.

\begin{equation} \label{eq4}
    IC = \sum_{i=1}^{p} \alpha_{i} IC_{i} (D_{si}, D_{ti})
\end{equation}

where $\alpha$ is further subdivided into two weights, $\alpha$$^L_{Ds}$ and $\alpha$$^L_{Dt}$, for $D_{s}$ and $D_{t}$ respectively at various locations \{\textit{$L_{1}$}, \textit{$L_{2}$}, \textit{$L_{3}$}, \dots, \textit{$L_{p}$}\}, as specified below.

\[ \alpha^L_{D_{s}}=\frac{\alpha^{Lp}_{D_{t}}}{\alpha^{Lp}_{D_{t}}} \ \ \textrm{and} 
\ \ \alpha^{Lp}_{D_{t}}=\textstyle\frac{\alpha^{Lp}_{D_{t}}}{\alpha^{Locations}} \]

Eq. \ref{eq5new} calculates the weighted factors for $D_{s}$ and $D_{t}$ with respect to their locations.

\begin{equation} \label{eq5new}
    \sum_{L=1}^{p} \alpha^{Lp}_{D_{s}} \alpha^{Lp}_{D_{t}} = 1 
\end{equation}

Where \textit{$L_{p}$} can be calculated using the Euclidean distance method defined in the algorithm \ref{alg:locationcalculate}.

The aggregate implicit confidence (previous and recent) of all trusted IoT devices is determined using Eq. \ref{eq5}.

\begin{equation} \label{eq5}
    IC^{L}_{D_s}, D_{t} =  \sum_{i=1}^p IC^{L_i}_{D_si}, D_{ti_(previous)} + IC^{L_i}_{D_si}, D_{ti_(recent)}
\end{equation}

Following the implicit confidence measurement, we measured the explicit confidence measurement when simple IoT devices request feedback \textit{fd} about the location of a trusted IoT device at a particular point. All IoT devices measure the location of a trusted IoT device and transmit the requested location as feedback to the requested IoT devices. For explicit confidence \textit{EC}, the \textit{p} trusted levels can be calculated as follows: $\left( \beta_{1}, \beta_{2}, \beta_{3}, \dots, \beta_{p}\right)$, where 0 $\leq$ $\beta_{i}$ $\leq$ 1. For IoT devices, the trusted level sequence can be implemented as $\beta_{1}$ $<$ $\beta_{2}$ $<$ $\beta_{3}$ $<$ \dots, $\beta_{p}$.

Individual feedback \textit{fd} from IoT devices can be calculated by measuring their position, which is expressed as \textit{fd}$_{D_{si}, D_{ti}}$. The explicit confidence  \textit{EC} for a trusted device $D_{t}$ in terms of its feedback $fd_{D_{t}}$ is measured using Eq. \ref{eq6}.

\begin{equation} \label{eq6}
  \delta (EC_{D_s}, D_{t}, fd) = \begin{cases}
  
  \sum_{i=0}^{p}, 0 \leq EC_{Ds_i}, D_{t_i} \leq 1 & \\
  \sum_{i=0}^{p}, 0 \leq fd_{D_t} \leq 1 
\end{cases} 
\end{equation}

The explicit confidence for the trusted IoT device is measured using Eq. \ref{eq7}.

\begin{equation} \label{eq7}
     EC_{D_{s}, D_{t}} = \frac{\sum_{i=1}^{FDs}(fd^{L_{p}}_{D_{s}, D_{t}} * \beta_{p})}{FDs}
\end{equation}

where \textit{$fd^{L_{p}}_{D_{s}}$} represents the individual feedback for trusted IoT devices regarding location, and \textit{FDs} represents the aggregate feedback for trusted IoT devices regarding the location that is obtained from all connected devices. $\beta_{p}$ denotes the weighted factors that the trusted IoT device possesses in the requested node. The overall weighted factor can be calculated by using Eq. \ref{eq:weightfactor}, taking into account input from all IoT devices.

\begin{equation}\label{eq:weightfactor}
    \beta_{p} =  \beta_{1} * fd1_{D_t} + \beta_{2} * fd2_{D_t} + \beta_{3} * fd3_{D_t} + \dots + \beta_{k} * fdk_{D_t}
\end{equation}
where $\beta_{1} + \beta_{2} + \beta_{3} + \dots$  = 1.

\section{Location Proof System} \label{sec:locationproof}

This section presents the LPS model following the context-aware modalities localisation technique. We developed several algorithms to explain the process of detecting clone node attack on LPS. Finally, this section discusses the execution flow of our proposed scheme between various components of the LPS model by using location proofs and batch verification concepts.

\begin{figure*}[hbt!]
\centering
\includegraphics[width=12cm, height=10cm]{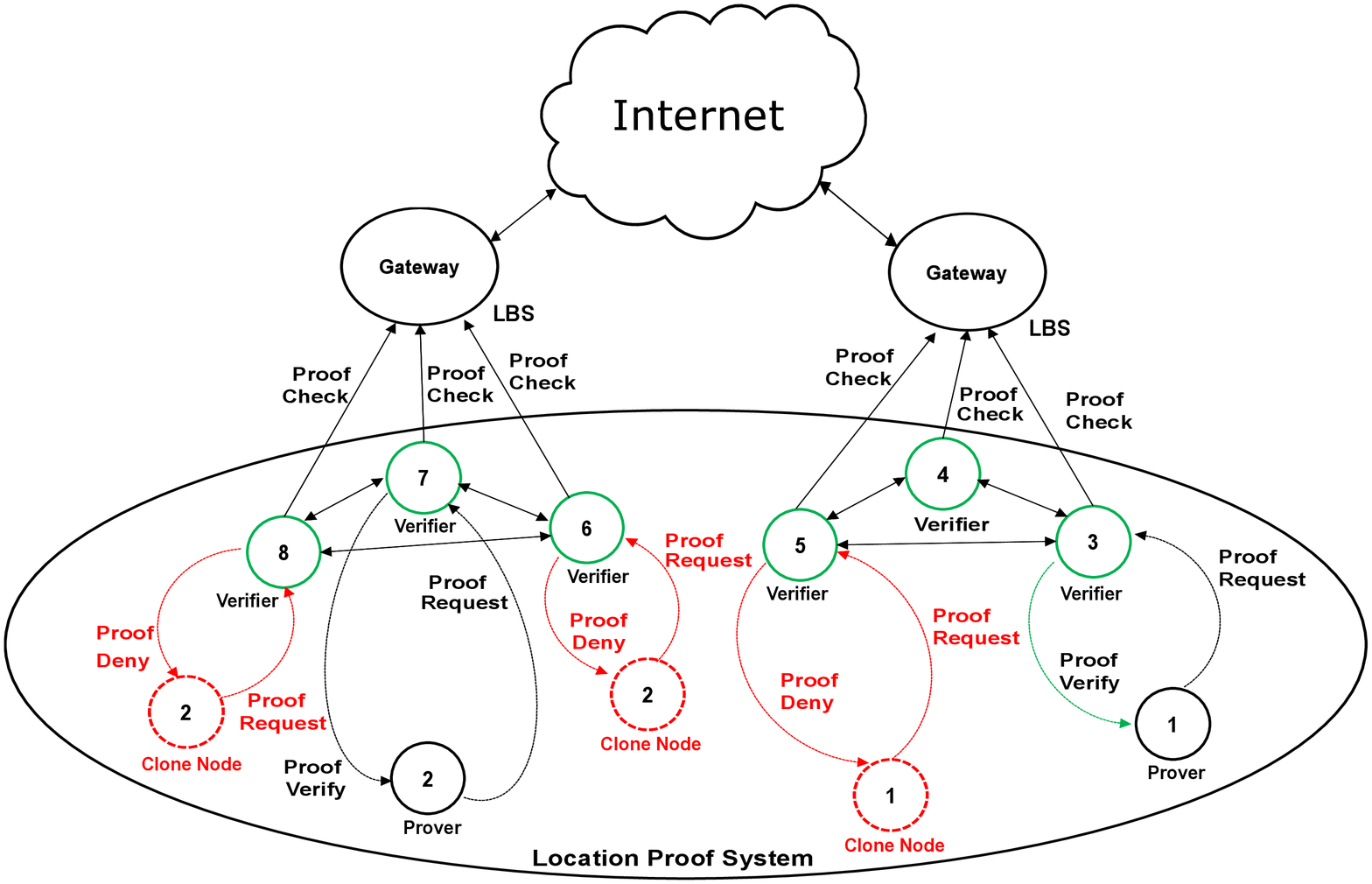}
\caption{Proposed Location Proof System}
\label{locationproofsystem}
\end{figure*}

\subsection{Localization Technique}
Localisation is an important concept in LPS, in which localisation and network/location infrastructure-independent methods are used to determine the location of the user's device. Localisation refers to how a device determines its position in relation to another device, satellite or maps, etc. Several software-based or hardware-based techniques have been used for localisation, including fingerprint, distance-bounding protocol, context-based modalities, proximity, triangulation, beaconing, and mobile network or tower-based approaches. 

We used context-aware modalities as a technique for localisation in our work because we used context-aware information to detect clone node attack on IoT networks. The basic idea behind context-based localisation is that it gathers various contextual values about the IoT device environment, such as ambient acoustic light, noise level, humidity, temperature, and Wi-Fi and Bluetooth signal power, and then generates proof of presence for physical device location by combining all of them to determine the device's location. Contextual information is collected simultaneously by the verifier and device. The device produces the proof of presence, which includes context information, and the verifier validates the context information to confirm the device's physical presence.

The network model of the LPS used to detect clone node attack on IoT networks via a context-aware information-based localisation technique is shown in Fig. \ref{locationproofsystem}. The network model of an LPS incorporates the following entities:

\begin{itemize}
    \item \textbf{Prover:} The provers are the IoT devices that want to demonstrate that adversaries have not compromised their identities and location. We referred to the simple IoT device as ``Prover'' for simplicity.
    \item \textbf{Clone Node:} The clone nodes are compromised IoT devices whose credentials, such as context information, have been compromised by the attacker.
    \item \textbf{Verifier:} The verifiers are the selected IoT devices that employ a trust model and communicate with the gateway to verify the evidence. We referred to the selected IoT devices as ``Verifiers'' for simplicity. 
    \item \textbf{LBS:} The gateway nodes serve as location-based services (LBS), with verifier IoT devices serving as clients.
\end{itemize}

In our proposed scheme, we consider a scenario in which a client of an LBS, referred to as verifiers, aims to demonstrate to the LBS the existence of provers at a specific location to detect clone nodes attack on an IoT environment. A prover and a verifier simultaneously collect contextual data through IoT devices to create an incident context. To validate the proofs, the verifier IoT devices compare the context information obtained from the prover IoT devices to their own context information to determine whether or not the IoT device has been compromised. For proof validation, we used the ECDSA* signature process, in which each party (prover and verifier) is assigned their public and private keys. We assume that an intruder from outside the context cannot detect context information.

\subsection{Proposed Algorithms}
Along with describing the network model, attacker model, and entire working mechanism of ECDSA*, we proposed several algorithms to demonstrate the execution of an LPS for detecting clone node attack in an IoT environment. These algorithms include a location calculation algorithm, a location proof generation algorithm, and a location proof-verification algorithm. The procedure of each proposed algorithm is defined in the algorithms \ref{alg:locationcalculate} - \ref{alg:proofverification}.  Table. \ref{tab1} shows the notations and descriptions used in the algorithms \ref{alg:locationcalculate} - \ref{alg:proofverification}.

\begin{table}[!htp]
\caption{Notations}
\begin{center}

\resizebox{\columnwidth}{!}{%

\begin{tabular}{|l|l|}
\cline{1-2}
 \textbf{Notations} & \textbf{Description}  \\ \cline{1-2}
 $\text{\textit{ID}}$& Provers' identification    \\ \cline{1-2}
 $\text{\textit{T}}$& Data sensing time   \\ \cline{1-2}
 $\text{\textit{{Loc}}}$&  Location on two dimensional space  \\ \cline{1-2}
 $\text{\textit{{Actv}}}$&  Prover's activity at specific time   \\ \cline{1-2}
  $\text{\textit{CI}}$ & Context Information  \\ \cline{1-2}
\{$CI_{1}$, $CI_{2}$, $CI_{3}$, \dots , $CI_{n}$\} & A set of stored context information on LBS\\\cline{1-2}
 H & Hash function (SHA256)  \\ \cline{1-2}
 $K_{Pr}$ & Prover's private key \\ \cline{1-2}
 $P$ &  Location Proof   \\ \cline{1-2}
 $P_{sign}$ & Location Proof Signature    \\ \cline{1-2}

\{$P_{1_sign}$, $P_{2_sign}$, $P_{3_sign}$, \dots, $P_{n_sign}$\} & A set of provers' signature \\ \cline{1-2}

\{$K_{1_Pb}$, $K_{2_Pb}$, $K_{3_Pb}$, \dots, $K_{n_Pb}$\} & A set of provers' public keys \\\cline{1-2}
\{$Ver_1$, $Ver_2$, $Ver_3$, \dots, $Ver_n$\} & A set of verifiers for batch verification\\\cline{1-2}
 Signature() & Elliptic Curve Digital Signature function  \\ \cline{1-2}
 
 isEmpty() & Checking the location proof request \\ \cline{1-2}

  isNotExisted() & Checking the existence of context information \\ \cline{1-2}
  
  revert() & Sent back the transaction \\\cline{1-2}
\end{tabular}
}
\label{tab1}
\end{center}
\vspace{-6mm}

\end{table}

\subsubsection{Location Calculation Algorithm}
Approximating distance is a significant obstacle when addressing a location in majority of the IoT networks. The algorithm \ref{alg:locationcalculate} demonstrates the process of determining the locations of network devices. We estimate the location of each device (such as the prover and verifier) as a key element of context information in our proposed location proof framework model by measuring the distance between the provers and verifiers in two-dimensional (2-D) space. A prover in two-dimensional space is represented by P = \{$x_{1}$, $y_{1}$\}, whereas a verifier in two-dimensional space is represented by V = \{$x_{1}$, $y_{1}$\}. Based on their distance estimation, we used the Euclidean distance algorithm to calculate the location of each prover with respect to the verifiers. The distance between verifier and prover is denoted by $d_{\left(V,P\right)}$.

The euclidean distance process starts by taking the prover and verifier's coordinates in two-dimensional space, such as P = \{$x_{1}$, $y_{1}$\} and V = \{$x_{1}$, $y_{1}$\}, as input values respectively. It calculates the distance between each verifier \{$Ver_1$, $Ver_2$, $Ver_3$, \dots, $Ver_n$\} by taking the square root of difference between the coordinates of each verifier and prover pair. After calculating the distance between their provers, each verifier maintains a list of the euclidean distances between their provers as determined locations at specific points, such as \{$d_{P1_{\left(v1,p1\right)}}$, $d_{P2_{\left(v1,p1\right)}}$, $d_{P3_{\left(v1,p1\right)}}$, \dots, $d_{Pn_{\left(v1,p1\right)}}$\}.

\begin{algorithm}[hbt!]
\caption{Calculate Location}\label{alg:locationcalculate}
\textbf{Input:} Two Points on 2-D space with their coordinates such as P = \{$p_{1}$, $p_{1}$\} and V = \{$v_{1}$, $v_{2}$\}   \\
\textbf{Output:} Distance \textit{d} 
 
\begin{algorithmic}[1]

\Procedure {Distance Calculate} {V, P}

\For{Each Verifiers \{$Ver_1$, $Ver_2$, $Ver_3$, \dots, $Ver_n$\}  }

\begin{align*}
d_{\left(V,P\right)}   = \sqrt {\sum_{i=1}^{n}  \left( p_{i}-v_{i}\right)^2} 
\end{align*}

\EndFor

\State \parbox[t]{\dimexpr\linewidth-\algorithmicindent}{%
Maintain a list of provers' distances \{$d_{P1_{\left(v1,p1\right)}}$, $d_{P2_{\left(v1,p1\right)}}$, $d_{P3_{\left(v1,p1\right)}}$, \dots, $d_{Pn_{\left(v1,p1\right)}}$\}}

\EndProcedure

\end{algorithmic}
 \end{algorithm}

\subsubsection{Generate Location Proof}
The algorithm \ref{alg:proofgeneration} demonstrates the process of generating location proofs for IoT devices showing their presence at a specific location in terms of LBS. The proof generation process begins with sensing the contextual information by both the prover and the verifier about their deployed environment. The context information includes the identification of IoT device \textit{ID}, data sensing time \textit{T}, its specific location \textit{Loc} and activity \textit{Actv}. The combination of such information is referred to as \textit{CI}, and it is maintained and stored at LBS as \{$CI_{1}$, $CI_{2}$, $CI_{3}$, \dots , $CI_{n}$\}.  A location \textit{Loc} between two IoT devices or between an IoT device and a selected IoT device is determined using the Euclidean distance algorithm, which determines the length of a segment connecting the two points and its location in 2-dimensional space or at a specific point in place. An IoT device's activity can be any operation, such as monitoring, sensing, or broadcasting at a specific time \textit{T}. To generate a location proof, a verifier first requests that the prover generate a proof using sensed context information such as \textit{CI}. The proof is generated by signing the context information \textit{CI} using the prover's private key $K_{Pr}$.  The signature $P_{sign}$ is generated using the ECDSA* signature generation algorithm (algorithm \ref{alg:signaturegeneration}).

\begin{algorithm}[hbt!]
\caption{Generate Location Proof}\label{alg:proofgeneration}
\textbf{Input:} Context Information (ID, Time, Location, Activity), Private Key $K_{Pr}$ and Hash Function \textit{H}   \\
\textbf{Output:} Prover's Signature $P_{sign}$ 
 
\begin{algorithmic}[1]
\Procedure{Proof Generation}{$Context Information$, $K_{Pr}$, \textit{H}}
\State \parbox[t]{\dimexpr\linewidth-\algorithmicindent}{%
Both prover and verifier sense a contextual information}
\State \textit{ID} $\leftarrow$ Identification of IoT device 
\State \textit{T} $\leftarrow$ Data sense time
\State \textit{Loc} $\leftarrow$ Location
\State \textit{Actv} $\leftarrow$ Activity
\State Context Information \textit{CI} = \{\textit{ID}, \textit{T}, \textit{Loc}, \textit{Actv}\}
\State \parbox[t]{\dimexpr\linewidth-\algorithmicindent}{%
Stored context information at LBS such as \{$CI_{1}$, $CI_{2}$, $CI_{3}$, \dots , $CI_{n}$\}}

\If  {Request\_Location\_Proof = isEmpty()  } 
    \State revert(``Reject Proof'')
        \Else
        \State Proof\_Generation(\textit{CI}, $K_{Pr}$)
        \State $P$ = Hash(\textit{CI})
        \State $P_{sign}$ = Signature\_$K_{Pr}$(P) \Comment{Algorithm \ref{alg:signaturegeneration}} \\
        \hspace{1cm} \Return $P_{sign}$
\EndIf

\EndProcedure
\end{algorithmic}
 \end{algorithm}
\subsubsection{Verify Location Proof}

The algorithm \ref{alg:proofverification} illustrates the process of verifying location proofs for IoT devices claiming to be at a specific location with context information \textit{CI}. The verification process begins with taking inputs such as provers' signatures as \{$P_{1_sign}$, $P_{2_sign}$, $P_{3_sign}$, \dots, $P_{n_sign}$\}, and their respective public keys as \{$K_{1_Pb}$, $K_{2_Pb}$, $K_{3_Pb}$, \dots, $K_{n_Pb}$\}. To validate the position proof obtained from the prover, the verifiers \{$Ver_1$, $Ver_2$, $Ver_3$, \dots, $Ver_n$\} analyze the contextual information \textit{CI} from the LBS and perform ECDSA* batch verification on the signatures (algorithm \ref{alg:verification}) after getting confirmation about the availability of stored information on the LBS. As the batch verification process is carried out by multiple verifiers chosen using the trust model, the LBS maintains and controls the list of verifiers. The verifiers used the prover's public key $K_{Pb}$ to validate the signature $P_{sign}$. After successfully verifying the signatures obtained from each selected verifier, the verifier will confirm the authenticity of the IoT device in the network and accept the proof with location confirmation and other credentials. However, if the signature is not successfully verified, the verifier notifies the LBS of the compromise of the prover in the IoT network.

\begin{algorithm}[hbt!]

\caption{Verify Location Proof}\label{alg:proofverification}
\textbf{Input:} A set of Provers' Signatures \{$P_{1_sign}$, $P_{2_sign}$, $P_{3_sign}$, \dots, $P_{n_sign}$\}, A set of Provers' Public Keys \{$K_{1_Pb}$, $K_{2_Pb}$, $K_{3_Pb}$, \dots, $K_{n_Pb}$\}, A set of Verifiers \{$Ver_1$, $Ver_2$, $Ver_3$, \dots, $Ver_n$\}  \\
\textbf{Output:} Accepted/Rejected

\begin{algorithmic}[1]
\Procedure{Proof Verification}{$P_{sign}$, $K_{Pb}$}
\State \parbox[t]{\dimexpr\linewidth-\algorithmicindent}{%
The verifiers \{$Ver_1$, $Ver_2$, $Ver_3$, \dots, $Ver_n$\} check the context information from LBS}

\If {CI = isNotExisted()   } 
   \State revert(``Information not existed'')
   
        \Else 
        \For{\parbox[t]{\dimexpr\linewidth-\algorithmicindent}{%
        Each verifiers \{$Ver_1$, $Ver_2$, $Ver_3$, \dots, $Ver_n$\} } } 
        \If{\State Extract CI }
        \State \parbox[t]{\dimexpr\linewidth-\algorithmicindent}{%
        Proof\_Verification ($P_{sign}$, $K_{Pb}$) 
        \Comment{Algorithm \ref{alg:verification}} }
        \State \Return ``Proof Accepted'' 
    
         \Else 
        \State revert(``Proof Rejected'')
        
        \EndIf
\EndFor        
\EndIf

\EndProcedure
\end{algorithmic}
 \end{algorithm}

\subsection{Execution Flow} \label{executionflow}

\begin{figure*}[hbt!]
\centering
\includegraphics[width=12cm, height=10cm]{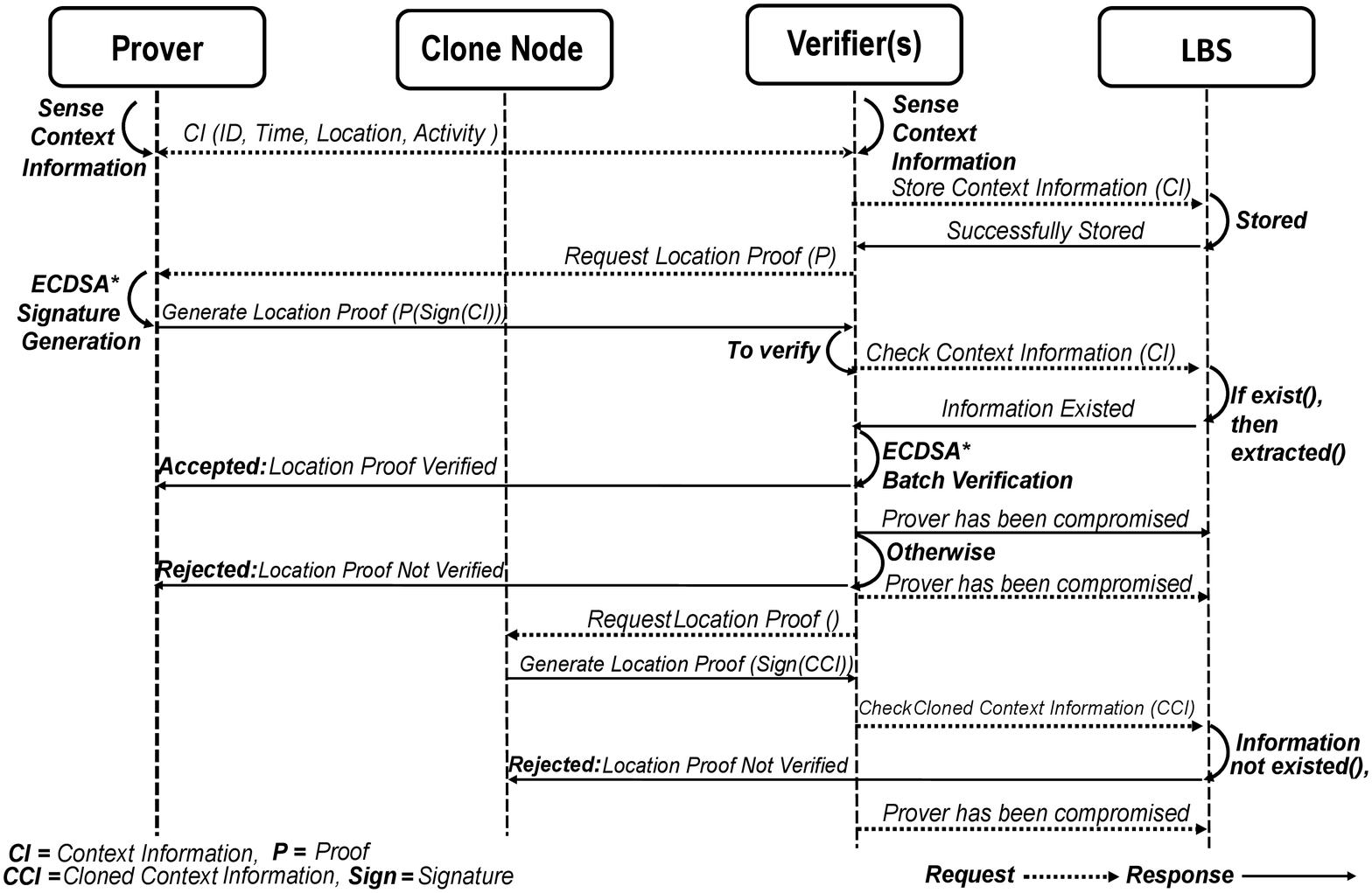}
\caption{Execution Flow of the Location Proof System}
\label{ExecutionFlow}

\end{figure*}

The detection of clone node attack on an IoT environment is accomplished by using the numerous interconnected modules included in the proposed network model.  Following the proposed network model, we proposed an LPS for detecting clones node in an IoT environment, which consists of the following entities: prover, clone nodes, verifiers, and LBS, all of which interact with one another to detect a clone node attack successfully. Fig. \ref{ExecutionFlow} depicts the execution flow of our proposed LPS model, which demonstrates the relationship between the entities (successful/failed). In addition, the execution flow diagram illustrates the method of implementing a proposed clone node attack detection scheme based on the LPS. 

The process of detecting clone nodes attack begins with the prover and verifiers sensing information about the deployed environment. The context information includes the identifier of the IoT device \textit{ID} the time stamp for data sensing \textit{Time}, the device's specific  \textit{location}, and the activity performed \textit{Activity}. The combination of these context information is grouped and represented as \textit{CI}. Following the process of sensing context information \textit{CI}, the verifiers save this information to the LBS entity and receive an acknowledgement from the LBS about the saved information.

After storing the context information in the LBS, the verifiers request that the prover give the location proof in order to validate the prover's authenticity and to ascertain whether or not the prover has been compromised. The prover generates a location proof of sensed context information \textit{CI} using the ECDSA* signature generation method and sends it to the verifiers for verification. Upon receiving the signed proof, the verifiers compare the existence of context information  \textit{CI} with stored context information  \textit{CI} to the LBS and receive acknowledgement of the existence of information to the LBS.

After obtaining the signed proof from the prover, the verifiers conduct the proof verification process on the received signed proof using the ECDSA* batch verification process and determine the provers liable for pretending to be at a particular location using context information. If the verifiers verify the signed proof successfully, it notifies the prover of the accepted location proof. Meanwhile, the verifiers notify the LBS when a prover is compromised in the network.

We designed two scenarios in the LPS to detect clone node attack in the IoT context, such as detecting the clone node attack by evaluating and comparing the details to the LBS and requesting the location proof to the clone node determining its compromise after verification. In the former case, if the signature is not successfully verified by the verifiers, then LBS indicates that the provers in the network have been compromised. In the latter case, the verifier requests the clone node's location proof in order to validate its authenticity and location modification. Since an intruder created clones of the provers and copied their context information \textit{CI} to cloned context information \textit{CCI}, the verifiers compare the cloned context information \textit{CCI} to the previously stored context information \textit{CI} after receiving the location proof from the cloned node. Upon receiving confirmation from LBS that context information did not exist, the verifiers rejected the location proof and informed LBS about the provers' compromise.

\section{Security Analysis}\label{sec:securityanalysis}
This section provides a security analysis of the proposed scheme for detecting clone nodes based on LPS. Given that our proposed scheme makes use of the ECDSA* principle for achieving security and validating proofs in an LPS, we conduct an in-depth security analysis of ECC, including domain parameter selection criteria and address implementation issues and attacks on them. By following this, we enhanced our security analysis by discussing the security criteria for key pair selection and describing the definition and theorem for security attacks on digital signatures and hash functions used in LPS.

\subsection{Selection of Domain Parameters}
To implement a secure cryptographic algorithm based on elliptic curves, a list of domain parameters is required to construct a concrete curve on which computations can be performed. These parameters are much more sophisticated than the algorithms of finite field methods such as DSA. The selection of such parameters emphasises the importance of implementing the algorithms from both a computational and a security standpoint, necessitating a focus on security attacks and execution errors. Further, the selection of particular domain parameters also resulted in the development of guidelines for designing secure protocols or algorithms, each of which has its own set of security conditions for the curves and frequently specifies concrete curves for certain security levels \cite{janvcarsecurity}.

The domain parameter selection in ECDSA is crucial to ensuring its overall security, which can be achieved for a single user or an entire group of users as interest grows in the application of ECDSA to the various groups of rational points on an elliptic curve over a finite field. As a result, the number of different types of attacks on ECDSA has been increased for some types of curves. For ECDSA, the domain parameters consist of an elliptical curve \textit{E} appropriately selected and determined by a finite field \(\mathbb{F}_p\) with characteristic \textit{p} and the base point \textit{G} = \textit{E} (\(\mathbb{F}_p\)).

A selection of domain parameters must follow the following security requirements inherent to the elliptic curves based explicitly on the discrete logarithm problem. These security requirements follow the standard of elliptic curve \textit{E}  over the prime field  \(\mathbb{F}_p\) due to its popularity in most cryptographic algorithms. 

\subsubsection{Generator Order}
An important requirement of selecting the domain parameters for an elliptic curve-based algorithm is the order of generation in which they are generating. As generating order is an important requirement, therefore in this regard, many algorithms have been proposed, such as Pollard's rho algorithm and Pohlig-Hellman that works for different cyclic groups. For example, the Pohlig-Hellman algorithm worked on the cyclic group for composite order, which divides the entire group into different subproblems to achieve the fastest computation. 
\begin{definition}
Let \textit{G} is a cyclic group of order \textit{n}, \textit{k} is a sub-problem size, \textit{P} is a prime order subgroup \{\textit{$p_{1}$}, \textit{$p_{2}$, \textit{$p_{3}$}, \dots, \textit{$p_{k}$}}\}, \textit{l} is a positive integer. 

Here is an example of how the Pohlig-Hellman algorithm works \cite{sommerseth2015pohlig}: Firstly compute prime factorization of \textit{n} prime numbers such as $\textit{p}_1^{\textit{e}_{1}}.
p_2^{e_{2}}. p_3^{e_{3}}, \dots, p_k^{e_{k}}$ as computed in Eq. \ref{eq:pohlig}.

\begin{equation} \label{eq:pohlig}
    \textit{n} = \prod_{i=1}^{k} p_i^{e_{i}}
\end{equation}

Then, compute \textit{l} as \textit{$l_{i}$} = \textit{l} $\mod$ \textit{$p_{k}^{e_{k}}$} for all 
1 $\leq$ \textit{i} $\leq$ k.

A Chinese remainder theorem is used to calculate the \textit{l} as a linear equation. Here, $p_{1}$, $p_{2}$, $p_{3}$, \dots, $p_{k}$ are mutually coprime set of prime order subgroup for which \textit{gcd($p_{i}$, $p_{j}$)}=1. The positive integer \textit{l} is obtained by computing the linear equation using the Extended Euclidean Algorithm.

Each $l_{k}$ can be computed with the base point \textit{p} as in Eq. \ref{eq:chinese}.

\begin{equation} \label{eq:chinese}
    \textit{l}_{k} = \textit{x}_{0} + \textit{x}_{1}\textit{p} + \textit{x}_{2}\textit{p}^2 + \dots + \textit{x}_{e-1}\textit{p}^{e-1}
\end{equation}

where \textit{$x_{i}$} $\in$ [0, \textit{p}-1].

For complexity point of view, the worst-case time for group of order \textit{n} is $\mathcal{O}(\sqrt{n})$ and for in relatively smooth case is $\mathcal{O}(logn+\sqrt{pk})$  if the order is prime.

\end{definition}

\subsubsection{Anomalous Curves}

Another important requirement for selecting the domain parameters is the selection of anomalous curve, which can also be computed as prime field curves in the following Eq. \ref{eq:anomalous}.

\begin{equation} \label{eq:anomalous}
    |\textit{E} (\mathbb{F}_p)| = \textit{p}
\end{equation}

To compute the discrete logarithm on an anomalous curve, a polynomial-time algorithm is a commonly used algorithm. In other words, a polynomial-time algorithm is also called additive transfer that allows converting the hard elliptic curve discrete logarithm problem into simple discrete logarithm problem over the additive group of \(\mathbb{F}_p\).  

\subsubsection{Multiplicative transfers}

A transfer is a process of converting the elliptic curve discrete logarithm problem into the simple discrete logarithm problem. As the anomalous curve is mostly worked on the concept of additive transfer of a group of \(\mathbb{F}_p\). However, in multiplicative transfer, elliptic curve discrete logarithm problem on every curve  $\dfrac{E}{\mathbb{F}_p}$ is converted into discrete logarithm problem over the group of \(\mathbb{F^{*}}_{p_t}\). A \(\mathbb{F^{*}}_{p_t}\) is an extension of the original curve \(\mathbb{F}_p\) over multiplication. The degree of multiplicative transfer can be defined as follows: 

\begin{definition}
Let \textit{e} is a embedding degree, \textit{l} is a generator order, \(\mathbb{F}_p\) is a finite field, and \(\mathbb{N}\) is a set of natural number, where \textit{e} $\in$ \(\mathbb{N}\) then it can be computed by using Eq. \ref{eq:multiplicative}:

\begin{equation}\label{eq:multiplicative}
    \textit{l} | \textit{p}^{x}-1
\end{equation}

\end{definition}

The embedding degree \textit{e} is also stated as the order of \textit{p} over \(\mathbb{Z}_l\). Unlike to anomalous curves, the probabilistic polynomial time is computed for multiplicative transfer work by defining the relation between $ \dfrac{E}{\mathbb{F}_p} $ subgroup of \textit{G} and subgroup of \textit{$l_{th}$} roots of \(\mathbb{F}_{p_{t}}\). 

By following this, there is an attack on the multiplicative transfers called MOV attack, which occurred due to sub-exponential times calculated for the index-calculus \cite{luca2004mov}. 
\subsubsection{Extension Field Curves}
Besides the binary fields \(\mathbb{F}(2^{m})\) and prime fields \(\mathbb{F}(p)\), there has been increased interest in use of Optimal extension fields \(\mathbb{F}(p^{m})\) that offers offer considerable computational advantages in software by selecting \textit{p}
and \textit{m} specifically to match the underlying hardware used to perform the arithmetic operations.  Besides, efficient methods have been devised for speeding up field arithmetic for elliptic curves over general extension fields \cite{zhang2012efficient}. The benefits of defining the extension on existing curves are to solve the ECDLP by using the index-calculus methods on the group of rational points. An example of multiplicative curve overextension of optimal field curve can be defined as follows:

\begin{definition}
Let \(\mathbb{F}(p^{m})\) is an optimal extension field, \textit{m} is size of the exponent of the field, \textit{p} is the characteristic of field, on which extension field for multiplicative group is as shows in Eq. \ref{eq:extension}.

\begin{equation} \label{eq:extension}
    Extension Field = \mathbb{F}_{m}^{*}
\end{equation}

\end{definition}

\subsubsection{Discriminant Size}

A discriminant size is an important parameter in selecting domain parameters that can greatly enhance the speedup process of the discrete logarithm problem over different curves. For example, a choice of small discriminant size is computationally efficient, which can speed up the scalar-multiplication algorithm. However, the choice of this discriminant can also lead to the computational complexity of the discrete logarithm computation that greatly affects the security of the ECDLP on the curve. A multiplication discriminant of an elliptic curve over prime field is defined as follows:

\begin{definition}

Let \textit{E} is a elliptic curve, \(\mathbb{F}_p\) is a prime field, \textit{M} is a non-negative integer representing multiplication discriminant and thus can be computed as in Eq. \ref{eq:discriminant}:

\begin{equation}\label{eq:discriminant}
    4\textit{p} = \textit{t}^2 - \textit{s}^2\textit{M}
\end{equation}

Where \textit{s} $\in$ \(\mathbb{N}\) and |\textit{E} ($\mathbb{F}_p$)| = \textit{p} + 1 - \textit{t}.

\end{definition}

\subsection{Attacks on Domain Parameters Implementation}

Apart from meeting security criteria when selecting domain requirements for different cryptographic algorithms based on ECDLP, there are still adherent complexity concerns when implementing ECDLP over different curves, resulting in a variety of different types of attacks; therefore, must be tackled in order to ensure secure implementation of cryptographic algorithms \cite{janvcarsecurity}. The following are the most common types of attacks on domain parameter implementation, each defined in detail.

\subsubsection{Side-channel Attack}
Side-channel attacks on the implementation of domain parameters result in data leakage through various side channels, the majority of which include the timing and power consumption factors of the underlying systems, for which direct data leakage is not possible.

In the former factor, the time required to complete the required cryptographic operations often creates the side channel issues that mostly leak the values of different secret primitives such as the private key used in operation. A timing attack is also possible if the attacker has access to sufficiently precise measurements of the time required to perform a given cryptographic operation, even over a network link. However, network latency makes the attack more difficult. However, timing attacks on systems based on elliptic curves are still in their infancy, as an attacker can typically leak only a few bits of the private key through timing \cite{hankerson2006guide}.

In the latter case, the power consumption of hardware devices such as sensors that conduct cryptographic operations may also serve as a side-channel, leaking significantly more information than the timing side-channel does. The rationale for impacting the power consumption side channel on elliptic curves is that, in most cases, a power trace of the execution of a cryptographic operation involves more details than just the time factor of the operation. Therefore, this attack is only possible with physical access to the device performing the cryptographic operation, such as a clone node attack. There are two types of power analysis attacks: simple power analysis (SPA) and differential power analysis (DPA). In simple power analysis, an attacker manually examines the power trace to discover a cryptographic secret that was employed; however, in differential power analysis, an attacker employs statistical techniques to retrieve minor differences in power usage associated with hidden data through a massive collection of power traces \cite{cohen2005handbook}.

\subsubsection{Fault Attack}

In a fault attack, an attacker gains physical access to devices in the deployed environment that are performing cryptographic operations, potentially modifying their physical and logical states, such as location and time, in order to insert incorrect code and perform malicious actions. In addition, this activity resulted in acquiring secret information, such as private keys, in committing further malicious acts on behalf of target devices. For example, an invalid curve attack is used as a fault attack on scalar multiplication over several curves in order to obtain the computation results as secret values \cite{biehl2000differential}.

\subsubsection{Small-subgroup Attack}

Small-subgroup attacks are primarily directed at the Diffie-Hellman protocol's implementation of cryptographic keys and thus apply to ECC. This attack is classified into two types: key confinement and key recovery.

The small-subgroup key confinement attack can happen if an implementation does not verify that untrusted points obtained during EC are located on the relevant subgroup of the generated curve. Additionally, this attack requires that the curve lacks a prime order and contains several small-order subgroups in which small points on the curve's order constrain the derived shared EC key to a small subset of keys \cite{valenta2017measuring}. 

The small-subgroup key recovery attack meets the same validity criteria as the small-subgroup key confinement attack, except that the victim's key cannot be amorphous. This attack is quite similar to an active Pohlig-Hellman attack in which an attacker computes various points on small-order subgroups to implement an elliptic curve. This attack employs the Chinese Remainder Theorem, which reconstructs the victim's key by repeatedly performing the small-subgroup key confinement attack on numerous different subgroups of the curve group \cite{lim1997key}.

\subsubsection{Invalid Curve Attack}
When implementing elliptic curves, affine coordinates and the short Weierstrass equation often ignored untrusted points on the group of the intended curve, resulting in invalid curve attacks. An adversary uses the curve point \textit{P} with different parameters as the public keys in this attack. As a result, the attacker exploits and compares the different public critical values on the compromised curve to the public keys on the original curves during the scalar multiplication process to determine the collection of private keys for the victim. Once an attacker has obtained all of the secrets, such as private keys, he may launch an attack against various secure protocols, such as SSL/TLS, using an elliptic curve on the targeted points \cite{biehl2000differential, antipa2003validation}.
\subsubsection{Twist Attack}

The addition formulas used to implement an elliptic curve apply to the original curve and all nontrivial quadratic twists. However, a nontrivial quadratic twist on a curve has a different number of points than the original curve and hence can have an entirely different group structure than the original curve \cite{cohen2005handbook}.

As with invalid-curve attacks, an attacker can use a twist attack only if the implementation does not verify that untrusted points are within the group of the intended curve. This shortcoming enables an attacker to supply twist points during the ECDH exchange, compelling the victim to compute the twist's scalar multiple. Consequently, if the twist's point security is less than the original curve's point security, the overall security of the implementation deteriorates to the twist's point security \cite{fouque2008fault}.

\subsubsection{Co-factor Computation Attack}
In a co-factor computation attack, an issue is raised when points on a curve over a prime order subgroup with a sufficiently large prime are calculated and improperly validated, posing severe challenges to the protocol or system's performance \cite{janvcarsecurity}.

\subsection{Selection of Key Pairs}

A pair of ECDSA keys is associated with a set of EC domain parameters. ECDSA key pairs can be guaranteed in the following steps: key pair generation, public key validation, and proof of private key ownership, each of which ensures the security mechanisms and meets the requirements necessary to secure the underlying system or protocol. 

\subsubsection{Key Pair Generation}

In our proposed LPS, the key pair for each device (e.g., prover, verifier, gateway) is associated with a specific set of EC domain parameters D = (q, FR, a, b, G, n, h). This association can be accomplished by cryptographically secure certificate authorities or context requirements requiring all entities to use the same domain parameters. In domain parameters, each individual must be certain that the domain parameters are acceptable to all before generating keys.

Each entity in the LPS, such as prover \textit{P}, verifier \textit{V}, and gateway \textit{G}, must generate their public key (\textit{Q}) and private key (\textit{d}) according to the steps below.
\begin{itemize}
    \item Choose a pseudo-random integer d from the range [1, n-1].
    \item Compute Q = dP
\end{itemize}

The mechanism by which keys are generated is also described in detail in algorithm \ref{alg:keygeneration}. 

\subsubsection{Validation of public key}
Public key validation guarantees that a public key has the necessary arithmetic properties but does not indicate that someone has computed the private key or has asserted ownership. Public key Validation should be performed to avoid inserting wrong keys and catch programming errors and omissions. Since the use of an invalid public key nullifies all intended security properties, several suggested methods for validating the public keys given to each individual in the LPS.

\begin{itemize}
    \item A public key \textit{Q} is generated for each entity, which is certified by a trusted authority \textit{TA}.
    \item A trusted authority \textit{TA} produces a public key \textit{Q} for each entity in the system, which is then transmitted through a secure communication channel.
    \item Each entity obtains confirmation from a trusted authority \textit{TA} that it has followed certain explicit public key validation protocols specified in \cite{antipa2003validation}.
    
\end{itemize}

\subsubsection{Proof-of-ownership of a private key}

Proof-of-ownership of a private key is a technique for establishing that the sender of a message holds a particular key. This process is used to determine that the intended recipient sent the message, assuming that the sender is the only one who possesses the key. 

This concept is demonstrated in our LPS in the following way: suppose an attacker compromises the  \textit{P} and creates a clone node \textit{C}, then a clone node \textit{C} can verify the prover \textit{P}'s public key \textit{Q} using its public key, then \textit{C} will claim that \textit{P}'s signed messages originated from \textit{C} as well. To avoid this from occurring in any system, especially those that use public-key cryptography, the trusted authority \textit{TA} should require all individuals involved in the system to demonstrate ownership of the private keys associated with their public keys before certifying the public key as belonging to an entity. There are two widely used methods for providing proof-of-ownership of a private key \textit{d}, which are as follows:
\begin{itemize}
    \item Each entity in the system is needed to demonstrate ownership of a private key \textit{d} through signatures; however, this method adds additional computation overhead to the system, especially for those using resource-constrained IoT devices.
    \item Additionally, zero-knowledge proofs can be used to establish ownership when the trusted authority \textit{TA} acquires no new knowledge about the entity's private key \textit{d}.
\end{itemize}

\subsubsection{Key-only Attack}
The key-only attack is a type of attack on the public key of an entity in which an attacker obtains the public key of the signer party, which is then used to verify the valid signatures of the signer party.

To explain the idea of a key-only attack, consider our suggested LPS. In this system, prover \textit{P}'s public key \textit{Q} is publicly accessible, and adversary $\adv$ takes advantage of this aspect by attempting to replicate prover \textit{P}'s signature $sig_o$ and sign messages ($m_1$, $m_2$, $m_3$, \dots, $m_k$) that prover \textit{P} does not intend to sign.

\subsection{Attacks on Signatures}
A digital signature is a cryptographic technique that combines the computational capabilities of hash functions to verify the message's authenticity and provide non-repudiation, meaning that the sender cannot deny endorsing the document. Although the digital signature is valuable for protecting sensitive information, it is unfortunately highly vulnerable to various attack vectors. For example, an adversary $\adv$ has the following goals associated with the digital signature mechanism to break any signature scheme.
\begin{itemize}
    \item \textbf{Total break:} In total break, an adversary $\adv$ successfully obtains secrets such as the public key \textit{Q} and the private key \textit{d} of any prover \textit{P}, allowing him to forge any signature $sig_{i}$ on any message $m_{i}$ of his choosing.

    \item \textbf{Selective Forgery:} In selective forgery, an adversary $\adv$ has a high probability of producing a valid signature $sig_{r}$ on any selective message $m_{r}$ chosen at random.
    \item \textbf{Existential Forgery:} In existential forgery, an adversary $\adv$ generates at least one message/signature pair, such as (message, signature) defined as (\textit{m}, \textit{sig}), where \textit{m} has never been signed by the valid signer. The adversary can freely choose \textit{m} from a list of messages ($m_1$, $m_2$, $m_3$, \dots, $m_k$) that have no valid meaning. The adversary has prevailed in creating an existential forgery as long as the pair (\textit{m}, \textit{sig}), is valid.
\end{itemize}

Attacks on digital signatures can be classified into three different types:

\subsubsection{Chosen-message Attack}
In a chosen-message attack, the adversary can choose the messages that the signer party wants to sign, and the adversary is aware of both the messages and their associated signatures. The chosen message attack is further divided into three different categories of attacks as described below.
\begin{itemize}
    \item \textbf{Generic Chosen-message Attack:} The adversary uses this approach to trick the signer party into digitally signing the messages without knowing the signer party's public key. The term ``generic'' refers to the fact that the attacker is unaware of the public keys, and an entire list of messages is created and chosen before digitally signing them.
    
    \begin{definition}
    Let $\adv$ is an adversary who has access to the prover's \textit{P} valid signatures for a list of messages ($m_1$, $m_2$, $m_3$, \dots, $m_k$) before attempting to crack the prover \textit{P}'s overall signature scheme. The entire set of messages for which digital signatures are produced is chosen randomly and is completely independent of public keys.
    \end{definition}

    \item \textbf{Directed Chosen-message Attack:} The directed chosen-message attack is very similar to the generic chosen message attack in which the adversary choose the messages before creating the signatures for the target entity. However, the only difference is that the adversary knows the target's public key before signature generation. Thus, the term  ``directed'' means that the attack is against a specific entity for which the public key is known. 
    
\begin{definition}
Let assume that adversary  $\adv$ has full knowledge of targeted prover \textit{P}’s public key \textit{Q} and obtains \textit{P}’s signature $sig_o$ on the list of messages ($m_1$, $m_2$, $m_3$, \dots, $m_k$) and tries to substitute the original message $m_o$ with the message $m_t$ that adversary $\adv$ intends to have targeted prover \textit{P} sign thus keeping \textit{P}’s signature unchanged.
\end{definition}
    
 \item \textbf{Adaptive Chosen-message Attack:} Like a directed chosen-message attack, an adaptive chosen-messages attack targets an entity's signature after obtaining his public key; however, the difference is that an adversary can select the message for signature after observing a list of messages and their associated signatures. Additionally, the adversary who obtains signatures for messages sent to his target cannot subsequently counterfeit the signature of a single message.

    \begin{definition}
       Let an adversary $\adv$ has targeted any prover \textit{P} after inspecting \textit{k} valid signatures for the message ($m_1$, $m_2$, $m_3$, \dots, $m_k$) chosen adaptively. An adversary $\adv$ also knows the public key \textit{Q} of prover \textit{P}. For an adaptive chosen-message attack, the adversary obtains a list of signatures ($sig_1$, $sig_2$, $sig_3$, \dots, $sig_k$) for its chosen list of messages ($m_1$, $m_2$, $m_3$, \dots, $m_k$), but an adversary $\adv$ can also request an additional signature $sig_{r_i}$ for message $m_{r_i}$ which is depend on previously signature $m_{r_i-1}$.

    \end{definition}

\end{itemize}

\subsubsection{Known-message Attack}
In a known-message attack, the adversary has access to specific existing signatures against a known collection of messages; however, unlike in a chosen message attack, the adversary does not choose the messages. In encryption, this attack is analogous to a well-known attack called a plain text attack.

\begin{definition}
Let an adversary $\adv$ has a limited number of previous messages  ($m_1$, $m_2$, $m_3$, \dots, $m_k$) and their corresponding signatures ($sig_1$, $sig_2$, $sig_3$, \dots, $sig_k$) of prover \textit{P}. Now adversary $\adv$ attempts to forge prover \textit{P}'s signature $sig_o$ on a message $m_o$ that prover \textit{P} does not wish to sign using the brute force method of examining previous messages ($m_1$, $m_2$, $m_3$, \dots, $m_k$) in order to replicate prover \textit{P}'s signature $sig_o$. 

\end{definition}

\subsection{Attacks on Hash Functions}

Hash functions are complex computational functions that are deliberately designed to be used in a one-way manner. For example, it is trivial to compute the hash value of any given messages; however, if someone has computed the hash value of some messages, it should be challenging to determine the original messages that produced the same hash value. Cryptographic hash functions are known to be one of the most complicated parts to break from a cryptography perspective \cite{rogaway2004cryptographic}. 

Let $\hash$ is defined as a hash function that is a member of the hash function family. Suppose, $\hash$: \textit{X} $\times$ \textit{Y} $\rightarrow$ \textit{O} where \textit{X} and \textit{O} are finite non-empty sets and \textit{Y} and \textit{O} are sets of strings. \textit{O} is the output of the hash function $\hash$ for which \textit{O} = $\{0,1\}^n$ where \textit{n} $>$ 0 and \textit{n} is expressed as the length of the hash function value. A random message \textit{m} is choose from the infinite set $\mathscr{S}$ for uniform distribution and expressed as \textit{m} $\leftarrow$ $\mathscr{S}$ . For random message \textit{m}, if \textit{m} $\in$ \textit{Y} then $\{0,1\}^{|\textit{m}|}$ $\subseteq$ \textit{Y}. A hash function $\hash$ for which hash value is calculated is expressed as $\hash_{x}(m)$ = $\hash$ (\textit{x}, \textit{m}) for all \textit{m} $\in$ \textit{Y} and \textit{x} $\in$ \textit{X}.

A cryptographically secure hash function must possess the following properties:
\begin{itemize}
    \item \textbf{Preimage Resistance:} Preimage resistance is a property of a hash function that specifies that it should be computationally infeasible to find an input that maps to a given element in the hash function's range and that it should be difficult to invert.
    
    \begin{definition}
    Let \textit{m\textquotesingle} is a pre-image such that \hash(\textit{m\textquotesingle}) = \textit{n} where any \textit{n} is given for which the corresponding input is unknown.
    
    \end{definition}

    \item \textbf{Second Preimage Resistance:} Second preimage resistance is another important property of a hash function that specifies that it is computationally impossible for someone to find a second input from a hash function that produces the same output as the first input. 
    \begin{definition}
    Let $m_{1}$ is a given input for which hash is determined as $\hash(m_{1})$; it should be difficult to find another input (i.e. second preimage) $m_{2}$, i.e. $m_{1}$ $\neq$  $m_{2}$, for which $\hash(m_{1})$ = $\hash(m_{2})$.
    \end{definition}

    \item \textbf{Collision Resistance:} Collision resistance in hash function implies that it is difficult to find two distinct inputs that produce identical hash values. For instance,

    \begin{definition}
    Let $m_{1}$ and $m_{2}$ are any two distinct input values, which hash to the same output, such that $\hash(m_{1})$ = $\hash(m_{2})$.
    \end{definition}
    
\end{itemize}

\subsubsection{Preimage Attack}
In a preimage attack, the attacker attempts to locate a message that contains a particular hash value. For instance, with a preimage attack, one tries to find a message with a particular hash, such as finding a hash that leads to a preimage for a hash.

\begin{theorem}
Let $\hash$: \textit{X} $\times$ \textit{Y} $\rightarrow$ \textit{O} be a hash function, \textit{n} be a number such that $\{0,1\}^n$ $\subseteq$ \textit{Y}  and $\adv$ be an adversary, then such as:

\begin{align*}
\adv_\hash^{Preimage} (P) = [ Preimage \textit{x} \leftarrow \textit{X}; m_1 \leftarrow \{0,1\}^n; \textit{O}  \leftarrow \hash_x(m_{1}) ;m_{2} \leftarrow P(x,O): \\ \hash_x(m_{2}) =  \textit{O} ]
\end{align*}

\begin{align*}
 \adv_\hash^{Preimage}(P) = \underset{\textit{O} \in \textit{Y} }{max} \{Second-Preimage[\textit{x} \leftarrow \textit{X}; m_{1} \leftarrow P(x): \hash_x(m_{1}) =  \textit{O} ]\}    
\end{align*}

\begin{align*}
  \adv_\hash^{Preimage} (P) = \underset{\textit{x} \in \textit{X} }{max} \{Second-Preimage[m_{1} \leftarrow \{0,1\}^n; \textit{O}  \leftarrow \hash_x(m_{1}); \\ m_{2} \leftarrow P(O): \hash_x(m_{2}) =  \textit{O}    
\end{align*}

\end{theorem}

Where \textit{P} is the prover in LPS, \textit{x} $\in$ \textit{X}: for any fixed hash function $\hash$: \textit{Y} $\rightarrow$ \textit{O} with $|$\textit{Y}$|$ $>$ $|$\textit{O}$|$ that generates the hash values for an input $m_{1}$  for which $\hash$ is used to determine its value.

The above three are the different definitions of second-preimage resistance representing as simple second-preimage resistance, everywhere second-preimage resistance, and always second-preimage resistance. 

\subsubsection{Second Preimage Attack}
In a second preimage attack, the adversary's goal is to find a second message for which the first message is given, and their hashes values are equal. In other words, an adversary already has an input and is attempting to determine the second input whose hash values are identical.

\begin{theorem}
Let $\hash$: \textit{X} $\times$ \textit{Y} $\rightarrow$ \textit{O} be a hash function, \textit{n} be a number such that $\{0,1\}^n$ $\subseteq$ \textit{Y}  and $\adv$ be an adversary, then such as:

\begin{align*}
   \adv_\hash^{Second-Preimage} (P) = Second-Preimage [\textit{x} \leftarrow \textit{X}; m_1 \leftarrow \{0,1\}^n; \\ m_2 \leftarrow P(x,m_1): (m_1 \neq m_2) \land (\hash_x(m_1) = (\hash_x(m_2)) ] 
\end{align*}

\begin{align*}
    \adv_\hash^{Second-Preimage} (P) = \underset{m \in \{0,1\}^n }{max} \{Second-Preimage[\textit{x} \leftarrow \textit{X}; \\ m_2 \leftarrow P(x):(m_1 \neq m_2) \land (\hash_x(m_1) = (\hash_x(m_2))]\} 
\end{align*}

\begin{align*}
   \adv_\hash^{Second-Preimage} (P) = \underset{\textit{x} \in \textit{X} }{max} \{Second-Preimage[m_1 \leftarrow \{0,1\}^n;\\  m_2 \leftarrow P(m_1): (m_1 \neq m_2) \land (\hash_x(m_1) = (\hash_x(m_2))]\}  
\end{align*}

\end{theorem}

Where \textit{P} is the prover in LPS. \textit{x} $\in$ \textit{X}: for any fixed hash function $\hash$: \textit{Y} $\rightarrow$ \textit{O} with $|$\textit{Y}$|$ $>$ $|$\textit{O}$|$ that compares the hash values for two inputs $m_{1}$ and inputs $m_{2}$ that determines their values are equal. 

The above three are the different definitions of second-preimage resistance representing as simple second-preimage resistance, everywhere second-preimage resistance, and always second-preimage resistance.

\subsubsection{Collision Attack}

A collision attack on a cryptographic hash function attempts to discover two inputs that produce the same hash value referred to as a hash collision. The collision attack is in contrast to a conventional hash attack, in which an attacker attempts to discover a single hash value and has no effect on previously computed hashes.

In our proposed LPS, the collision attack is defined as follows: Let an adversary $\adv$ discover two distinct messages. $m_{1}$ and $m_{2}$ in such a way that \hash($m_{1}$) = \hash($m_{2}$).

\begin{theorem}
Let $\hash$: \textit{X} $\times$ \textit{Y} $\rightarrow$ \textit{O} be a hash function and $\adv$ be an adversary, then such as:

\begin{align*}
    \adv_\hash^{Collision} (P) = Collision-Resistance[\textit{x} \leftarrow \textit{X}; (m_{1}, m_{2}) \leftarrow P(x): \\ (m_{1} \neq  m_{2}) \land (\hash_x(m_1) = (\hash_x(m_2))]
\end{align*}

\end{theorem}

Where \textit{P} is the prover in LPS, \textit{x} $\in$ \textit{X}: for any fixed hash function $\hash$: \textit{Y} $\rightarrow$ \textit{O} with $|$\textit{Y}$|$ $>$ $|$\textit{O}$|$ that generates the hash values for two inputs $m_{1}$ and inputs $m_{2}$ that collide under $\hash$.

\section{Evaluation Framework and Performance Analysis} \label{simulations}

To examine the applicability and robustness of our proposed system for detecting cloned node attack on the IoT environment, we built the evaluation framework of our proposed scheme by using a C language. We deployed a cluster system having ten machines including one master and nine computing machines. Each machine has 20 cores and each core run with the 2.11 GHz and 8 GB of memory. We deployed a cluster system comprised of ten machines, one of which served as the master and nine of which served as computing nodes. Each machine contains twenty cores, each of which operates at a CPU of 2.11 GHz and has eight GB memory capacity. In addition, we created a simulation of each machine using the 40 IoT devices. Thus, to simulate the concept of multiple IoT devices, we used multi-threading to simulate the environment of multiple IoT devices, in which multi-threading divides a single processor into multiple required threads. Since each thread corresponds to one sensor node, mainly defined for the required task. We communicate with other machines in the cluster system via the MPI library available for C. For batch verification using ECDSA*, we utilised the prime curve (P-256), in which all parameters evaluated for experimentation are NIST-recommended traditional values.

We evaluated a network with between 100 and 500 IoT devices in our experiment. From these devices, we choose 350 that need to be verified as clone nodes in the network. We allocate 30 verifiers to the batch verification tasks. The remaining devices are deliberately deployed as clone nodes to determine the detection rate for various test cases. We conducted a series of experiments to assess the performance of our proposed scheme for detecting cloned nodes and analysed the findings using a variety of evaluation parameters. These parameters include the attack detection probability, detection time, computational overhead, communication overhead, and storage overhead. We compared our experimental findings, and overhead with two existing state-of-the-art mechanisms \cite{shanmugam2020two} and \cite{anitha2021intelligent}. The results analysis shows that our proposed scheme outperforms detecting clone node attack on IoT networks with a high detection rate in a reasonable amount of time. Additionally, our proposed scheme has a lower computational, communication, and storage overhead than existing techniques \cite{shanmugam2020two} and \cite{anitha2021intelligent}.

\subsection{Attack Detection Analysis}

\begin{figure}[!htp]
   \begin{minipage}{0.48\textwidth}
     \centering
     \includegraphics[width=1.0\linewidth]{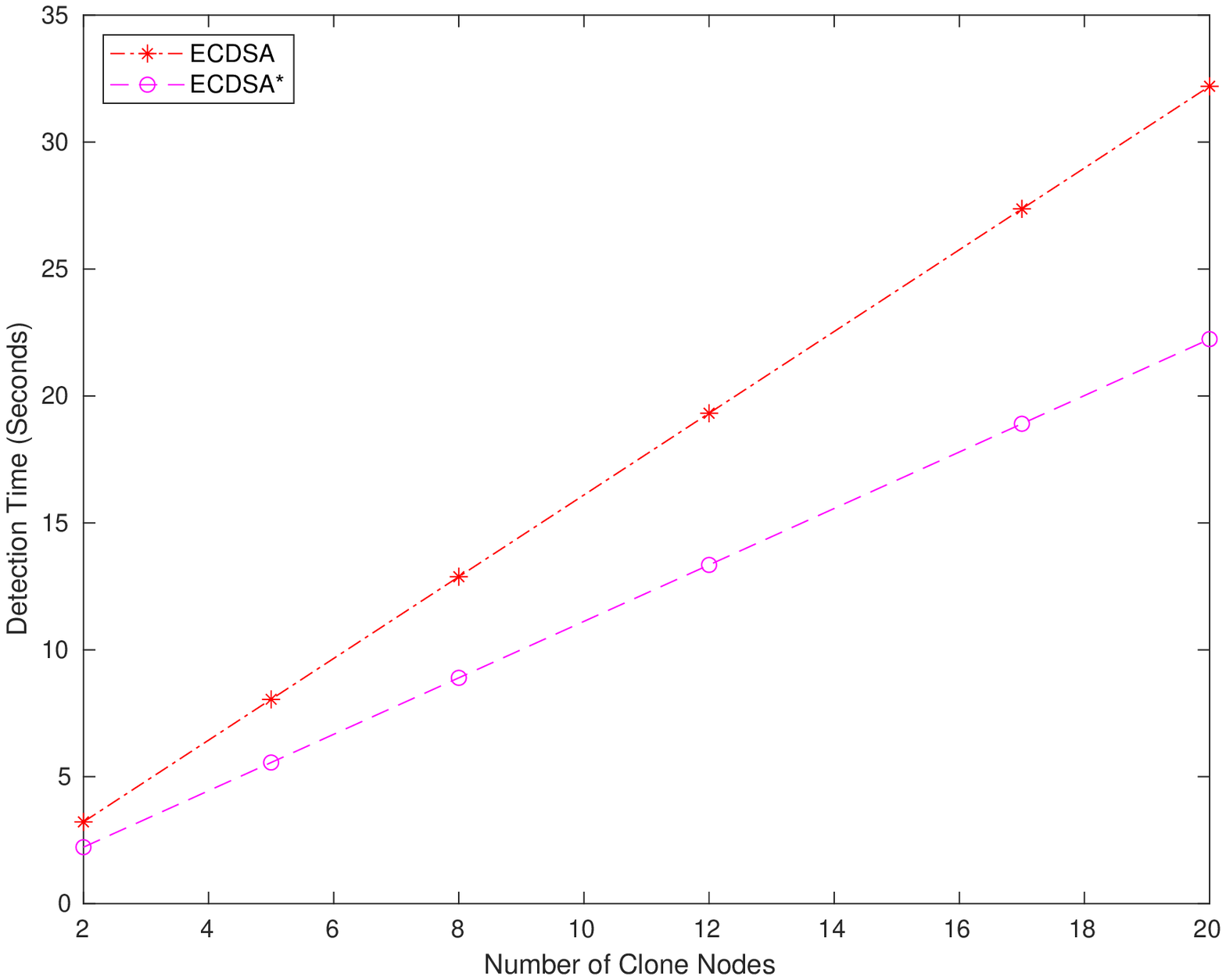}
     \caption{Detection Time - Sparse Environment}\label{clone1}
   \end{minipage}\hfill
   \begin{minipage}{0.48\textwidth}
     \centering
     \includegraphics[width=1.0\linewidth]{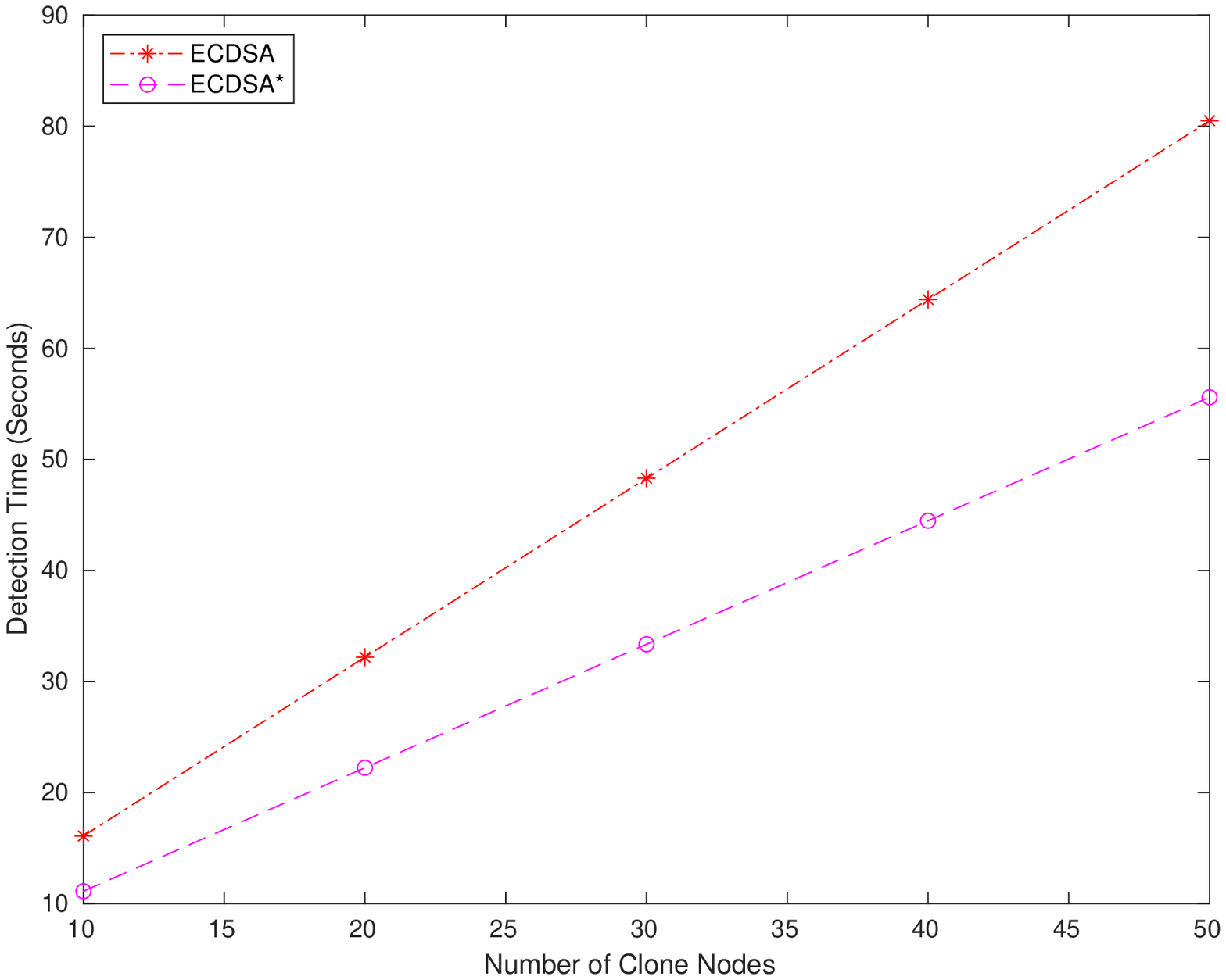}
      \caption{Detection Time - Dense Environment}\label{clone2}
   \end{minipage}
\end{figure}

We analysed the detection of clone node attack in our proposed scheme using two parameters: detection probability and detection time. The detection probability estimates the probability that cloned or replica nodes will be accurately detected. In contrast, the detection time indicates the time required to detect clone nodes attack on our network successfully. Each of which is discussed and measured in more detail in the following sub-sections:

Further, we considered two different types of environments when configuring the clone nodes setup.

\begin{itemize}
\item \textbf{Sparse Environment:} The network is built and operated using a limited number of IoT devices in a sparse device environment. We choose 20 devices from the total number of available devices on the network to operate as clone devices in our setup.
\item \textbf{Dense Environment:} The number of clone devices in a dense setup ranges between 25 and 50.
\end{itemize}

\subsubsection{Detection Probability}
The detection probability can be described as the proportion of successfully detected cloned nodes divided by the total number of cloned nodes in the deployed IoT ecosystem.

Since our proposed scheme detects clone node attack based on context information collected by both the prover and the verifier and then stored in LBS for verification purposes. Each prover must prove its location by creating location proofs at varying intervals, which are then validated by the network's selected trusted nodes known as verifiers. The verification step verifies the sensed context information to the data stored at the LBS. The proposed scheme determines whether or not a device has been compromised by successfully verifying and matching such information. The detection of clone nodes resiliency in our proposed scheme is based on two scenarios:

\begin{itemize}
\item \textbf{Case I:} An adversary $\adv$ compromises a random prover \textit{P}, replicates it and creates a clone node \textit{P'}  of it by extracting all of the prover's credentials, including the context information \textit{CI}, and deploying it in several locations throughout the network. In a network model, all network nodes are represented as an undirected graph \textit{G} = (\textit{V}, \textit{E}), where \textit{V} and \textit{E} denote a collection of nodes and edges, respectively. To detect clone node attack, suppose a verifier \textit{V} generates and sends a location proof request to a prover \textit{P}. When the adversary $\adv$ receives a location proof request at clone node textit{P}, it generates the signature for the location proof $P_{sign}$ and sends it to the verifier \textit{V} for verification. Once the verifier \textit{V} receives the $P_{sign}$, it performs the ECDSA* on the location proof and, if the signature does not include context information, \textit{V} determines that the prover \textit{P} has been compromised.

\item \textbf{Case II:} An adversary $\adv$ compromises a random prover \textit{P}, replicates it and creates a clone node \textit{P'} of it by collecting all of the prover's credentials, including the context information \textit{CI}, and deploying it in various locations throughout the network.  To identify clone nodes, consider that a verifier \textit{V} makes a location proof request and sends it to the cloned node \textit{P'}. When the clone node \textit{P'} received a location proof request, the verifier \textit{V} compared the cloned context information \textit{CCI} to the \textit{CI} stored at LBS. However, because the location information of the original nodes been changed during the cloning process, \textit{CI} $\neq$ \textit{CCI}, the verifier \textit{V} can infer that prover \textit{P'} is the compromised node in the network.

\end{itemize}

By considering both cases for clone node attack detection on our proposed LPS, we demonstrate that our proposed detection scheme achieves the maximum detection resiliency and provides the highest probability of clone node attack detection. However, the latter situation is more computationally efficient because it simply requires verifying and matching the information with the information stored in the LBS, as opposed to the other case, which requires verifiers to perform ECDSA* batch verification to validate the signatures. As a result, we can confidently assert that our proposed scheme achieves the highest detection probability of 100\% in both cases. Finally, we compare the detection probability of our proposed scheme to that of the existing schemes \cite{shanmugam2020two} and \cite{anitha2021intelligent}, which have detection rates of 90\% and 92.5\%, respectively.

\subsubsection{Detection Time}

The detection time is required for the protocol or system to identify the network's clone nodes successfully. The detection time in our proposed scheme is the time consumed from generation to verification of location proofs to detect clone node attack. We determine the detection time for both sparse and dense environments. Fig. \ref{clone1} illustrates the detection time of a clone node attack in a sparse environment, demonstrating the significantly faster detection of an attack when employing the ECDSA*. Fig. \ref{clone2} illustrates the detection time for clone nodes in a dense environment.

The detection time analysis reveals that our proposed scheme delivers a more efficient detection rate for clone node attack detection when compared to conventional ECDSA.

\subsection{Computational Overhead}

Another significant factor in determining the efficiency of our proposed scheme is the computation time. A computational time is defined in our proposed scheme as the time required to perform various operations such as key generation for provers and verifiers, signature generation for provers to generate location proofs, and verification time for verifiers to verify the location proofs. Our proposed scheme is based on the assumption that LBS is a trustworthy entity capable of performing cryptographic operations without processing or storage constraints. The computation time required to perform each operation in the proposed scheme is detailed below.

\subsubsection{Key Generation}
In our proposed scheme, we generate a pair of ECDSA keys with the P-256 (secp256k1) curve for signing and verifying location proofs to detect clone node attack in the IoT environment. Even though our LPS is based on the batch verification process for ECDSA*, the key generation process for both ECDSA and ECDSA* is identical. The process of generating ECDSA* keys is described in detail in the algorithm \ref{alg:keygeneration}. The length of keys is specified in bits, and the time required to generate them is calculated in seconds. Thus, the security requirements of ECDSA are met proportionately by a shorter key length, and similar levels of security are achieved by other cryptographic algorithms such as RSA. A significant benefit of a smaller key size is that computations can be performed more quickly, reducing storage space, processing power, power consumption, and bandwidth.

\begin{figure}[h]
    \centering
    \includegraphics[width=10cm, height=6cm]{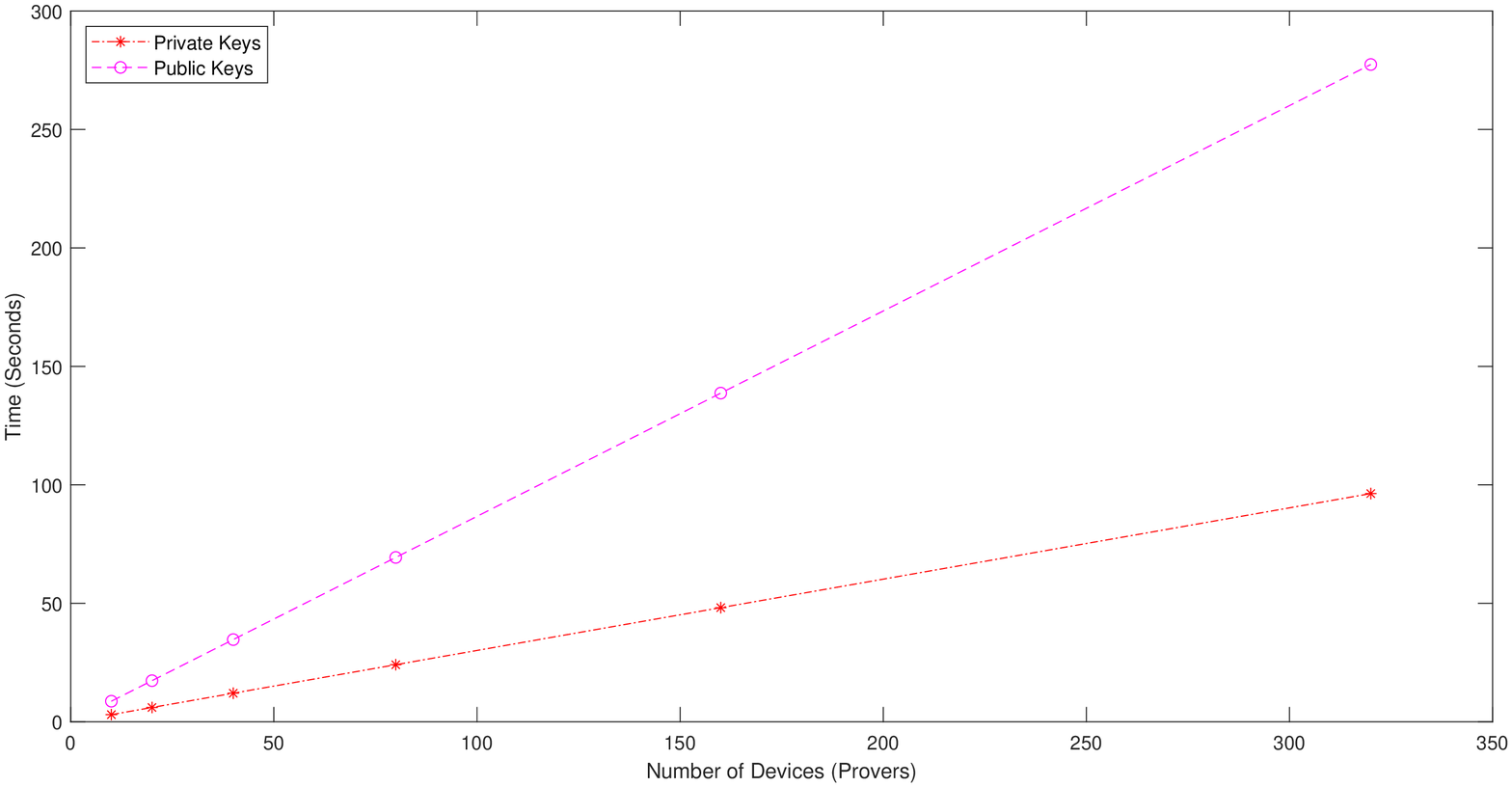}
    \caption{Public and Private Keys (Provers)}\label{proverskeys}
\end{figure}

\begin{figure}[!htb]
    \centering
      \includegraphics[width=10cm, height=6cm]{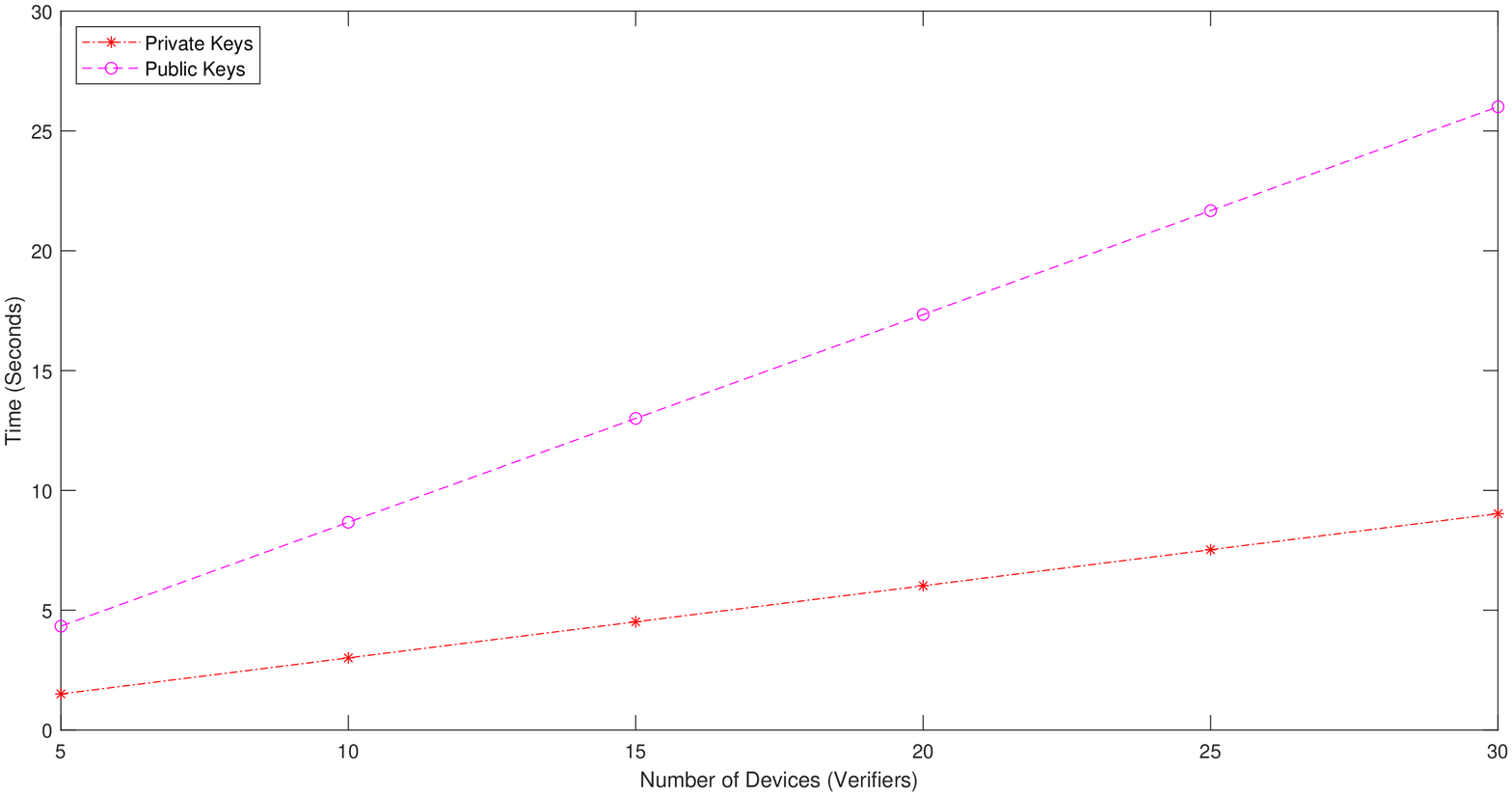}
      \caption{Public and Private Keys (Verifiers)}
\label{verifierskeys}
\end{figure}

We calculate the time required to generate the public and private keys for the provers and verifiers in LPS. Each device is required to generate and verify location proofs. For example, the prover's private key is used to generate location proofs as a signature, verified using the prover's public key. Fig. \ref{proverskeys} specifies the time in seconds required to generate the public and private keys of provers. The analysis demonstrates that private keys take less time to generate than public keys. Public keys are generated by multiplying the respective private key with an elliptic curve generator point. Similarly to the generation of public and private keys of provers, Fig. \ref{verifierskeys} illustrates the time in seconds required to generate public and private keys of verifiers. Similarly to the generation of public and private keys of provers, Fig. \ref{verifierskeys} illustrates the time required in seconds to generate the public and private keys of verifiers.

Additionally, we determine the cost of crucial generation in terms of the total number of \textit{N} provers and verifiers, which is $\mathcal{O}($N$)$.

\subsubsection{Location Proof Generation} 

\begin{figure}[!htp]
   \begin{minipage}{0.48\textwidth}
     \centering
     \includegraphics[width=1.1\linewidth]{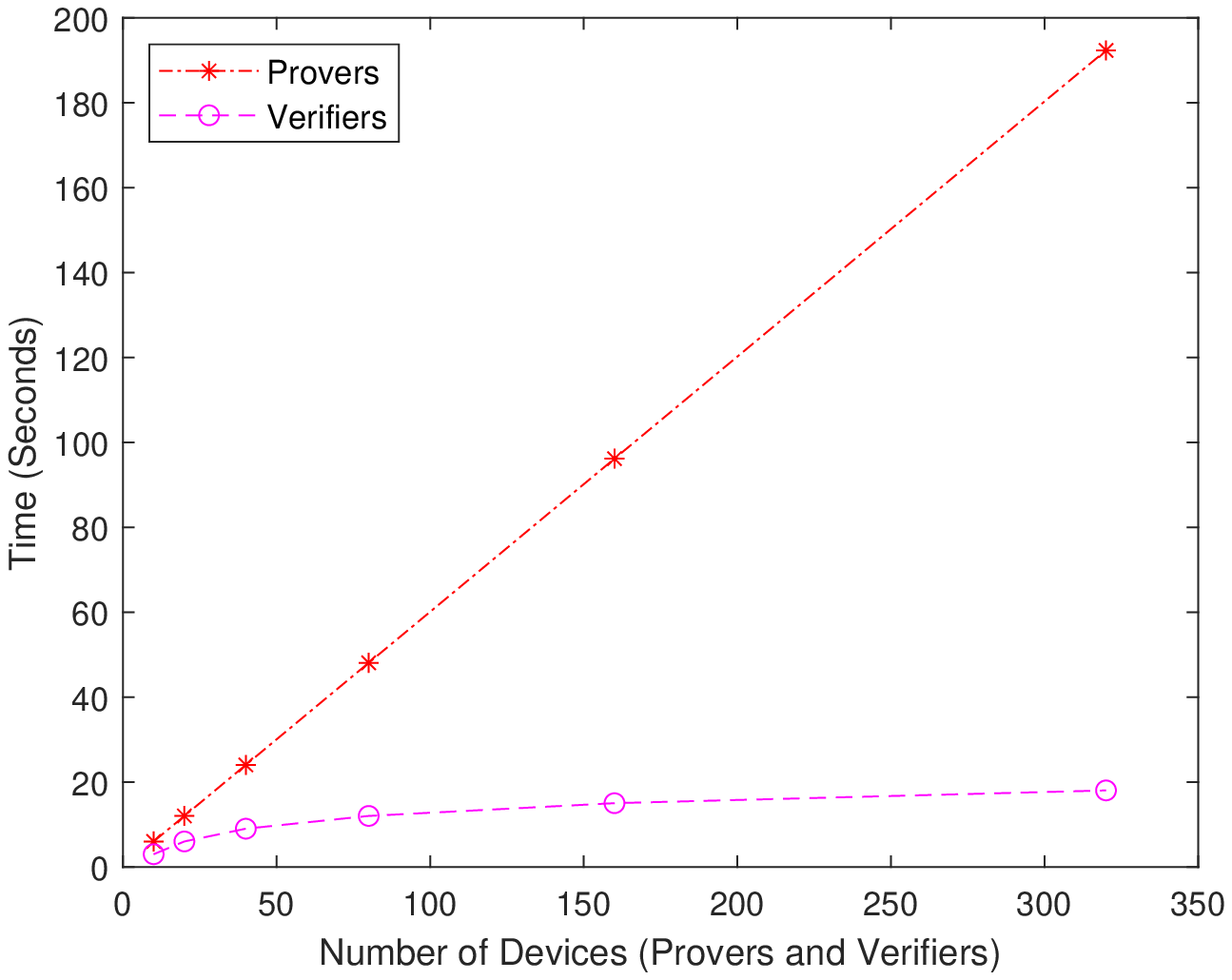}
     \caption{Generating location proofs for Provers and Verifiers}
    \label{signatures}
   \end{minipage}\hfill
   \begin{minipage}{0.48\textwidth}
     \centering
     \includegraphics[width=1.1\linewidth]{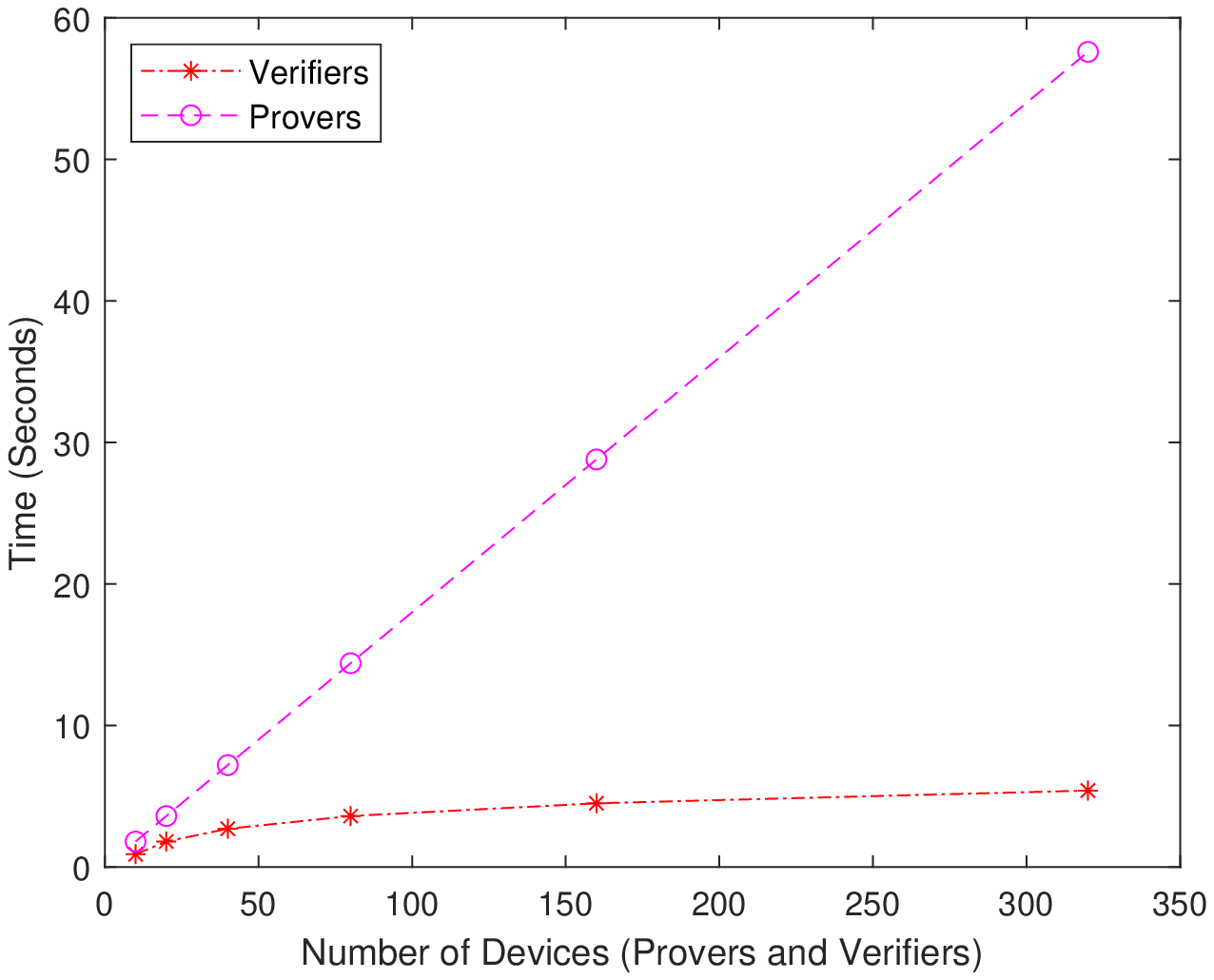}
    \caption{Location Estimation of Provers and Verifiers}
    \label{location}
   \end{minipage}
\end{figure}

The process of generating location proofs in an LPS is intimately linked to the process of generating signatures, as illustrated and explained in the algorithm \ref{alg:signaturegeneration}. To detect clone node attack on IoT networks using an LPS, each verifier requests proof of location from the provers to determine whether or not the prover has been compromised. To formulate the location proof, each prover uses its private key to sign its sensed data, referred to as context data, and sends it to the verifier. The signature process converts the sensed context information into a hash value using various hash functions. We used the secp256k1 curve in ECDSA* to generate the cryptographic keys; we generated the hash value of the context information using the SHA256 hash function. The computational cost of generating the signatures for provers and verifiers is depicted in Fig. \ref{signatures}. However, the goal of our proposed scheme is to validate the location proofs given by provers. Therefore, we are only concerned with the computational cost of provers' signatures.

Additionally, as indicated in the working mechanism of LPS, the location of each device is essential for identifying between original and clone nodes, and it is computed using the Euclidean distance specified in the location calculation algorithm \ref{alg:proofgeneration}. We estimated the average computational time required to calculate the locations of deployed devices such as provers and verifiers as part of their context information in LPS. However, because the devices in the mobility network constantly changed their locations, we believe that the computational cost continually increases as the number of devices in the network increases. The cost of estimating the location of provers and verifiers in the LPS is illustrated in Fig. \ref{location}.

We analyse the computational complexity of generating a location proof by estimating its location between devices using the Euclidean distance technique and then calculating the cost of each prover's signature generation.

We analyse the computational complexity of generating a location proof by estimating its location between devices using the Euclidean distance technique and then calculating the cost of each prover's signature generation. For example, since each of the provers and verifiers has a coordinate ($x_{i}$, $y_{i}$) in a two-dimensional space, the distance difference between them for a \textit{N} devices take linear time, followed by squares, additions, and square root, each of which also requires the linear amount of time. As a result, the total computational complexity of \textit{N} provers and verifiers is $\mathcal{O}($N$)$. Additionally, each prover generates its signature independently in response to a request from the verifiers for location proofs. Thus, similar to location estimation, the signature generation procedure takes linear time, and the overall cost of generating signatures is $\mathcal{O}($N$)$ for \textit{N} provers. To summarise, when the computational difficulty of location estimation $\mathcal{O}($N$)$ and signature generation $\mathcal{O}($N$)$ are added together, the overall computational complexity for location proof generation is $\mathcal{O}($N$)$.

\subsubsection{Location Proof Verification}

After obtaining location proofs in the form of signatures from provers, the verifiers verify the proofs to detect clone node attack. Since our scheme used the ECDSA* batch verification approach to validate several digital signatures at once rather than individually. The working mechanism of location proof verification using batch verification is explained and discussed in detail in algorithm \ref{alg:proofverification}.

We define several batch sizes in our experiments for batch verification of location proofs, which correspond to the number of signatures to be verified in each batch. As stated earlier, the verification process in an LPS is performed by trusted verifiers; therefore, we assigned each signature to each verifier for the sake of simplicity rather than having the base station verify each signature individually. In addition, the ECDSA* batch verification methodology includes operations such as point addition and scalar multiplication, which may affect the time required to verify signatures.

We specify the total five batches for location proof verification in our experiment based on the number of verifiers in the LPS. The batch size is increased by five in each simulation; for example, 5, 10, 15, 20, 25, the computational time in seconds is determined. We compared the batch verification time of ECDSA* with the simple ECDSA for each batch size and concluded that batch verification of ECDSA* is substantially more computationally efficient than the ordinary ECDSA.

\begin{figure}[!htb]
     \begin{center}
        \subfigure[Batch Verification - Size 5]{%
            \label{batch1}
            \includegraphics[width=0.5\textwidth]{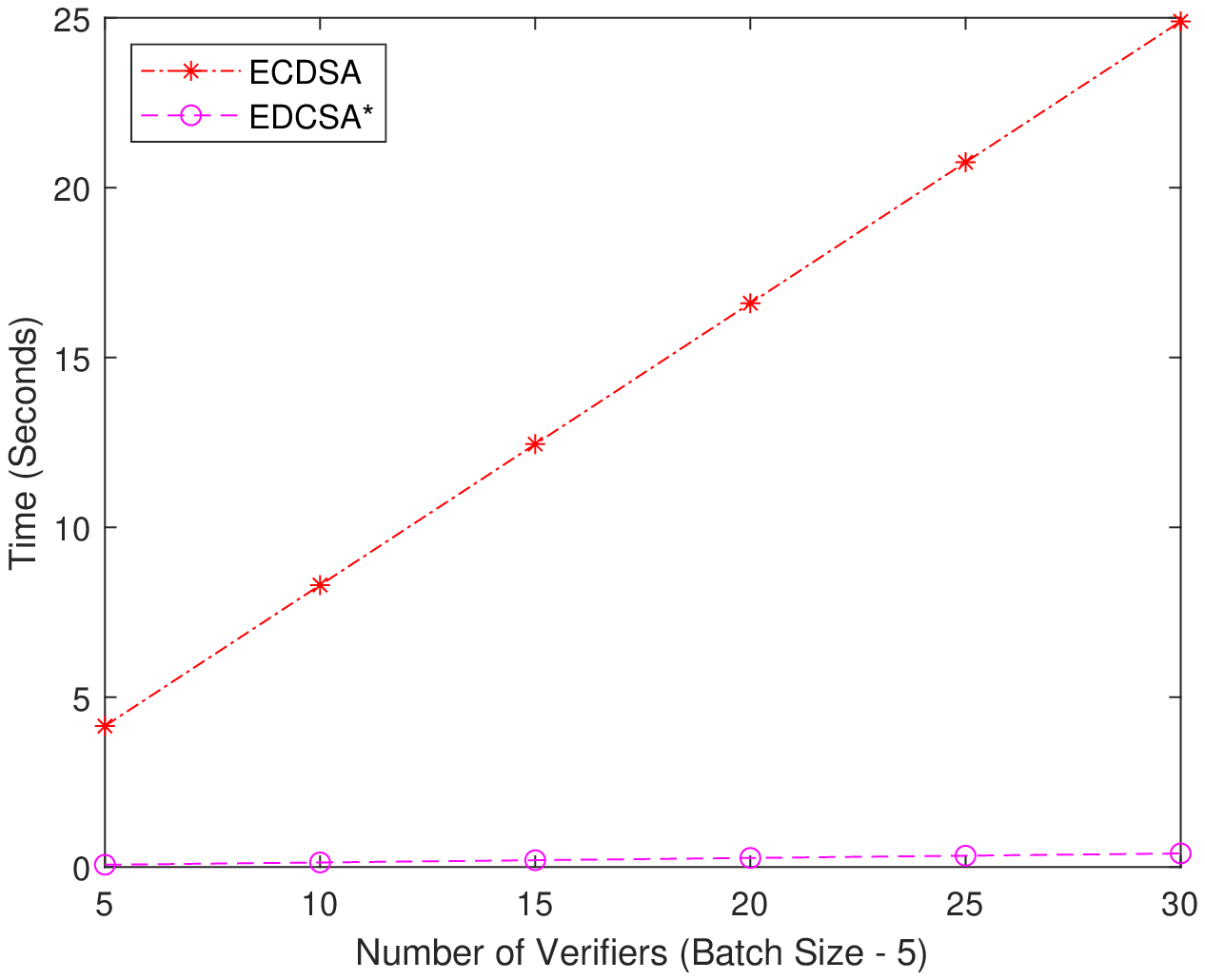}
        }%
        \subfigure[Batch Verification - Size 10]{%
           \label{batch2}
           \includegraphics[width=0.5\textwidth]{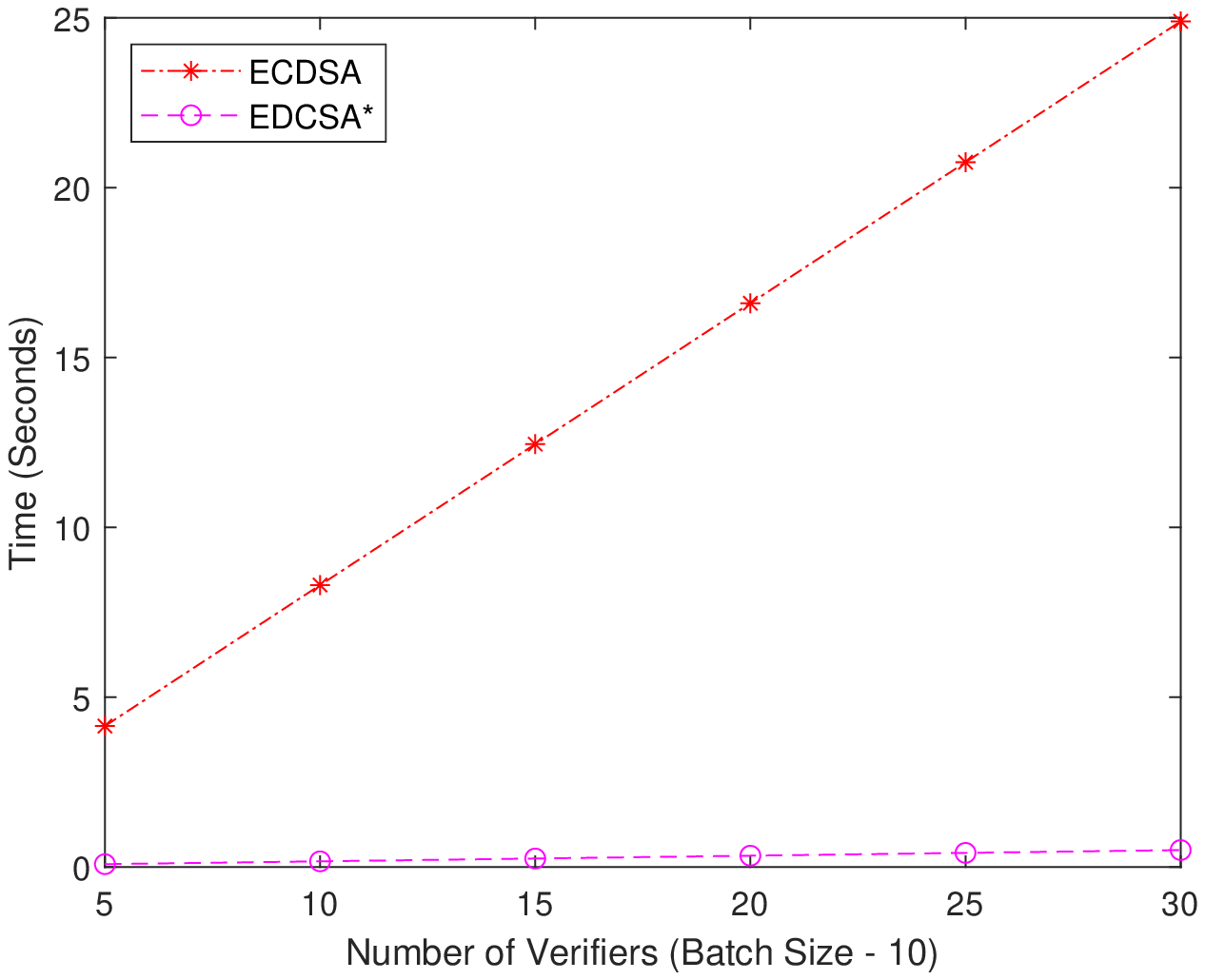}
        }\\ 
        \subfigure[Batch Verification - Size 15]{%
            \label{batch3}
            \includegraphics[width=0.5\textwidth]{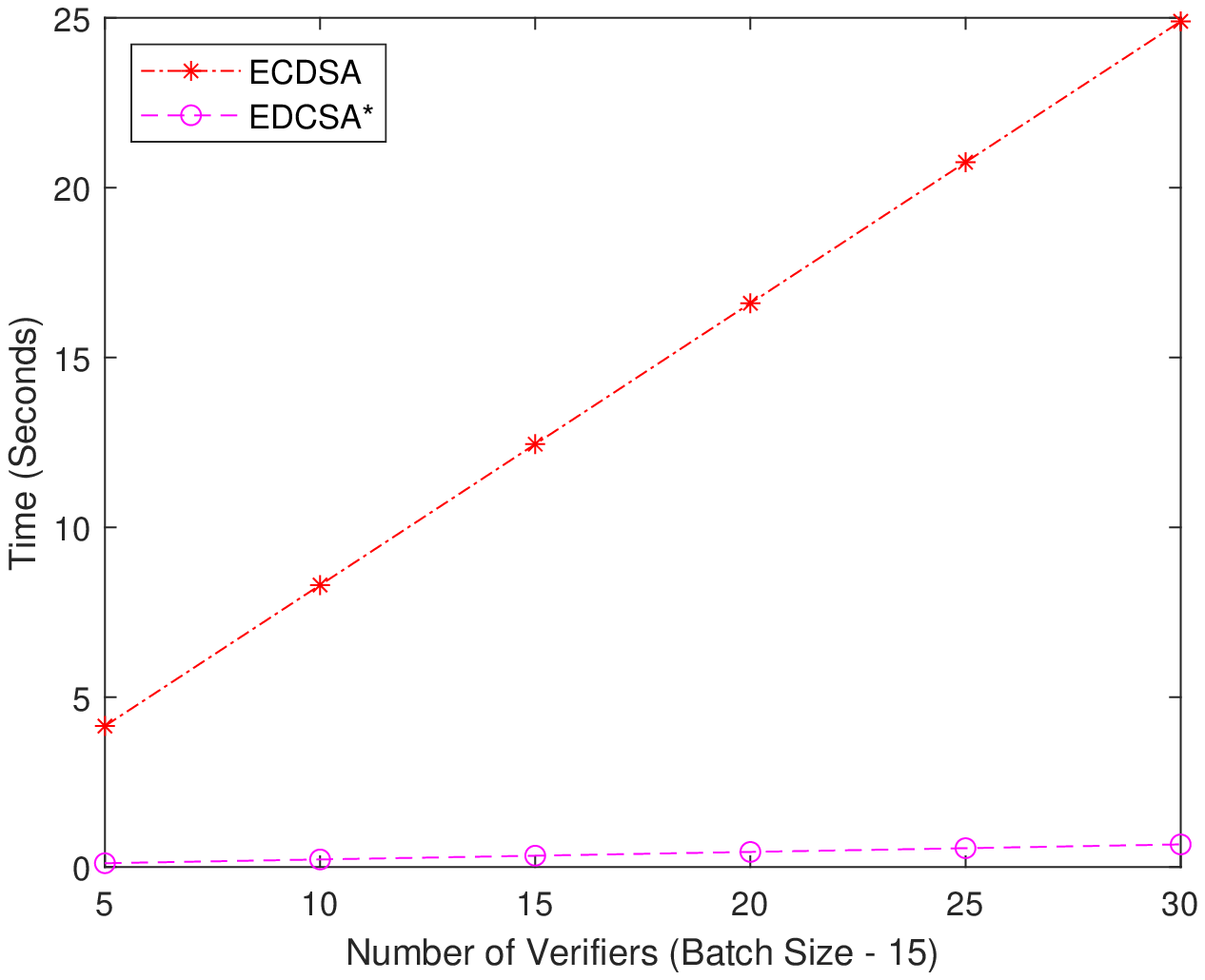}
        }%
        \subfigure[Batch Verification - Size 20]{%
            \label{batch4}
            \includegraphics[width=0.5\textwidth]{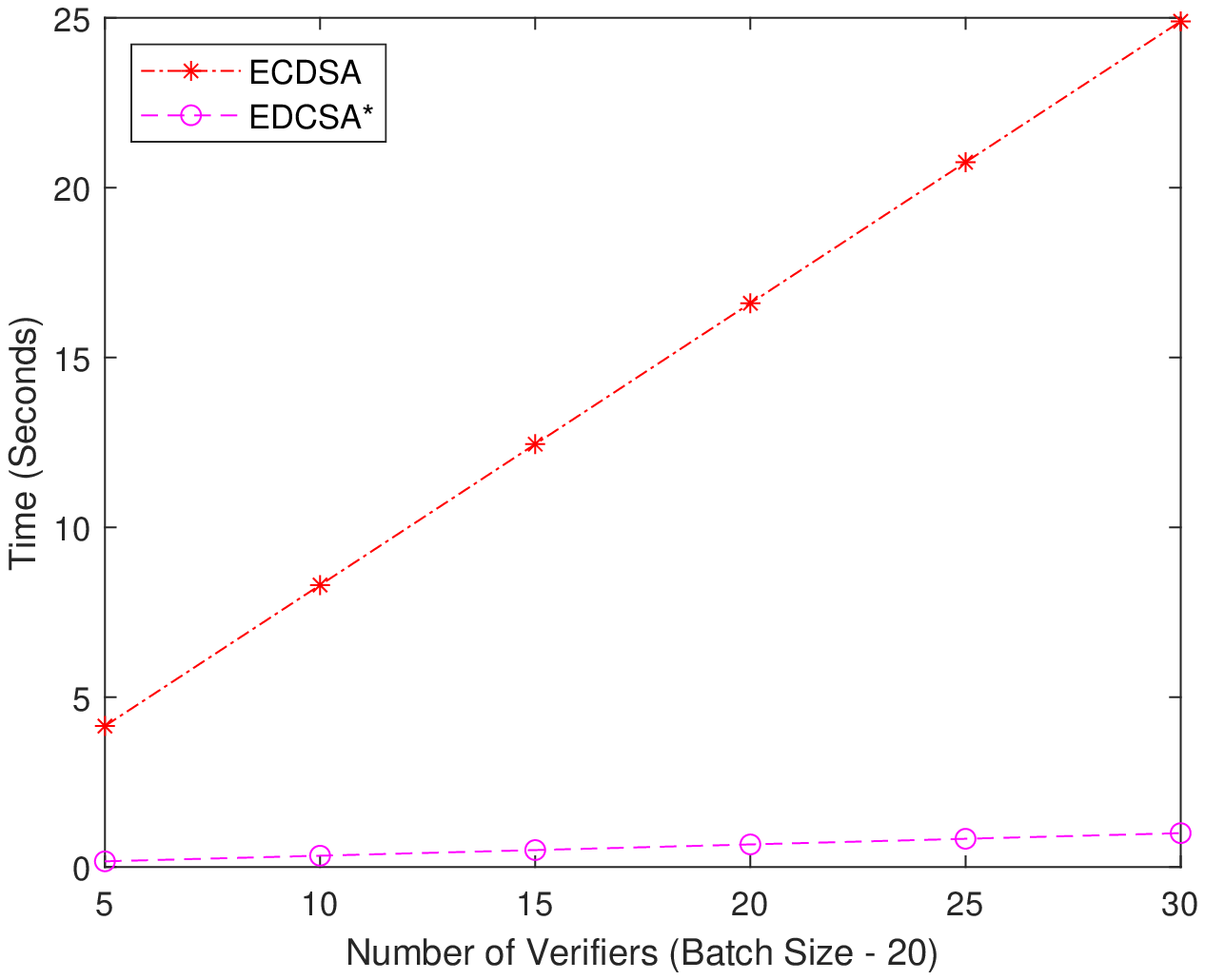}
        }\\%
        \subfigure[Batch Verification - Size 25]{%
            \label{batch5}
            \includegraphics[width=0.5\textwidth]{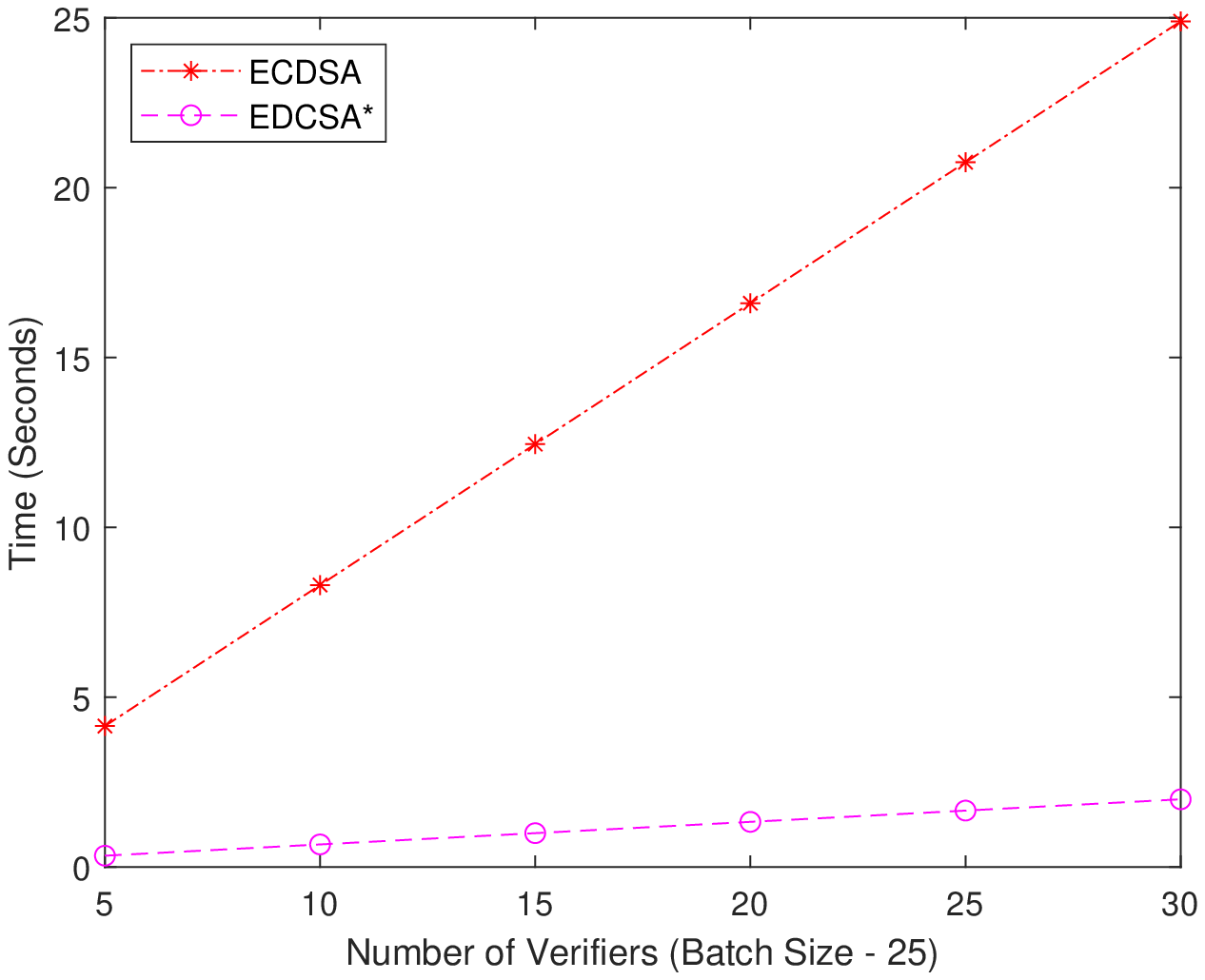}
        }%
        \subfigure[Batch Verification - All Sizes]{%
            \label{overallbatch}
            \includegraphics[width=0.5\textwidth]{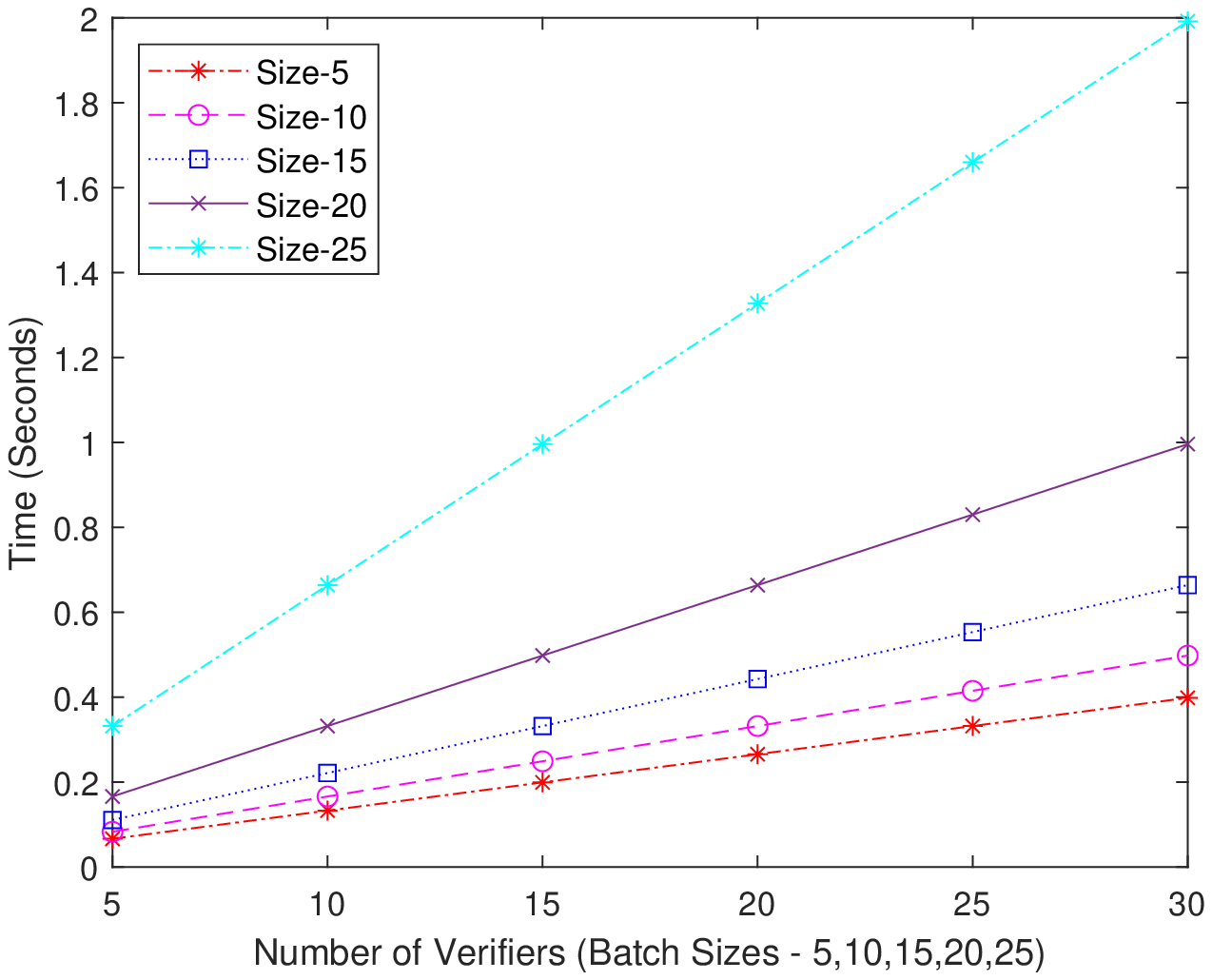}
        }%
    \end{center}
    \caption{%
        Batch Verification of Difference Sizes (5,10,15,20,25)
     }%
   \label{fig:subfigures}
\end{figure}

The computational time taken in seconds for verification of location proofs utilising batch verification of ECDSA* concerning ordinary ECDSA for batch size five validated by the verifiers is depicted in Fig. \ref{batch1}.

Fig. \ref{batch2} illustrates the computing time in seconds required to verify location proofs using ECDSA* versus traditional ECDSA on a batch size ten.

Figs. \ref{batch3}, \ref{batch4} and \ref{batch5} illustrate the computing times in seconds required to verify location proofs utilising batch verification of ECDSA* in comparison to standard ECDSA on batch sizes of 15, 20, and 25.

Fig. \ref{overallbatch} illustrates the computational time in seconds required to verify location proofs in the proposed scheme using the ECDSA* and conventional ECDSA for all specified batch sizes. We conclude that batch verification is significantly more efficient than conventional ECDSA and thus highly recommended for resource-constrained IoT devices.

Additionally, we determine the computational complexity of verifiers verifying location proofs in the LPS. For example, to detect clone nodes, each verifier requests that the prover provide location proofs for verification. Each verifier requests the total number of  $\sqrt{N}$ tracked provers, thus if the LPS has \textit{N} verifiers, the total computational cost of verifying each verifier's location proofs is O($\sqrt{N}$).

\subsection{Storage Overhead}
We analyse the performance of our proposed scheme in terms of storage overhead and compare it to the most applicable works. The storage overhead associated with the mechanism for detecting clone node attack is expressed as the average number of bytes required to store values in each node. It is defined in our proposed scheme as the average number of detected data, such as context information stored on IoT devices, that any detection protocol must collect to detect clone node attack in an IoT environment successfully. To demonstrate the feasibility of our proposed scheme from a storage overhead perspective, we consider IoT devices to be resource-constrained devices. 

Furthermore, we analyse the storage overhead of our proposed scheme concerning LPS from two perspectives: individual device storage overhead and total scheme storage overhead. Each of which is discussed in detail and measured in the subsections that follow.

\subsubsection{Storage Overhead of Individual Device}

We measure the storage overhead associated with each device in our proposed LPS as the average number of bytes required to store the context information \textit{CI} in each device, such as the prover \textit{P} and verifier \textit{V}. As previously mentioned, the context information is composed of the unique device identification \textit{ID}, the time  \textit{T}, the position \textit{Loc}, and the activity \textit{Actv}, each of which requires two bytes, two-byte, four bytes, and eight-byte, respectively. As a result, the context information requires eight bytes to store in each prover and verifier. For each prover and verifier's public and private keys, we use the 256-bit elliptic curve (also known as secp256k1). The private key is 32 bytes in size, while the compressed public key is 33 bytes. These public and private keys are unique to each device and are created only once they are compromised. However, context information is generated each time the prover device changes position, and the verifier must identify it. Table. \ref{storagerequirement} specifies the storage requirements for an individual device in terms of context information and key pairs, which reflects the storage overhead associated with the prover and verifier in the proposed scheme. Each device stores an average of 73 bytes, which is used for iterations to detect clone nodes in the LPS. Since each device, regardless of the number of inputs, follows the same steps to sense the same information, the objective complexity of storage overhead for both prover and verifier is $\mathcal{O}(1)$.

\begin{figure}[!htb]
     \begin{center}
        \subfigure[Storage Overhead of Provers]{%
            \label{storageoverhead1}
            \includegraphics[width=0.5\textwidth]{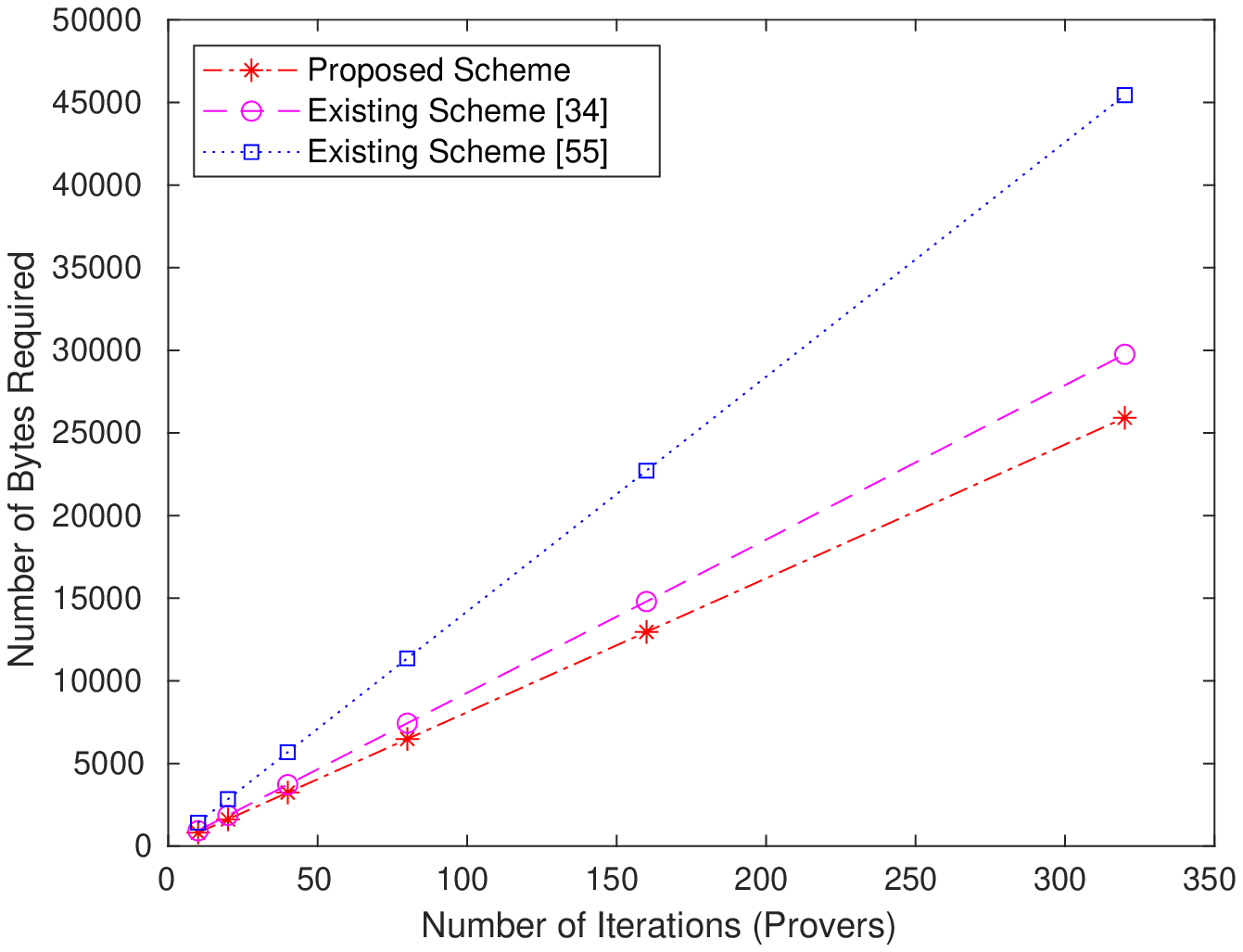}
        }%
        \subfigure[Storage Overhead of Verifiers]{%
           \label{storageoverhead2}
           \includegraphics[width=0.5\textwidth]{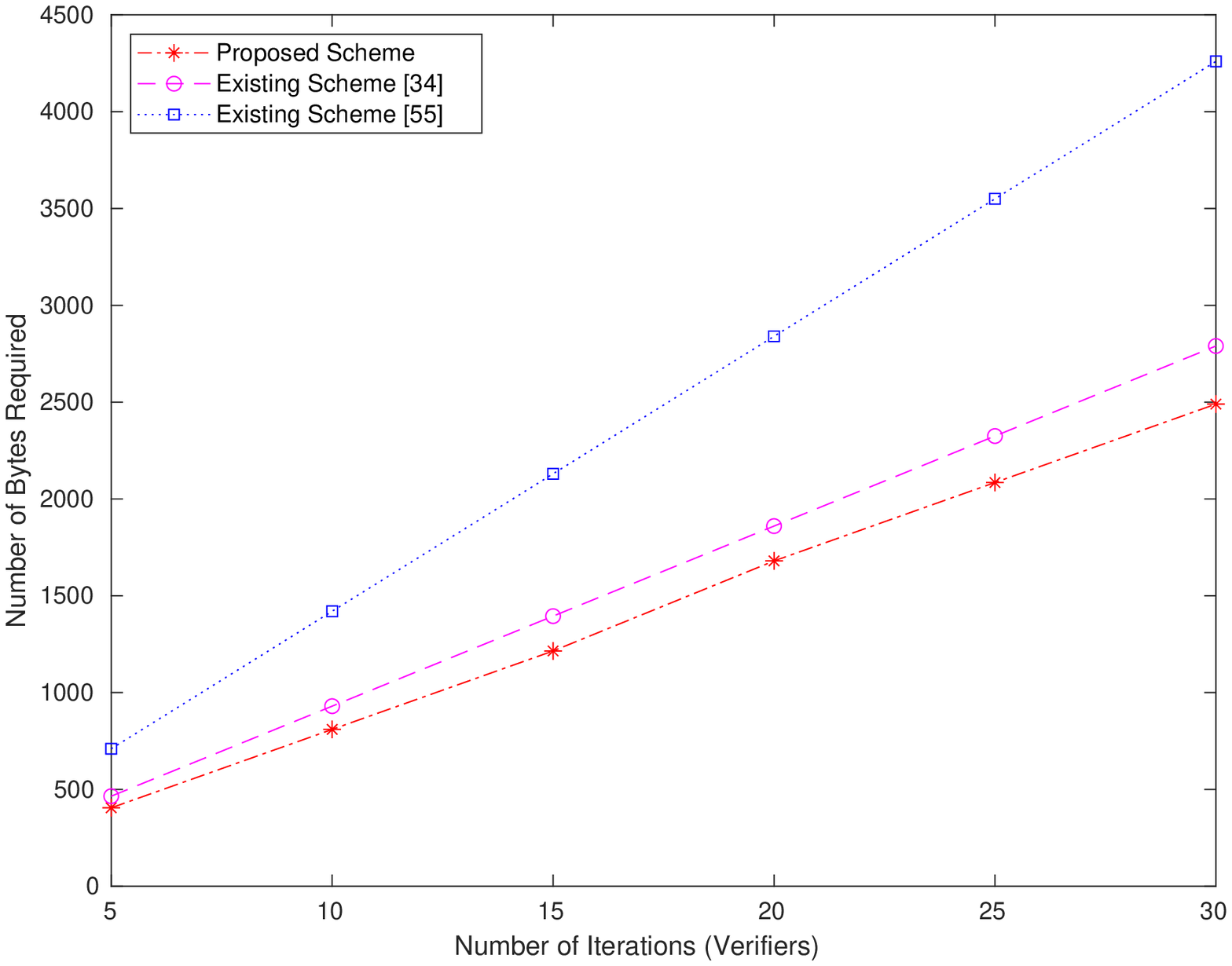}
        }\\ 
        \subfigure[Storage Overhead of Overall Scheme]{%
            \label{storageoverhead3}
            \includegraphics[width=0.5\textwidth]{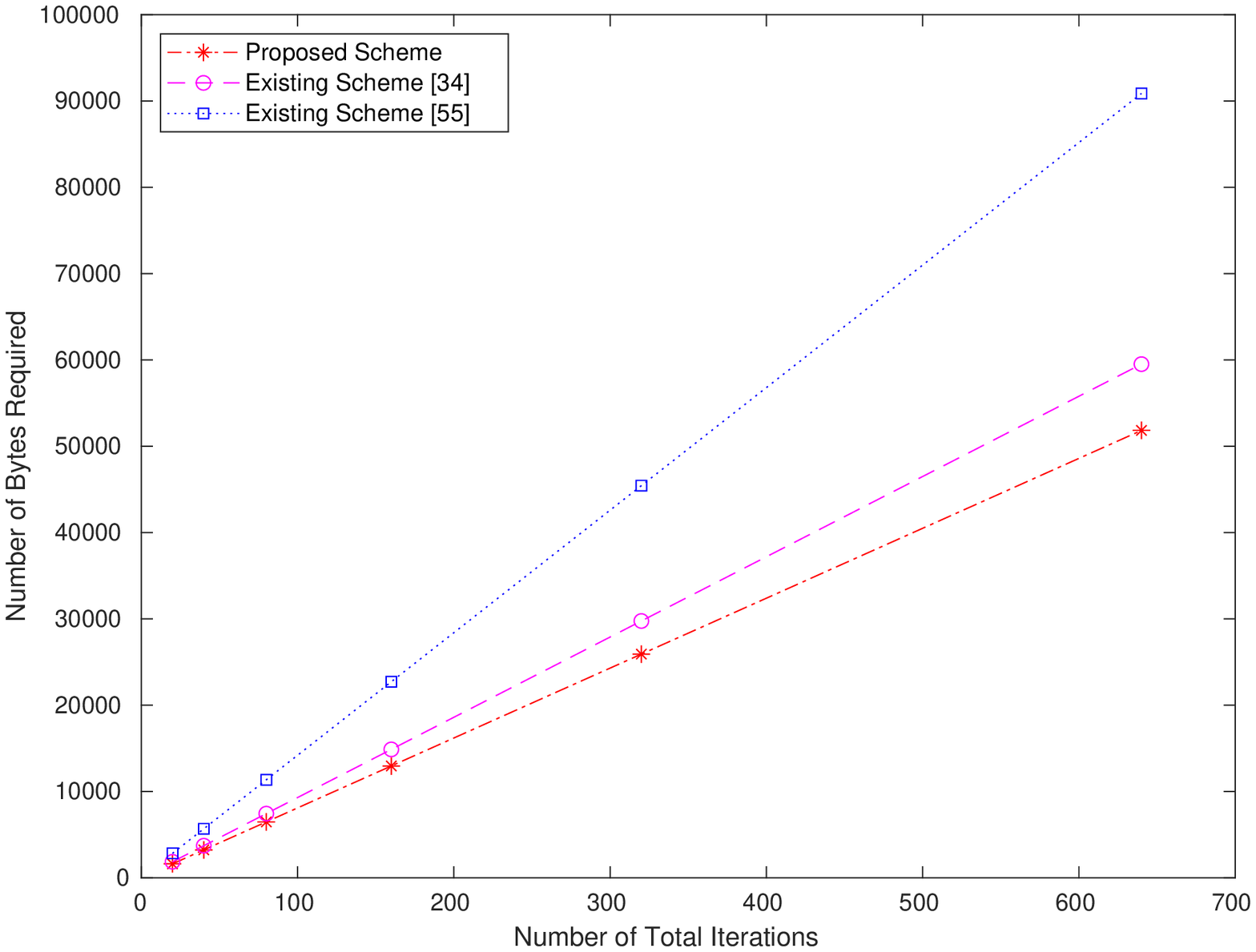}
        }%
    \end{center}
    \caption{%
        Storage Overhead of Proposed Scheme (Provers, Verifiers, Overall Scheme)
     }%
   \label{fig:subfigures}
\end{figure}

\begin{table}[!htp]
\centering
\caption{Storage Requirement of Each Device}
\begin{tabular}{|c|c|}
\hline
\multicolumn{2}{|c|}{\textbf{Context Information}} \\ \hline
     ID      &     2 bytes      \\ \hline
      Time     &   2 byte        \\ \hline
        Location   &     4 bytes      \\ \hline
        Activity   &    8 byte       \\ \hline
\multicolumn{2}{|c|}{\textbf{ECDSA Key-Pair}} \\ \hline
     Private Key     &     32 bytes      \\ \hline
         Public Key (compressed)  &     33 bytes      \\ \hline

\end{tabular}
\label{storagerequirement}
\end{table}

Figs. \ref{storageoverhead1} and \ref{storageoverhead2} illustrate the analysis of the storage overhead of provers and verifiers respectively in our proposed scheme in comparison to existing schemes \cite{shanmugam2020two} \cite{anitha2021intelligent}. The analysis reveals that the storage overhead of each device increases as the number of iterations (in bytes) required to detect clone node attack in the system increases.

\subsubsection{Storage Overhead of Overall Scheme}
We measure the storage overhead of our proposed scheme with the existing schemes \cite{shanmugam2020two} \cite{anitha2021intelligent} concerning several iterations performed by both prover and verifier, and an average number of bytes takes in the system to detect clone node attack.

To detect clone node attack, a verifier must initiate the location proof mechanism and send it to the prover to verify whether or not it has been compromised. Thus, to detect clone nodes, each system (prover and verifier) must contribute and perform iterations following the other. Fig. \ref{storageoverhead3} depicts the proposed scheme's overall storage overhead to the existing schemes \cite{shanmugam2020two} \cite{anitha2021intelligent} in terms of the number of iterations (in bytes) performed by devices to detect clone node attack. According to the analysis, our proposed scheme is significantly more efficient in terms of storage overhead and requires less space on devices than existing schemes.

\begin{figure}[!htb]
     \begin{center}
        \subfigure[Communication Overhead of Sensing Context Information]{%
            \label{Newcommunicationoverhead1}
            \includegraphics[width=0.4\textwidth]{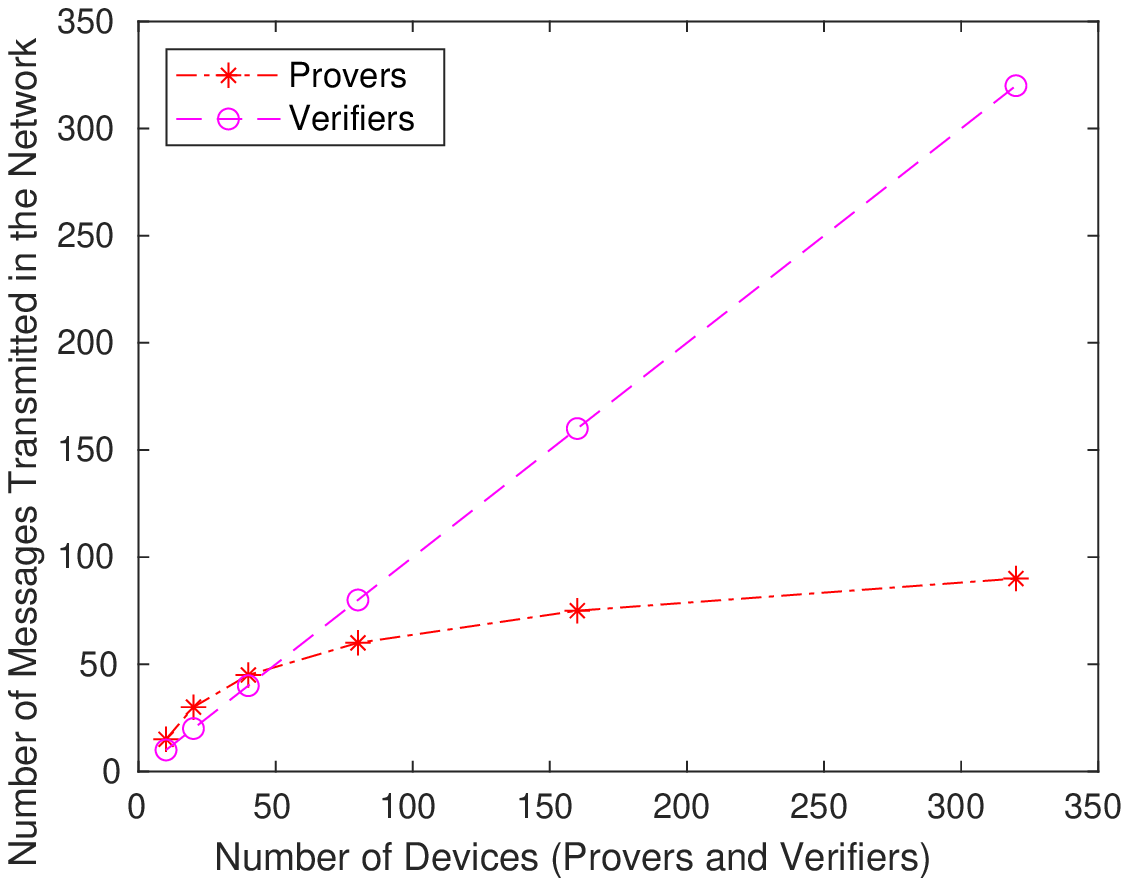}
        }%
        \subfigure[Communication Overhead of Storing Context Information]{%
           \label{Newcommunicationoverhead2}
           \includegraphics[width=0.4\textwidth]{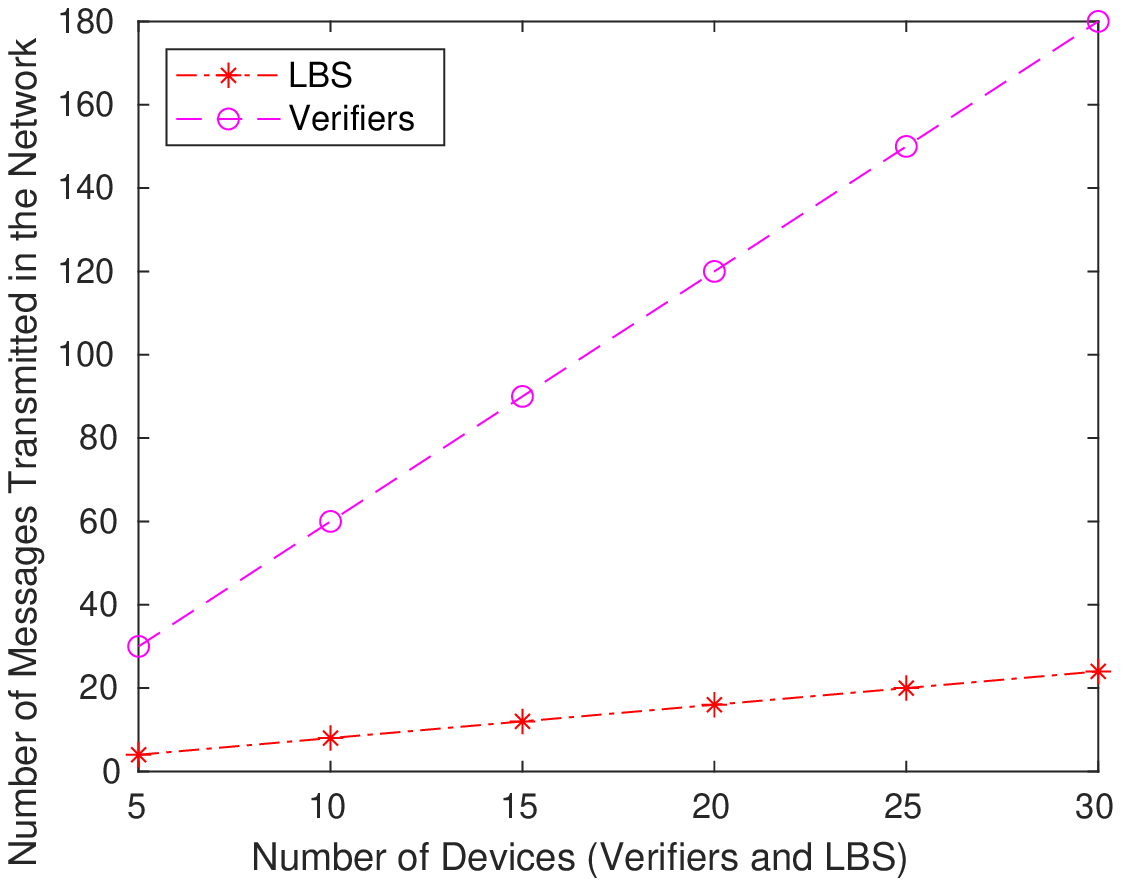}
        }\\ 
        \subfigure[Communication Overhead of Verifying Location Proof]{%
           \label{Newcommunicationoverhead3}
            \includegraphics[width=0.4\textwidth]{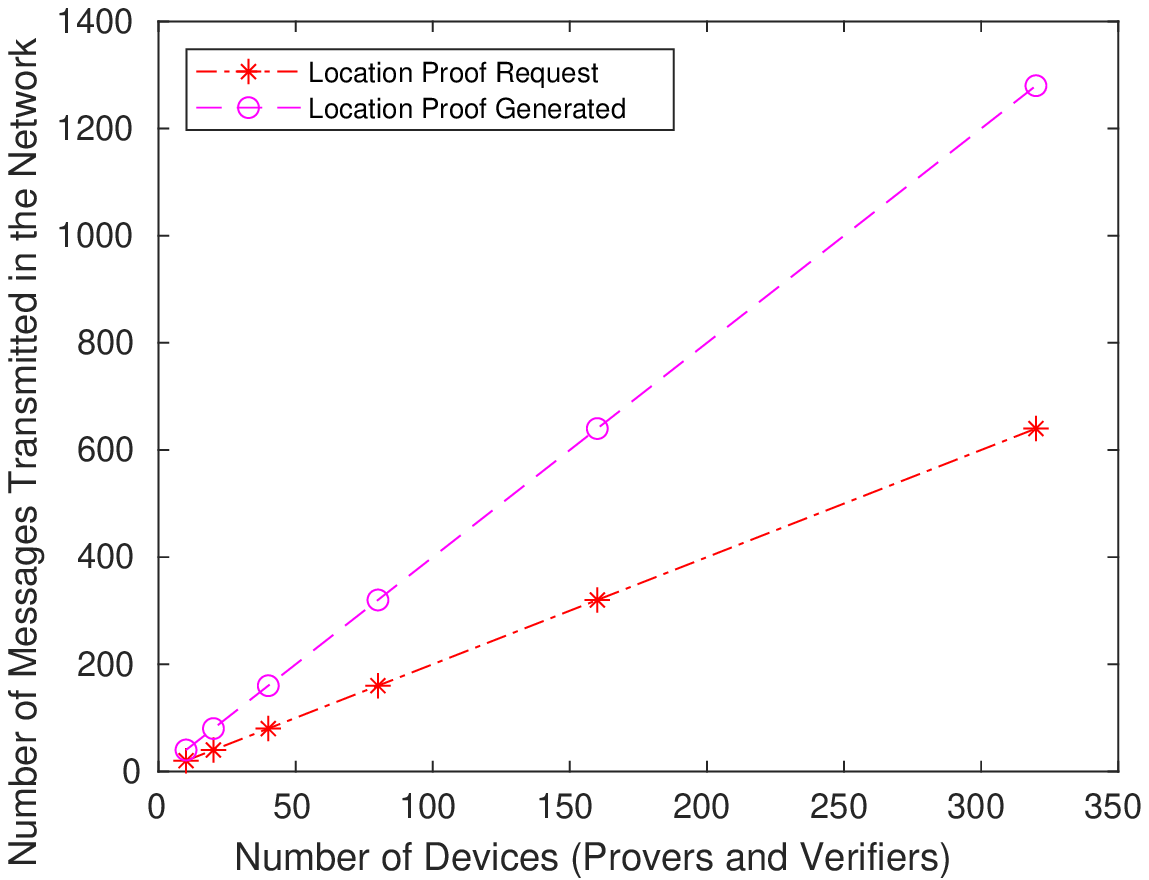}
        }%
        \subfigure[Communication Overhead of Detecting Clone Nodes]{%
           \label{Newcommunicationoverhead4}
            \includegraphics[width=0.4\textwidth]{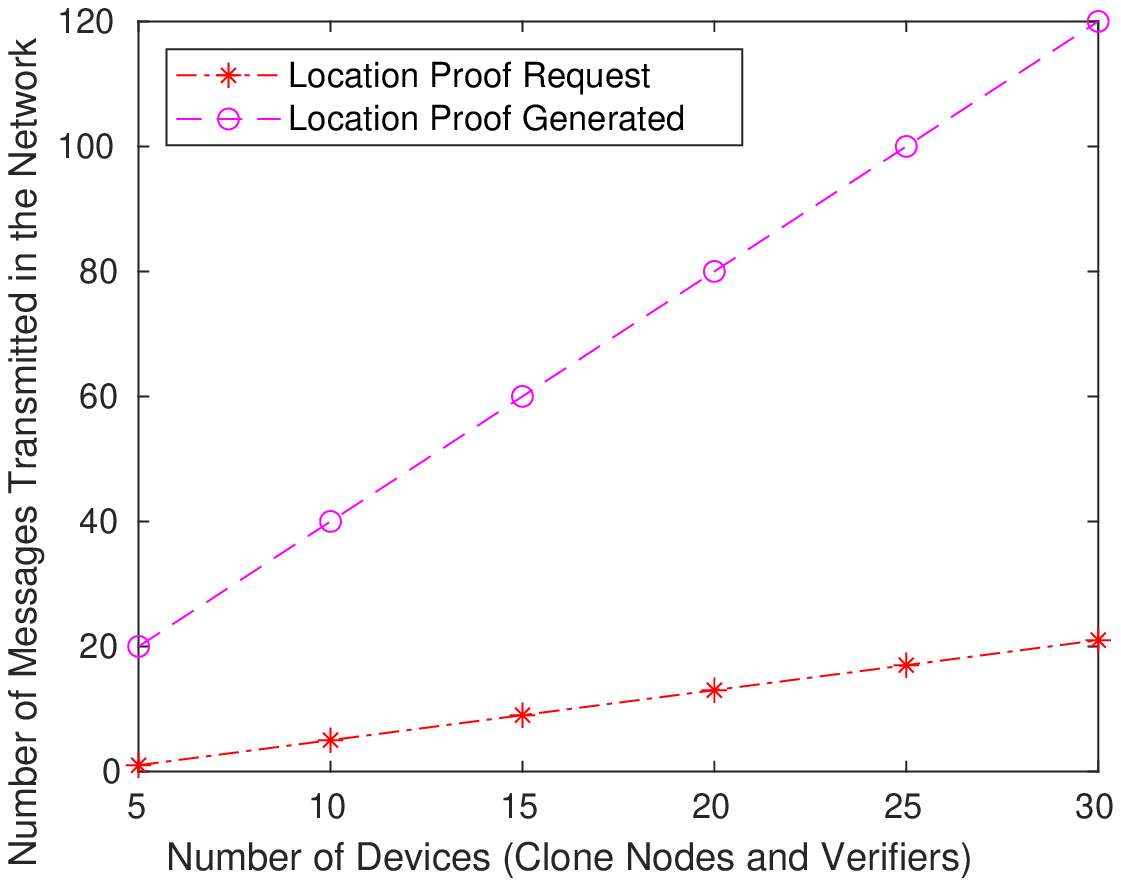}
        }\\%
        
        \subfigure[Overall Communication Overhead of All Devices]{%
          \label{Newcommunicationoverhead5}
            \includegraphics[width=0.4\textwidth]{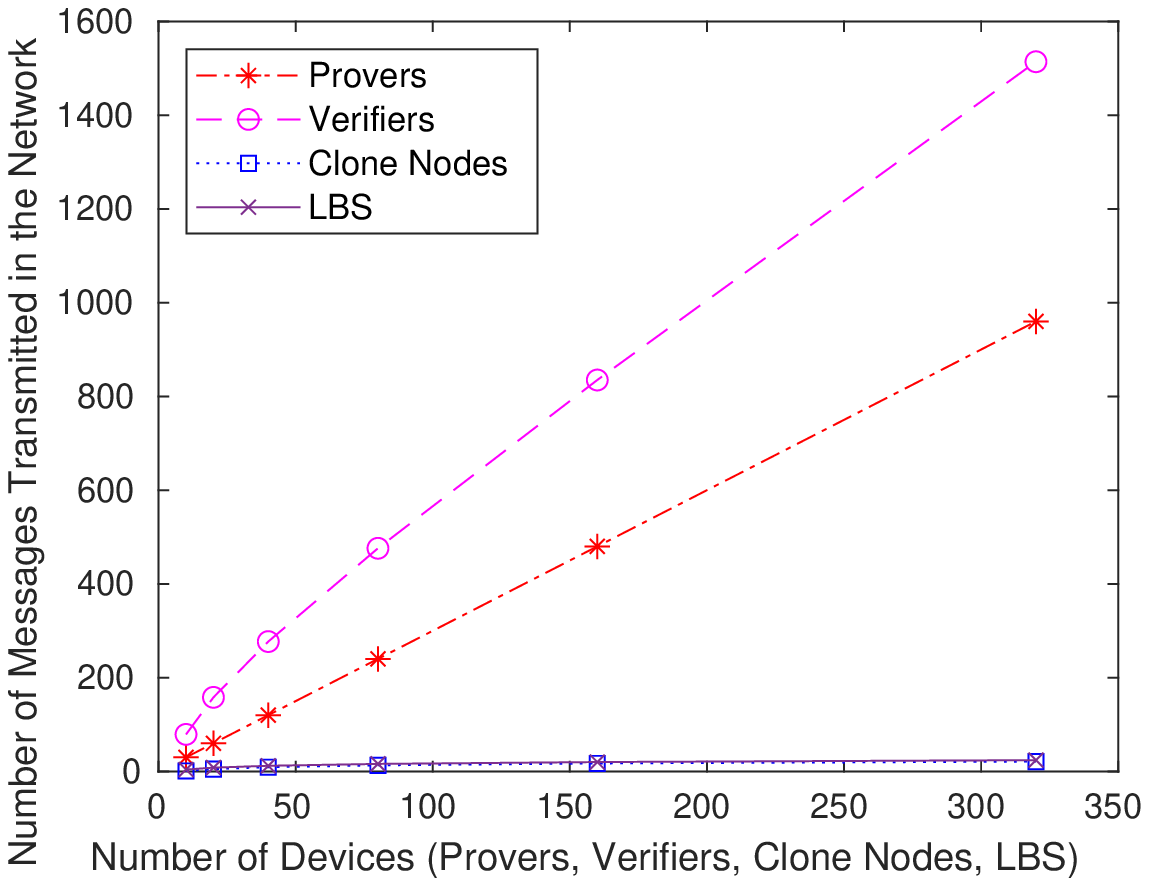}
        }
    \end{center}
    \caption{%
       Communication Overhead (No. of Messages Transmitted)
     }%
   \label{fig:subfigures}
\end{figure}

Additionally, we determine the objective complexity of our proposed scheme in terms of its overall storage overhead. For example, in proposed LPS, each prover \textit{P} sense and store the context information \textit{CI} such as unique \textit{ID}, time \textit{T}, location \textit{Loc}, and activity \textit{Actv} which is used to generate the location proof \textit{LP} to be verified by the verifiers for detection of clone nodes in the network. The context information is also called the detection information that needs to be stored in a prover when the protocol for detecting node clone attacks works in networks. Each verifier \textit{V} also need to sense the context information \textit{CI} used to compare with the context information \textit{CI} obtain from the prover \textit{P}. 
For detection of clone nodes, each verifier \textit{V} need to obtain a location proof \textit{LP} from prover \textit{P} which is computed from algorithm \ref{alg:proofgeneration}. Upon receiving a latest location proof \textit{$LP_{i}$} from prover \textit{P}, the former location proof \textit{$LP_{i-1}$} is ignored, only latest location proof \textit{$LP_{i}$} is stored to wait for the next location proof \textit{$LP_{i+1}$}. To store all the location proofs from the prover \textit{P}, a first come first served (FCFS) queue is maintained. Therefore, a fixed length of storage space including a location proof and queue is required for each prover. Each verifier \textit{V} has the $\sqrt{N}$ tracked provers and every prover \textit{P} has $\sqrt{N}$ verifiers, thus the total storage cost of each node (prover and verifier) is  O($\sqrt{N}$).

The storage overhead measurement for IoT devices shows that the increase in the number of iterations performed during the clone node detection attack is the reason for the increase in the system's total storage overhead.

\subsubsection{Communication Overhead}
In IoT-based Networks, the term ``communication overhead'' refers to the data sent to or from the base station to IoT devices to conduct various network operations. The communication overhead in our proposed scheme is defined as the average number of iterations each node transmits and receives during various network operations to detect clone node attack in LPS.

The networks operations included in our proposed scheme are sensing the context information, storing the context information on some server in location-based services,  requesting and accepting location proofs between verifier and provers and between verifiers and clone nodes, and reporting of different iterations among verifiers, provers, clone nodes, and LBS as illustrated in detail in subsection \ref{executionflow}. The communication cost of these operations can be measured by the estimated number of messages transmitted.

Fig. \ref{Newcommunicationoverhead1} illustrating the communication overhead associated with provers and verifiers sensing context information in order to initiate the process of LPS for detecting clone node attack in the network. This analysis of communication overhead is highly dependent on the number of iterations and the number of devices activated during the overall initial setup. Since these verifiers are the primary devices in the LPS, interacting with the provers and LBS to verify the location proofs, they perform more iterations and have a higher communication overhead.

Fig. \ref{Newcommunicationoverhead2} shows the communication overhead associated with the verifiers storing context information in the LBS. As with the analysis in Fig. \ref{Newcommunicationoverhead1}, the analysis demonstrates that verifiers have a higher communication overhead than the LBS because verifiers perform more network activities than any other device in the network.

Fig. \ref{Newcommunicationoverhead3} demonstrating the communication overhead in terms of iterations required to request and generate location proofs for detecting clone nodes attack between provers and verifiers. The analysis demonstrates that the communication overhead for location proof generated on the verifiers' side is high, as verifiers must generate the proof request and verify it with the LBS, in addition to notifying provers about the device breach.

Fig. \ref{Newcommunicationoverhead4} illustrates the communication overhead between clone nodes and verifiers in terms of iterations required to detect clone nodes in the network. As with the analysis in Fig. \ref{Newcommunicationoverhead3}, this analysis demonstrates that the communication overhead for verifiers increases as the number of iterations on the various network operations increases.

Fig. \ref{Newcommunicationoverhead5} demonstrates the overall communication overhead of our proposed LPS in terms of iterations between the system's devices, which include provers, verifiers, clone nodes, and LBS. The results indicate that the verifier performs many iterations since it is considered the main element for verifying network operations such as sensing information, storing it in the LBS, generating and validating proofs, and detecting clone node attack in the system.

\begin{figure}[!htb]
     \begin{center}
        \subfigure[Communication Overhead of Sensing Context Information]{%
           \label{communicationoverhead1}
            \includegraphics[width=0.45\textwidth]{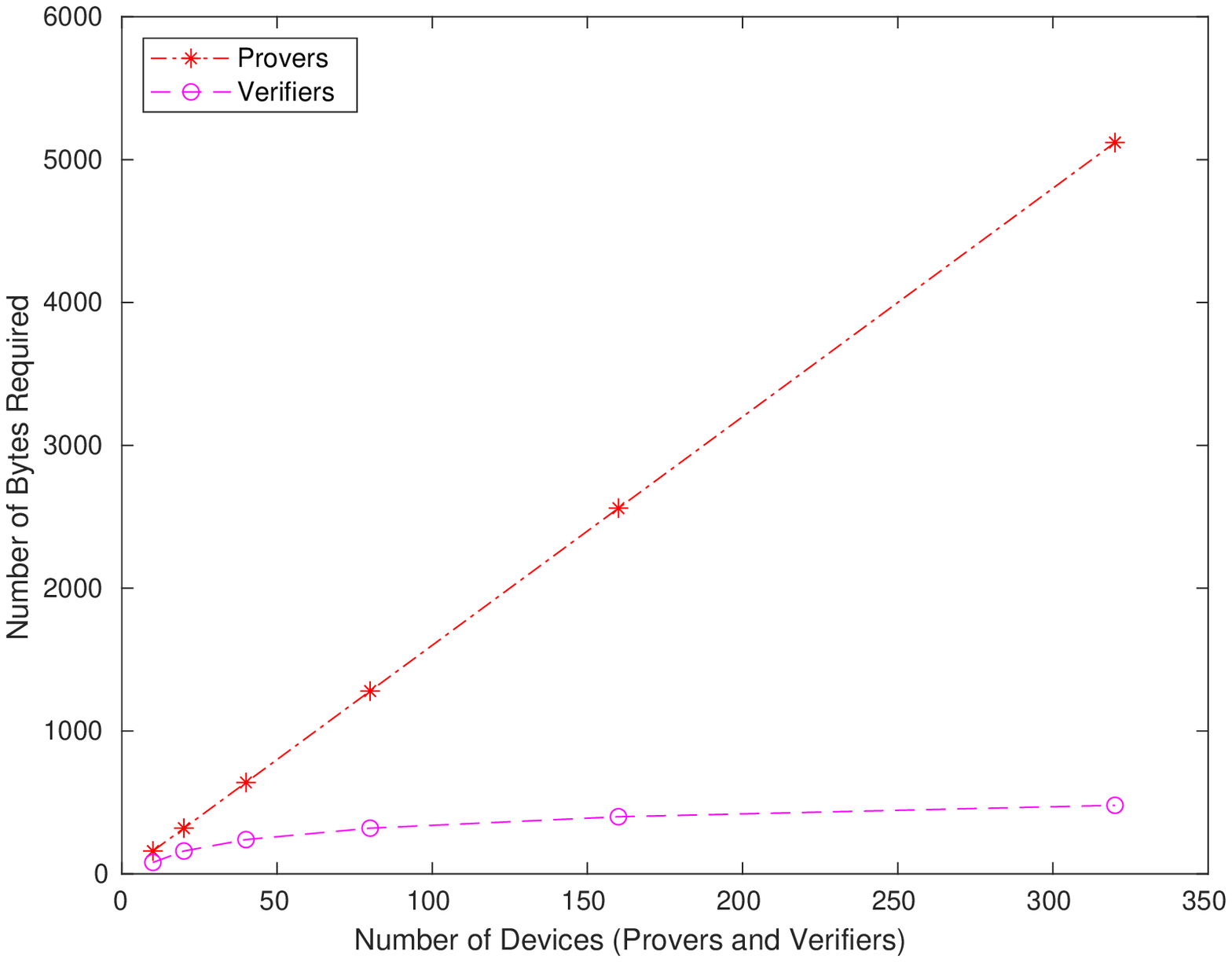}
        }%
        \subfigure[Communication Overhead of Storing Context Information]{%
           \label{communicationoverhead2}
           \includegraphics[width=0.45\textwidth]{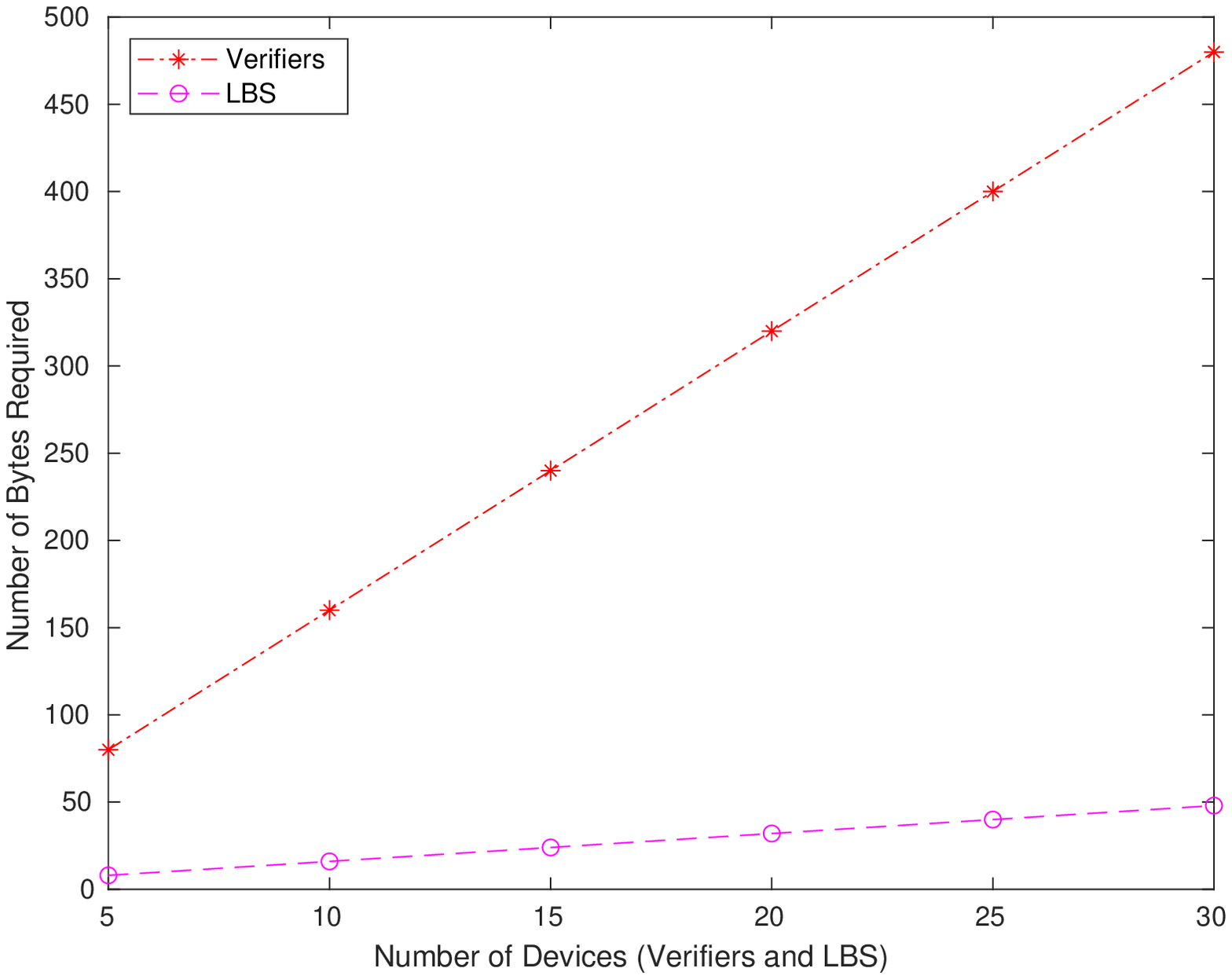}
        }\\ 
        \subfigure[Communication Overhead of Verifying Location Proof]{%
            \label{communicationoverhead3}
            \includegraphics[width=0.45\textwidth]{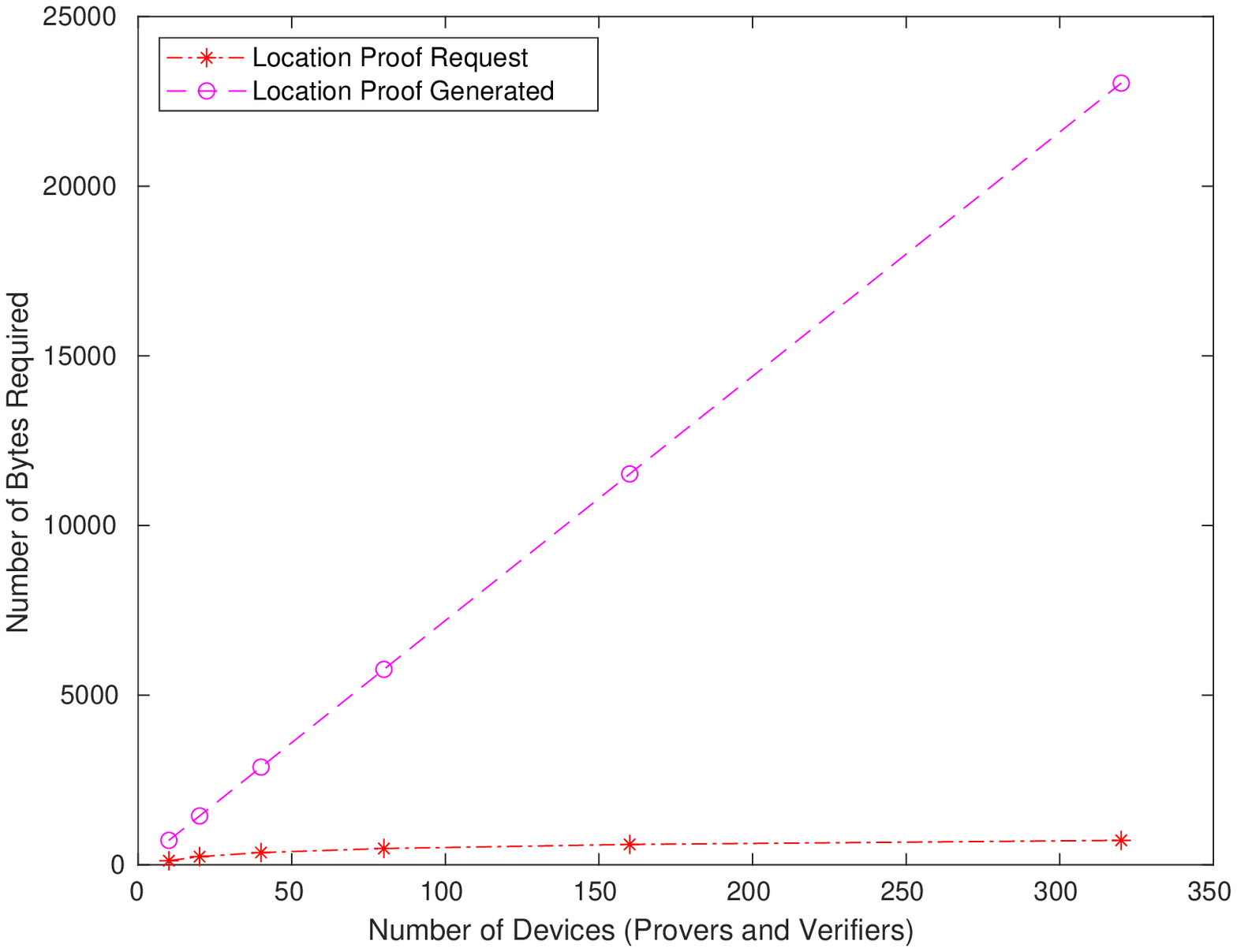}
        }%
        \subfigure[Communication Overhead of Detecting Clone Nodes]{%
            \label{communicationoverhead4}
            \includegraphics[width=0.45\textwidth]{Communication2.eps}
        }\\%
        \subfigure[Overall Communication Overhead of All Devices]{%
            \label{communicationoverhead5}
            \includegraphics[width=0.45\textwidth]{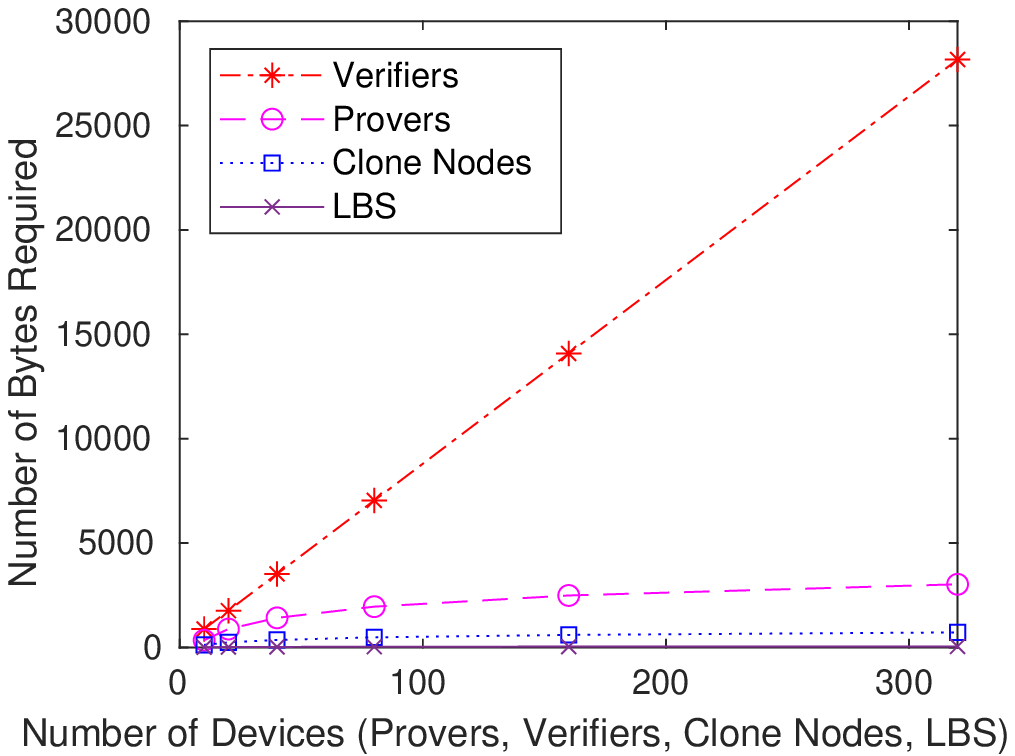}
        }
    \end{center}
    \caption{%
    Communication Overhead (No. of Bytes Required)}%
   \label{fig:subfigures}
\end{figure}

Apart from measuring computation overhead in terms of iterations, we also analysed the computation overhead of our proposed scheme in terms of bytes used by the average number of iterations when conducting various network operations to detect clone node attack. For example, Fig. \ref{communicationoverhead1} depicts the computation overhead of sensing context information by provers and verifiers in terms of bytes, which clearly shows the increase in computation overhead with the number of devices connected to the network. Furthermore, since the number of provers in the network is more than the number of verifiers, the computation cost is likewise more significant for the provers.

Fig. \ref{communicationoverhead2} demonstrating the communication overhead associated with storing sensing data to a server once it is obtained from the deployed environment. This occurs between the verifiers and the LBS authority. The analysis demonstrates that, in addition to storing context information, verifiers are also involved in various reporting actions to other devices in the network, such as provers. As a result, the higher the number of iterations, the greater the communication cost in terms of bytes.

Fig.  \ref{communicationoverhead3} demonstrates the communication overhead associated with validating the location proof process through iterations of generating and verifying the location proof between the provers and verifiers to detect clone node attack. Furthermore, as verifiers make location proof requests, they entail some additional processes for LBS, such as checking the store context information and adding proof generation, which resulted in a more significant increase in bytes than generating the location proof just by the prover.

The communication overhead in terms of bytes for conducting iterations of detecting clone nodes in an LPS is illustrated in Fig. \ref{communicationoverhead4}. The result analysis demonstrates that verifier communication overhead increases as additional steps for validating proofs with LBS are performed. It also varies on the number of devices.

Fig. \ref{communicationoverhead5} illustrates the overall communication overhead of our proposed LPS when the average number of iterations occurs between different devices such as provers, verifiers, clone nodes, and LBS. The communication overhead between such devices is measured in bytes. This analysis demonstrates that the communication overhead of verifiers is more significant than that of all other devices because verifiers are the core devices responsible for conducting the network's primary activities such as sensing information, generating and verifying location proof requests, and also providing a means of communicating with the LBS for confirming context information. Additionally, the communication cost for provers and clone nodes is significantly lower than the cost for verifiers, which is the primary objective of the proposed scheme.

Along with analysing the communication overhead of our proposed scheme, we perform a worst-case analysis to determine its objective complexity in terms of communication overhead. For example, provers and verifiers incur a network cost of $\mathcal{O}($N$)$ message transmission when sensing context information. Similarly to sensing context information, storing such information to the LBS incurs $\mathcal{O}($N$)$ message transmission costs across the network. We consider a tree to generate location proofs requests for provers, with the root acting as a verifier that propagates location proofs to the provers functioning as leaf nodes. The amount of location proofs sent to provers is proportional to the size of a tree.  Assume that \textit{P} trees are generated for the network's provers and have a moderate degree of $\textit{P}_d$.  As a result, the total number of iterations executed by the tree is calculated using Eq. \ref{eqcommunication1}.

\begin{equation} \label{eqcommunication1}
    \centering
     \sum_{i=1}^{h} P^i_d = \frac{P^{h+1}_d-1}{P_{d}-1}
\end{equation}

Where \textit{h} is the height of the tree for the provers and is denoted as:

\[  \floor{logP_d(\frac{N}{KP})} \]

\textit{N} is the total number of devices, \textit{K} denotes the average size of a subset of the tree, and the network generates a total of $\frac{N}{K}$ sub-trees. As a result, the location proof message transmission takes $\frac{N}{KP}$ in the tree structure, resulting in a total message transmission rate of $\mathcal{O}(\frac{N}{K})$ in the network.

The transmission of iterations in a tree is identical to the creation of the tree, except that each root incurs an additional cost for transmitting a final subset iteration to the root node, such as a verifier. Thus, the iteration transmission overhead in the network is $\mathcal{O}( P\sqrt{N} +\frac{N}{K})$. 

As a result, the objective complexity of our proposed scheme in terms of network communication overhead is estimated by combining the costs of all iterations required to perform various network operations necessary to detect a clone node in the network using Eq. \ref{eqcommunication2}.

\begin{equation} \label{eqcommunication2}
    \centering
     \mathcal{O}(N) + \mathcal{O}(N) + \mathcal{O}(\frac{N}{K}) + \mathcal{O}(P\sqrt{N}+\frac{N}{S}) = \mathcal{O}(N)
\end{equation}

In contrast to other schemes that constantly collect and relay huge amounts of data to the base station, our proposed scheme has a lower communication overhead on the network since it is only needed to check the neighbour node's position evidence. Since communication cost in an LPS is proportional to the number of trusted verifiers, we argue that the proposed scheme is appealing due to its mobile nature and lower communication overhead.

Table. \ref{comparisontable} compares the computational complexity of evaluation parameters such as computation overhead, storage overhead, and communication overhead in our proposed system to existing state-of-the-art schemes. According to the analysis results, our scheme has a lower overhead in all computational complexities.

\begin{table}[!htp]
\centering
\caption{Comparison of Evaluation Parameters with Existing Schemes}
\begin{tabular}{|c|c|c|c|}
\hline
 & \multicolumn{3}{c|}{\textbf{Evaluation Parameters}}    \\ \hline
\textbf{Schemes} & \textbf{\makecell{Computation \\ Overhead}} & \textbf{\makecell{Storage \\ Overhead}} & \textbf{\makecell{Communication \\ Overhead}} \\ \hline
\cite{shanmugam2020two}   &  O(N)  + O(N)          &    N/A      & O($N^{2}$)          \\ \hline
\cite{anitha2021intelligent}  &     N/A      & O(N)          &   O(N)        \\ \hline
 \cite{parno2005distributed}  &     O(N)      &      N/A     &      O($N^{2}$)     \\ \hline
  \makecell{Our Proposed\\ Scheme}  &  O(N)  + O(N)  +  O($\sqrt{N}$)      &   O($\sqrt{N}$)        &    O(N)       \\ \hline
\end{tabular}
\label{comparisontable}
\end{table}

\section{Conclusion and Future Work} \label{sec:conclusion}

The majority of sensors embedded into IoT devices lack tamper-resistant hardware, which is the primary cause of IoT device compromise or hijacking and frequently results in device cloning and replication. Considering these security concerns, it is vital to have a robust detection mechanism to secure against clone node attack. This paper proposes an efficient mechanism for detecting clone node attack in IoT networks to leverage their context-aware information. We develop a secure LPS that uses the context information of IoT devices as location proofs in conjunction with batch verification of ECSDA* to accelerate the verification process at the proposed selected trustworthy nodes model. Using the algorithms and sequence diagram, we discussed each component of the LPS and illustrated the working of the proposed scheme.
Furthermore, we conducted an extensive security analysis, outlining the prerequisites and several security needs for ECC and the possibility of several types of attacks on the signatures and hashes used in our proposed scheme. Finally, we design a prototype of an LPS in order to validate the performance and overhead of our various proposed algorithms. The experimental results are compared to existing schemes and conclude that our system provides a robust and considerable attack detection rate for clone node attack in IoT networks while minimising computing, storage, and communication overhead. In our future work, we intend to develop our prototype in a real-time IoT-based scenario to assess its applicability for large industrial setups, focusing on energy usage, network latency, and message drop ratio, etc.

\section*{Declaration of Competing Interest}

The authors of this paper declare that they have no known competing financial interests or personal relationships that could have influenced the work reported in this paper.

\section*{Funding}

This research did not receive any specific grant from funding agencies in the public, commercial, or
not-for-profit sectors







\bibliography{Main}

\begin{thebibliography}{10}
\expandafter\ifx\csname url\endcsname\relax
  \def\url#1{\texttt{#1}}\fi
\expandafter\ifx\csname urlprefix\endcsname\relax\def\urlprefix{URL }\fi
\expandafter\ifx\csname href\endcsname\relax
  \def\href#1#2{#2} \def\path#1{#1}\fi

\bibitem{al2015internet}
A.~Al-Fuqaha, M.~Guizani, M.~Mohammadi, M.~Aledhari, M.~Ayyash, Internet of
  things: A survey on enabling technologies, protocols, and applications, IEEE
  communications surveys \& tutorials 17~(4) (2015) 2347--2376.

\bibitem{diaz2016state}
M.~D{\'\i}az, C.~Mart{\'\i}n, B.~Rubio, State-of-the-art, challenges, and open
  issues in the integration of internet of things and cloud computing, Journal
  of Network and Computer applications 67 (2016) 99--117.

\bibitem{alaa2017review}
M.~Alaa, A.~A. Zaidan, B.~B. Zaidan, M.~Talal, M.~L.~M. Kiah, A review of smart
  home applications based on internet of things, Journal of Network and
  Computer Applications 97 (2017) 48--65.

\bibitem{zanella2014internet}
A.~Zanella, N.~Bui, A.~Castellani, L.~Vangelista, M.~Zorzi, Internet of things
  for smart cities, IEEE Internet of Things journal 1~(1) (2014) 22--32.

\bibitem{guerrero2015integration}
J.~A. Guerrero-Ibanez, S.~Zeadally, J.~Contreras-Castillo, Integration
  challenges of intelligent transportation systems with connected vehicle,
  cloud computing, and internet of things technologies, IEEE Wireless
  Communications 22~(6) (2015) 122--128.

\bibitem{qadri2020future}
Y.~A. Qadri, A.~Nauman, Y.~B. Zikria, A.~V. Vasilakos, S.~W. Kim, The future of
  healthcare internet of things: a survey of emerging technologies, IEEE
  Communications Surveys \& Tutorials 22~(2) (2020) 1121--1167.

\bibitem{wollschlaeger2017future}
M.~Wollschlaeger, T.~Sauter, J.~Jasperneite, The future of industrial
  communication: Automation networks in the era of the internet of things and
  industry 4.0, IEEE industrial electronics magazine 11~(1) (2017) 17--27.

\bibitem{yang2017survey}
Y.~Yang, L.~Wu, G.~Yin, L.~Li, H.~Zhao, A survey on security and privacy issues
  in internet-of-things, IEEE Internet of Things Journal 4~(5) (2017)
  1250--1258.

\bibitem{frustaci2017evaluating}
M.~Frustaci, P.~Pace, G.~Aloi, G.~Fortino, Evaluating critical security issues
  of the iot world: Present and future challenges, IEEE Internet of things
  journal 5~(4) (2017) 2483--2495.

\bibitem{parno2005distributed}
B.~Parno, A.~Perrig, V.~Gligor, Distributed detection of node replication
  attacks in sensor networks, in: 2005 IEEE Symposium on Security and Privacy
  (S\&P'05), IEEE, 2005, pp. 49--63.

\bibitem{numan2020systematic}
M.~Numan, F.~Subhan, W.~Z. Khan, S.~Hakak, S.~Haider, G.~T. Reddy, A.~Jolfaei,
  M.~Alazab, A systematic review on clone node detection in static wireless
  sensor networks, IEEE Access 8 (2020) 65450--65461.

\bibitem{becher2006tampering}
A.~Becher, Z.~Benenson, M.~Dornseif, Tampering with motes: Real-world physical
  attacks on wireless sensor networks, in: International Conference on Security
  in Pervasive Computing, Springer, 2006, pp. 104--118.

\bibitem{raza2013svelte}
S.~Raza, L.~Wallgren, T.~Voigt, Svelte: Real-time intrusion detection in the
  internet of things, Ad hoc networks 11~(8) (2013) 2661--2674.

\bibitem{xing2008real}
K.~Xing, F.~Liu, X.~Cheng, D.~H. Du, Real-time detection of clone attacks in
  wireless sensor networks, in: 2008 The 28th International Conference on
  Distributed Computing Systems, IEEE, 2008, pp. 3--10.

\bibitem{lou2012single}
Y.~Lou, Y.~Zhang, S.~Liu, Single hop detection of node clone attacks in mobile
  wireless sensor networks, Procedia Engineering 29 (2012) 2798--2803.

\bibitem{alamer2020efficient}
A.~Alamer, An efficient group signcryption scheme supporting batch verification
  for securing transmitted data in the internet of things, Journal of Ambient
  Intelligence and Humanized Computing (2020) 1--18.

\bibitem{kittur2017batch}
A.~S. Kittur, A.~R. Pais, Batch verification of digital signatures: approaches
  and challenges, Journal of information security and applications 37 (2017)
  15--27.

\bibitem{hu2019autonomous}
J.-W. Hu, L.-Y. Yeh, S.-W. Liao, C.-S. Yang, Autonomous and malware-proof
  blockchain-based firmware update platform with efficient batch verification
  for internet of things devices, Computers \& Security 86 (2019) 238--252.

\bibitem{naccache1994can}
D.~Naccache, D.~M'Ra{\"I}hi, S.~Vaudenay, D.~Raphaeli, Can dsa be
  improved?—complexity trade-offs with the digital signature standard—, in:
  Workshop on the Theory and Application of of Cryptographic Techniques,
  Springer, 1994, pp. 77--85.

\bibitem{harn1998batch}
L.~Harn, Batch verifying multiple rsa digital signatures, Electronics Letters
  34~(12) (1998) 1219--1220.

\bibitem{karati2012batch}
S.~Karati, A.~Das, D.~Roychowdhury, B.~Bellur, D.~Bhattacharya, A.~Iyer, Batch
  verification of ecdsa signatures, in: International Conference on Cryptology
  in Africa, Springer, 2012, pp. 1--18.

\bibitem{abowd1999towards}
G.~D. Abowd, A.~K. Dey, P.~J. Brown, N.~Davies, M.~Smith, P.~Steggles, Towards
  a better understanding of context and context-awareness, in: International
  symposium on handheld and ubiquitous computing, Springer, 1999, pp. 304--307.

\bibitem{perera2013context}
C.~Perera, A.~Zaslavsky, P.~Christen, D.~Georgakopoulos, Context aware
  computing for the internet of things: A survey, IEEE communications surveys
  \& tutorials 16~(1) (2013) 414--454.

\bibitem{de2020context}
E.~de~Matos, R.~T. Tiburski, C.~R. Moratelli, S.~Johann~Filho, L.~A. Amaral,
  G.~Ramachandran, B.~Krishnamachari, F.~Hessel, Context information sharing
  for the internet of things: A survey, Computer Networks 166 (2020) 106988.

\bibitem{sezer2017context}
O.~B. Sezer, E.~Dogdu, A.~M. Ozbayoglu, Context-aware computing, learning, and
  big data in internet of things: a survey, IEEE Internet of Things Journal
  5~(1) (2017) 1--27.

\bibitem{zafar2020location}
F.~Zafar, A.~Khan, A.~Anjum, C.~Maple, M.~A. Shah, Location proof systems for
  smart internet of things: Requirements, taxonomy, and comparative analysis,
  Electronics 9~(11) (2020) 1776.

\bibitem{krishna2020location}
M.~B. Krishna, P.~Lorenz, Location, context, and social objectives using
  knowledge-based rules and conflict resolution for security in internet of
  things, IEEE Internet of Things Journal 8~(1) (2020) 407--417.

\bibitem{sun2017efficient}
G.~Sun, V.~Chang, M.~Ramachandran, Z.~Sun, G.~Li, H.~Yu, D.~Liao, Efficient
  location privacy algorithm for internet of things (iot) services and
  applications, Journal of Network and Computer Applications 89 (2017) 3--13.

\bibitem{zafar2020mobchain}
F.~Zafar, A.~Khan, S.~U.~R. Malik, A.~Anjum, M.~Ahmed, Mobchain: Three-way
  collusion resistance in witness-oriented location proof systems using
  distributed consensus, arXiv preprint arXiv:2011.08538 (2020).

\bibitem{brooks2007}
R.~Brooks, P.~Y. Govindaraju, M.~Pirretti, N.~Vijaykrishnan, M.~T. Kandemir, On
  the detection of clones in sensor networks using random key predistribution,
  IEEE Transactions on Systems, Man, and Cybernetics, Part C (Applications and
  Reviews) 37~(6) (2007) 1246--1258.
\newblock \href {https://doi.org/10.1109/TSMCC.2007.905824}
  {\path{doi:10.1109/TSMCC.2007.905824}}.

\bibitem{choi2007set}
H.~Choi, S.~Zhu, T.~F. La~Porta, Set: Detecting node clones in sensor networks,
  in: 2007 Third International Conference on Security and Privacy in
  Communications Networks and the Workshops-SecureComm 2007, IEEE, 2007, pp.
  341--350.

\bibitem{alsaedi2017detecting}
N.~Alsaedi, F.~Hashim, A.~Sali, F.~Z. Rokhani, Detecting sybil attacks in
  clustered wireless sensor networks based on energy trust system (ets),
  Computer communications 110 (2017) 75--82.

\bibitem{rikli2016lightweight}
N.-E. Rikli, A.~Alnasser, Lightweight trust model for the detection of
  concealed malicious nodes in sparse wireless ad hoc networks, International
  Journal of Distributed Sensor Networks 12~(7) (2016) 1550147716657246.

\bibitem{shanmugam2020two}
A.~Shanmugam, J.~Paramasivam, A two-level authentication scheme for clone node
  detection in smart cities using internet of things, Computational
  Intelligence 36~(3) (2020) 1200--1220.

\bibitem{yu2013localized}
C.-M. Yu, Y.-T. Tsou, C.-S. Lu, S.-Y. Kuo, Localized algorithms for detection
  of node replication attacks in mobile sensor networks, IEEE transactions on
  information forensics and security 8~(5) (2013) 754--768.

\bibitem{zhou2016improved}
Y.~Zhou, N.~Xiong, M.~Tan, R.~Huang, J.~Kleonbet, An improved mobility-based
  control protocol for tolerating clone failures in wireless sensor networks,
  Sensors 16~(11) (2016) 1955.

\bibitem{lee2018mdsclone}
P.-Y. Lee, C.-M. Yu, T.~Dargahi, M.~Conti, G.~Bianchi, Mdsclone:
  multidimensional scaling aided clone detection in internet of things, IEEE
  Transactions on Information Forensics and Security 13~(8) (2018) 2031--2046.

\bibitem{shaukat2014node}
H.~R. Shaukat, F.~Hashim, A.~Sali, M.~F. Abdul~Rasid, Node replication attacks
  in mobile wireless sensor network: a survey, International Journal of
  Distributed Sensor Networks 10~(12) (2014) 402541.

\bibitem{khalil2021identification}
U.~Khalil, A.~Ahmad, A.-H. Abdel-Aty, M.~Elhoseny, M.~W.~A. El-Soud, F.~Zeshan,
  Identification of trusted iot devices for secure delegation, Computers \&
  Electrical Engineering 90 (2021) 106988.

\bibitem{altaf2020robust}
A.~Altaf, H.~Abbas, F.~Iqbal, M.~M. Z.~M. Khan, M.~Daneshmand, Robust, secure
  and adaptive trust-oriented service selection in iot-based smart buildings,
  IEEE Internet of Things Journal (2020).

\bibitem{qolomany2020trust}
B.~Qolomany, I.~Mohammed, A.~Al-Fuqaha, M.~Guizani, J.~Qadir, Trust-based cloud
  machine learning model selection for industrial iot and smart city services,
  IEEE Internet of Things Journal (2020).

\bibitem{kittur2020trust}
A.~S. Kittur, A.~R. Pais, A trust model based batch verification of digital
  signatures in iot, Journal of Ambient Intelligence and Humanized Computing
  11~(1) (2020) 313--327.

\bibitem{janvcarsecurity}
J.~Jan{\v{c}}{\'a}r, Security considerations for elliptic curve domain
  parameters selection, Tech. rep., Bachelor’s Thesis, Masaryk University
  Faculty of Informatics (2015).

\bibitem{sommerseth2015pohlig}
M.~L. Sommerseth, H.~Hoeiland, Pohlig-hellman applied in elliptic curve
  cryptography, Tech. rep., Technical Report, University of California Santa
  Barbara (2015).

\bibitem{luca2004mov}
F.~Luca, D.~J. Mireles, I.~E. Shparlinski, et~al., Mov attack in various
  subgroups on elliptic curves, Illinois Journal of Mathematics 48~(3) (2004)
  1041--1052.

\bibitem{zhang2012efficient}
X.~Zhang, K.~Wang, D.~Lin, On efficient pairings on elliptic curves over
  extension fields, in: International Conference on Pairing-Based Cryptography,
  Springer, 2012, pp. 1--18.

\bibitem{hankerson2006guide}
D.~Hankerson, A.~J. Menezes, S.~Vanstone, Guide to elliptic curve cryptography,
  Springer Science \& Business Media, 2006.

\bibitem{cohen2005handbook}
H.~Cohen, G.~Frey, R.~Avanzi, C.~Doche, T.~Lange, K.~Nguyen, F.~Vercauteren,
  Handbook of elliptic and hyperelliptic curve cryptography, CRC press, 2005.

\bibitem{biehl2000differential}
I.~Biehl, B.~Meyer, V.~M{\"u}ller, Differential fault attacks on elliptic curve
  cryptosystems, in: Annual International Cryptology Conference, Springer,
  2000, pp. 131--146.

\bibitem{valenta2017measuring}
L.~Valenta, D.~Adrian, A.~Sanso, S.~Cohney, J.~Fried, M.~Hastings, J.~A.
  Halderman, N.~Heninger, Measuring small subgroup attacks against
  diffie-hellman., in: NDSS, 2017.

\bibitem{lim1997key}
C.~H. Lim, P.~J. Lee, A key recovery attack on discrete log-based schemes using
  a prime order subgroup, in: Annual International Cryptology Conference,
  Springer, 1997, pp. 249--263.

\bibitem{antipa2003validation}
A.~Antipa, D.~Brown, A.~Menezes, R.~Struik, S.~Vanstone, Validation of elliptic
  curve public keys, in: International workshop on public key cryptography,
  Springer, 2003, pp. 211--223.

\bibitem{fouque2008fault}
P.-A. Fouque, R.~Lercier, D.~R{\'e}al, F.~Valette, Fault attack on elliptic
  curve montgomery ladder implementation, in: 2008 5th Workshop on Fault
  Diagnosis and Tolerance in Cryptography, IEEE, 2008, pp. 92--98.

\bibitem{rogaway2004cryptographic}
P.~Rogaway, T.~Shrimpton, Cryptographic hash-function basics: Definitions,
  implications, and separations for preimage resistance, second-preimage
  resistance, and collision resistance, in: International workshop on fast
  software encryption, Springer, 2004, pp. 371--388.

\bibitem{anitha2021intelligent}
S.~Anitha, P.~Jayanthi, V.~Chandrasekaran, An intelligent based healthcare
  security monitoring schemes for detection of node replication attack in
  wireless sensor networks, Measurement 167 (2021) 108272.

\end{thebibliography}


\end{document}